\documentclass[prd,twocolumn,nofootinbib,showpacs,10pt]{revtex4}
\usepackage{textcomp}
\usepackage{amsmath}
\usepackage{amsfonts}
\usepackage{natbib}
\usepackage{calc}
\usepackage{mathrsfs}
\usepackage{amssymb}
\usepackage{amsmath}
\usepackage{color}
\usepackage{bm}
\usepackage[final,dvips]{graphicx}

\newcommand{\diff}[2]{\frac{d #1}{d #2}}
\newcommand{\pdiff}[2]{\frac{\partial #1}{\partial #2}}
\newcommand{\ddiff}[2]{\frac{d^2 #1}{\ d #2^2}}
\newcommand{\pddiff}[2]{\frac{\partial^2 #1}{\ \partial #2^2}}

\newcommand{\e}{\varepsilon}
\newcommand{\A}[2]{\hat A_{#1}\coeff{#2}}
\newcommand{\B}[2]{\hat B_{#1}\coeff{#2}}
\newcommand{\C}[2]{\hat C_{#1}\coeff{#2}}
\newcommand{\D}[2]{\hat D_{#1}\coeff{#2}}

\newcommand{\F}[2]{\hat F_{#1}\coeff{#2}}

\renewcommand{\H}[2]{\hat H_{#1}\coeff{#2}}
\newcommand{\I}[2]{\hat I_{#1}\coeff{#2}}
\newcommand{\K}[2]{\hat K_{#1}\coeff{#2}}
\newcommand{\nhat}{\hat{n}}
\newcommand{\rad}{\mathcal{R}}
\newcommand{\retr}{r_{\text{ret}}}
\newcommand{\advr}{r_{\rm adv}}

\newcommand{\tail}{h^{\rm tail}}
\newcommand{\order}[1]{O\!\left(#1\right)}
\newcommand{\etide}{\mathcal{E}}
\newcommand{\ein}{\widetilde{\mathcal{E}}}
\newcommand{\btide}{\mathcal{B}}
\newcommand{\bin}{\widetilde{\mathcal{B}}}

\newcommand{\hmn}[2]{h^{(#2)}_{#1}}

\newcommand{\del}[1]{\nabla_{\!\!#1}}
\newcommand{\Lie}[1]{\pounds_{\!#1}}

\DeclareMathOperator{\STF}{STF}
\newcommand{\trans}{\psi}
\newcommand{\map}{\varphi}
\newcommand{\man}{\mathcal{M}}
\newcommand{\exact}[1]{\mathsf{#1}}
\newcommand{\expand}{\Phi}

\newcommand{\coeff}[1]{^{(#1)}}
\newcommand{\dcoeff}[1]{_{(#1)}}
\newcommand{\an}[1]{a^{(#1)}}

\renewcommand{\t}{\tilde t}
\newcommand{\R}{\tilde R}

\begin{document}
\title{Singular perturbation techniques in the gravitational self-force problem} 
\author{Adam Pound} 
\affiliation{Department of Physics, University of Guelph, Guelph, Ontario, N1G 2W1}
\pacs{04.20.Cv, 04.25.Nx} 
\date{\today}

\begin{abstract}
Much of the progress in the gravitational self-force problem has involved the use of singular perturbation techniques. Yet the formalism underlying these techniques is not widely known. I remedy this situation by explicating the foundations and geometrical structure of singular perturbation theory in general relativity. Within that context, I sketch precise formulations of the methods used in the self-force problem: dual expansions (including matched asymptotic expansions), for which I identify precise matching conditions, one of which is a weak condition arising only when multiple coordinate systems are used; multiscale expansions, for which I provide a covariant formulation; and a self-consistent expansion with a fixed worldline, for which I provide a precise statement of the exact problem and its approximation. I then present a detailed analysis of matched asymptotic expansions as they have been utilized in calculating the self-force. Typically, the method has relied on a weak matching condition, which I show cannot determine a unique equation of motion. I formulate a refined condition that is sufficient to determine such an equation. However, I conclude that the method yields significantly weaker results than do alternative methods. 
\end{abstract}
\maketitle

\section{Introduction}\label{introduction}
Perturbation theory is a venerable field of study in general relativity (GR). In fact, because of the complexity of the Einstein field equation (EFE), most physically relevant analytical results in GR rely on perturbing away from a known solution. However, most of the foundational work in this area has focused only on descriptions of regular perturbation problems---problems in which a regular power-series expansion yields a uniform asymptotic approximation to a true solution. The underlying formalism of regular perturbation theory in GR has been studied extensively \cite{Geroch_limits, Stewart_limits, Bruni_limits}, and it has been shown that any regular asymptotic expansion of the field equations yields a perturbative solution that approximates at least one exact solution, at least locally \cite{perturbation_theory1, perturbation_theory2, Rendall_review}.

However, many physically interesting systems must be modeled as \emph{singular} perturbation problems---problems in which a regular power-series expansion fails to provide a uniform asymptotic approximation. Indeed, one of the most successful areas of research in GR, post-Newtonian theory, centers on a singular perturbation problem. As one would expect, the foundations of that particular problem have been studied extensively \cite{Rendall,PN_existence_review}, and it is now known that there exist a large class of exact solutions possessing post-Newtonian expansions \cite{PN_existence,PN_existence1,PN_existence2,PN_existence3}. But general discussions of singular perturbation theory in GR are lacking; Kates has provided the only such discussion~\cite{Kates_structure}, and his emphasis was on providing an overview of the geometrical structure of singular problems, foregoing any discussion of particular methods.

Along with the post-Newtonian expansion, another problem of great significance is also singular: the motion of an asymptotically small body through an external spacetime. This problem will be the primary focus of the present paper. Study of the point particle limit is less well developed than that of the Newtonian limit, and it has typically focused on proving that at leading order, a small, uncharged body behaves as a test mass, moving on a geodesic of the external background spacetime (see, e.g., Refs.~\cite{Infeld, Kates_motion, DEath_paper, DEath, Geroch_particle1, Geroch_particle2}). However, in recent years, the advent of gravitational wave astronomy has enjoined a need to go beyond the test particle approximation. Specifically, more accurate approximations are required to model extreme mass-ratio inspirals (EMRIs), in which a compact body (such as a neutron star or black hole) of mass $m\sim M_\odot$ spirals into a supermassive Kerr black hole of mass $M\sim (10^4\text{--}10^9)M_\odot$ lying at the center of a galaxy; see Refs.~\cite{Drasco_review, EMRI_review} for an overview of these systems. Extreme mass-ratio inspirals are a potentially important source of wave-signals for the planned gravitational-wave detector LISA \cite{LISA}, with a predicted rate of several to one hundred detectable events per year \cite{event_rate, event_rate2}.

For an EMRI, an expansion in the point particle limit roughly corresponds to an expansion in powers of the mass ratio $m/M\sim\e$. (For the remainder of this paper, I assume all variables have been scaled by a global lengthscale such as $M$, such that I can write, e.g., $m\sim\e$, where $\e$ is dimensionless.) At leading order in a regular expansion, the small body behaves as a point particle and moves on a geodesic of the spacetime of the large body. At sub-leading order, the metric perturbation generated by the body exerts a force on it, the dissipative part of which causes it to very slowly spiral into the large body. The acceleration of this inspiraling worldline, caused by the body's interaction with its own gravitational field, is called the gravitational self-force. A general, analytical expression for the self-force in an arbitrary vacuum background spacetime was first derived  by Mino, Sasaki, and Tanaka \cite{Mino_Sasaki_Tanaka} and Quinn and Wald \cite{Quinn_Wald}; that expression is now known as the MiSaTaQuWa equation.

This is a singular perturbation problem for two reasons. First, near the body, the first-order metric perturbation is that of a point particle, behaving as a delta function with support on the particle's worldline. If the body is anything other than a black hole, the error in the approximation (i.e., the difference between the exact and approximate metrics) will then be unbounded in a small neighborhood of the body; and even in the case of a black hole, the expansion will fail, since the second-order Einstein tensor will contain products of delta functions and hence be ill-defined as a distribution. Therefore, the approximation breaks down due to rapid changes, on the lengthscale $\sim\e$, near the body. The second reason the problem is singular is the existence of secular errors (i.e., ones that accumulate over time). If we use the first-order solution with a point particle source, then the Bianchi identity constrains the particle to move on a geodesic; at higher order in the expansion, this geodesic worldline is corrected by small deviation vectors \cite{Gralla_Wald, my_paper}. But the true path of the small body spirals into the large black hole, eventually deviating by a very large amount from the leading-order, geodesic worldline; generically, the error in the regular expansion will be unbounded on any time-interval $[0,1/\e^p]$, $p>0$. Therefore, the approximation also breaks down due to slow, cumulative changes over the radiation-reaction timescale $t_{rr}\equiv1/\e$. (In an EMRI, this is the time required for the particle's energy and angular momentum to undergo an order-$1$ change, since their rate of change is proportional to the self-force, itself of order $\e$.)

In singular perturbation theory, in order to overcome these types of errors, one makes use of \emph{general expansions},\footnote{In Ref.~\cite{my_paper}, I instead used the term ``singular expansions".} of the form
\begin{equation}\label{singular_expansion}
f(x,\e)=\sum_{n=0}^N\e^nf\coeff{\emph{n}}(x,\e) +\order{\e^{N+1}}.
\end{equation}
Unlike a regular power-series expansion, here the coefficients $f\coeff{\emph{n}}(x,\e)$ are allowed to depend on $\e$; but they are nevertheless constrained to be of order 1, in the sense that there exist positive constants $k$ and $\e_0$ such that $|f\coeff{\emph{n}}(x,\e)|\le k$ for $0\le\e\le\e_0$, but $\lim_{\e\to0} f\coeff{\emph{n}}(x,\e) \not\equiv 0$ (unless $f\coeff{\emph{n}}(x,\e)$ is identically zero). Put simply, the goal of a general expansion is to expand only \emph{part} of a function's $\e$-dependence, while holding fixed some specific $\e$-dependence that captures one or more of the function's essential features. There are two common types of general expansions: \emph{composite expansions} and \emph{multiscale expansions}. The first of these is designed to overcome the failure of a regular expansion near some submanifold (such as the position of the small body in an EMRI), while the second is designed to overcome secular errors (such as the deviation of the true motion from the leading-order, geodesic approximation).

Composite expansions are patched together from a finite number of regular expansions. For example, in the EMRI problem, if we use a rescaled radial coordinate $\tilde r\equiv r/\e$ near the body, then a regular expansion at fixed $\tilde r$ could be accurate on the scale $r\sim\e$, where an expansion at fixed $r$ fails; this \emph{inner expansion} can then be combined with an \emph{outer expansion} valid for $r\sim 1$, yielding a composite expansion that is uniformly accurate both near and far from the body. I will refer to a pair of inner and outer expansions such as this as \emph{dual expansions}. One can make use of dual expansions in a variety of ways, but historically, they have been used most often in the method of matched asymptotic expansions, in which the two expansions are first partially determined in their respective domains of validity, and then any remaining freedom in them is removed by insisting that they agree in a domain of mutual validity.\footnote{My nomenclature is not standard. In most of the literature on singular perturbation theory, the term ``matched asymptotic expansions" has the meaning that I have here assigned to dual expansions. However, I have opted to follow the usage in recent literature on the self-force, which has used the term to refer specifically to the procedure in which the field equations in the inner and outer expansions are first solved and then the solutions are made to match.} They were first used in fluid dynamics to analyze the behavior of a low-viscosity fluid near a boundary. In the context of GR, since the pioneering work of Burke~\cite{Burke}, who studied the effect of radiation-reaction on a post-Newtonian binary, and D'Eath~\cite{DEath_paper,DEath}, who studied the motion of black holes, dual expansions have mostly been utilized for two purposes: determining waveforms by matching wave-zone expansions to near-zone expansions (see, e.g., the review \cite{Blanchet_review}), and determining equations of motion for small bodies by matching an inner expansion near a body to an outer expansion in a larger region (see, e.g., \cite{Kates_motion,Kates_PN, Kates_Lorenz_force, Thorne_Hartle, PN_matching, Futamase_EM, Futamase_review}). The latter method, in various forms, has often been used to derive the MiSaTaQuWa equation \cite{Mino_Sasaki_Tanaka, Eric_matching, Gralla_Wald, my_paper, Fukumoto}. (See Ref.~\cite{my_paper} for a more thorough review.)

Composite expansions are suitable only when the different lengthscales dominate in different regions---e.g., the metric varies on the short lengthscale $\sim\e$ near the small body, while it varies on the background lengthscale $\sim\e^0$ everywhere else. In contrast, multiscale expansions are suitable in situations where multiple lengthscales are relevant everywhere in the region of interest. These expansions begin directly with the generalized form \eqref{singular_expansion}, where the coefficients $f\coeff{n}(x,\e)$ depend on some specific function of $\e$---for example, in a two-timescale expansion, $f\coeff{n}$ is assumed to depend on a time coordinate $t$ and a \emph{slow-time} coordinate $\tilde t=\e t$, which allows the expansion to capture both short-term and long-term effects. Recently, Hinderer and Flanagan \cite{Hinderer_Flanagan} have suggested a two-timescale expansion of the EFE and equation of motion in the EMRI problem; the method has also been applied to more restricted expansions of self-forced equations of motion \cite{other_paper, Mino_Price}.

In addition to these standard expansions, there has been one further general expansion utilized in the self-force problem: a self-consistent expansion with a fixed worldline. In this expansion, rather than allowing a functional dependence on an $\e$-dependent function from spacetime to $\mathbb{R}$, such as a slow-time coordinate, one allows a functional dependence on a function from $\mathbb{R}$ \emph{to} spacetime. This allows one to consider a metric perturbation that is a functional of an exact, $\e$-dependent worldline, bypassing the constraint that the worldline must be a geodesic at leading order. This idea underlies the methods of post-Newtonian theory, and it has been assumed in some form in much of the literature on the self-force, including in the earliest derivations of the MiSaTaQuWa equation \cite{Mino_Sasaki_Tanaka, Quinn_Wald}. In Ref.~\cite{my_paper}, this approach was first formalized in terms of a systematic approximation scheme. That scheme makes use of two general expansions: an inner expansion accurate near the small body, and an outer expansion accurate in the external background spacetime. In the outer expansion, the metric is treated as a functional of the exact worldline $\gamma$, and it is then expanded while holding that dependence fixed: $\exact{g}=g+\e\hmn{}{1}[\gamma]+\e^2\hmn{}{2}[\gamma]+...$. See Ref.~\cite{my_paper} for further details. Like a two-timescale expansion, this fixed-worldline approach promises to eliminate the secular errors of a regular expansion; however, the precise relationship between the two methods remains to be explored.

In this paper, I seek to accomplish two goals. My first goal is to extend the geometrical description of singular perturbation techniques. The geometrical picture of regular perturbation theory in GR is well known; the analogous description of singular perturbation theory, provided by Kates \cite{Kates_structure}, is less well known. In Sec.~\ref{perturbations_in_GR}, I review Kates' work and then go beyond it by describing in some detail the three techniques used in the self-force problem: dual expansions, multiscale expansions, and the self-consistent expansion presented in Ref.~\cite{my_paper}. I focus on making precise statements pertaining to these methods, with the aim of clarifying previous work. With regard to dual expansions, I identify two matching conditions that may be used in the method of matched asymptotic expansions---a strong matching condition analogous to the one used in traditional singular perturbation theory, and a weak matching condition that arises when working in multiple coordinate systems. With regard to multiscale expansions, I provide a covariant description. With regard to the self-consistent expansion, I extend the discussion of Ref.~\cite{my_paper} by providing an exact formulation of the self-force problem and a brief discussion of the gauge freedom in the expansion.

My second goal is to provide a precise formulation of the derivations of the gravitational self-force using matched asymptotic expansions. Section~\ref{matching} consists of a new version of this derivation, along with detailed analysis and discussion. Matching was used in some of the earliest derivations of the MiSaTaQuWa equation \cite{Mino_Sasaki_Tanaka,Eric_matching}. Because it can determine the first-order equation of motion from a first-order outer expansion, it is technically far less involved than the method of Refs.~\cite{Kates_motion, Gralla_Wald, my_paper}, which determine the $n$th-order equation of motion by solving the field equations to $n+1$st order in a region around the body. However, as I will discuss, the method as it has been utilized in the self-force problem has relied on the weak matching condition, which is too weak to actually determine an equation of motion. In order to determine an equation of motion, additional assumptions must be made. I pinpoint these assumptions and formulate a refined matching condition. Because of the variety of assumptions required, I conclude that matching yields weaker results than one would expect from traditional singular perturbation theory; and it yields weaker results than those obtainable through the more laborious approach of Refs.~\cite{Kates_motion, Gralla_Wald, my_paper}. The method can surely be improved, the number of assumptions reduced, but such improvement may prove more difficult than using alternative methods.

Before proceeding to these two goals, I begin in Sec.~\ref{traditional} with a review of traditional singular perturbation methods in applied mathematics. Such a review is warranted both to develop the concepts required in the later sections and to make the basics of the theory more widely known. More detailed overviews of the subject can be found in numerous textbooks (e.g., Refs.~\cite{Verhulst, Holmes, Kevorkian_Cole, Lagerstrom,Eckhaus}). Among these, the text by Kevorkian and Cole \cite{Kevorkian_Cole} covers the broadest range of topics, and the text by Eckhaus \cite{Eckhaus} provides the most rigorous treatment.

\section{Traditional singular perturbation theory}\label{traditional}
I begin by defining some useful notation. First, I define the following order symbols: for $x\in\mathbb R^n$,
\begin{itemize}
\item $f(x,\e)=\order{\zeta(\e)}$ if there exist positive constants $k$ and $\e^*$ such that $|f(x,\e)|\le k|\zeta(\e)|$ for fixed $x$ and $0\le\e\le\e^*$.
\item $f(x,\e)=o(\zeta(\e))$ if $\displaystyle\lim_{\e\to0}\frac{f(x,\e)}{\zeta(\e)}=0$ at fixed $x$.
\item $f(x,\e)=O_s(\zeta(\e))$ if $f(x,\e)=O(\zeta(\e))$ and $f(x,\e)\ne o(\zeta(\e))$.
\end{itemize}
For example, $5\e+2\e^{3/2}=\order{\e}=2\e^{3/2}$, $5\e+2\e^{3/2}\ne o(\e)=2\e^{3/2}$, and $5\e+2\e^{3/2}=O_s(\e)\ne 2\e^{3/2}$.

In general, we are concerned with the asymptotic behavior of functions, rather than the behavior of functions evaluated at particular locations. This means we need a norm appropriate for a function. Also, a central issue in perturbation theory is whether or not an approximation is uniformly accurate in a region of interest, where uniformity is defined as follows:
\begin{itemize}
\item $f(x,\e)=\order{\zeta(\e)}$ \emph{uniformly} in a region $D$ if there exist positive constants $k$ and $\e^*$ such that $||f(x,\e)||_D\le k|\zeta(\e)|$ for $0\le\e\le\e^*$, where $||\cdot||_D=\sup_{x\in D}|\cdot|$.
\end{itemize}
Analogous definitions hold for $o$ and $O_s$. These definitions provide a more useful measure of the asymptotic behavior of a function.

Finally, I define several important asymptotic quantities: relative to an asymptotic sequence $\lbrace\zeta_n(\e)\rbrace$, where $\zeta_{n+1}(\e)=o(\zeta_n(\e))$,
\begin{itemize}
\item $f(x,\e)$ is an $N$th-order \emph{asymptotic approximation} of $\exact{f}(x,\e)=O_s(1)$ if $f(x,\e)-\exact{f}(x,\e)=o(\zeta_N(\e))$,\footnote{In the case $\exact{f}(x,\e)=O_s(\zeta_k(\e))$, functions would be rescaled by $\zeta_k(\e)$ before making comparisons.}
\item $f(x,\e)$ is an $N$th-order \emph{asymptotic solution} to a differential equation $\exact{D}[\exact{f}(x,\e)]=0$ if $\exact{D}[f(x,\e)]=o(\zeta_N(\e))$,
\item $f(x,\e)=\displaystyle\sum_{n=0}^N \zeta_n(\e) f\coeff{\emph{n}}(x,\e)$, where $f\coeff{\emph{n}}(x,\e)=O_s(1)$, is an $N$th-order \emph{asymptotic series}. If $f\coeff{\emph{n}}$ is independent of $\e$, then the series is said to be \emph{regular} (sometimes called Poincar\'e-type); if not, then it is \emph{general}. If, in addition, $f$ is an asymptotic approximation to $\exact{f}$, then it is an $N$th-order \emph{asymptotic expansion} of $\exact{f}$.
\end{itemize}
The most common asymptotic sequence is $\lbrace\e^n\rbrace$, which I will use almost exclusively in this paper. Note that we are typically uninterested in whether or not an asymptotic series converges as $N\to\infty$. In fact, even if $f$ is an asymptotic series that both converges and asymptotically approximates $\exact{f}$, the function that it converges to might not be $\exact{f}$.

In any given perturbation calculation, one almost always calculates an asymptotic solution to an equation. Determining whether or not an asymptotic solution is also an asymptotic approximation to an exact solution is typically far more difficult. It is, however an essential step in proving the reliability of an expansion, because an asymptotic solution may not be an asymptotic approximation to an exact solution (more pathologically, an asymptotic approximation may not be an asymptotic solution \cite{Verhulst}).

General asymptotic expansions are a powerful tool for solving singular perturbation problems, which are defined by the failure of a regular expansion to provide a uniform approximation. This failure is often signaled by a change of character in the governing differential equation when $\e\to0$: for example, a hyperbolic equation might degenerate into a parabolic equation. The inaccuracy of a regular expansion is also frequently signaled by its failure to satisfy a given boundary condition, or by the expansion growing without bound in a system that we have reason to believe should be bounded. Typically, the underlying origin of the failure is the presence of distinct length scales, one of which appears only for $\e>0$. Often, this means that the exact solution to a problem is singular at $\e=0$. Hence, in singular perturbation problems, we assume that $\e\in(0,\e^*]$, which allows us to take the limit $\e\to 0$, but which prevents us from setting $\e=0$. In this section, I consider two types of systems: first, systems in which the exact solution undergoes a rapid change near a submanifold; second, systems in which rapid changes occur throughout the region of interest. In the first type of system, dual expansions can be used to construct a uniform general expansion; in the second type, a multiscale expansion can be used.

Before proceeding, I define two final pieces of notation. $\trans_*$ and $\trans^*$ denote, respectively, the push-forward and pull-back corresponding to a map $\trans$. So, for example, if $f$ is a function of coordinates $x$, and $\tilde x=\trans(x)$, then $\trans_* f$ is the function rewritten in terms of $\tilde x$. $\expand^N_\e\exact{f}$ denotes the $N$th-order regular asymptotic expansion of $\exact{f}$ in the limit of small $\e$, holding fixed the coordinates of which $\exact{f}$ is a function. 

\subsection{Dual expansions}
\begin{figure}
\begin{center}
\includegraphics[scale=0.83]{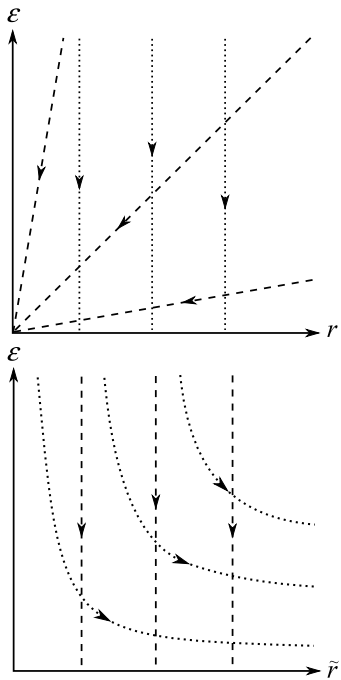}
\end{center}
\caption[Inner and outer limits]{Upper: inner limit (dashed curves) and outer limit (dotted curves) in the $(r,\e)$-plane. Lower: the same limits in the $(\tilde r, \e)$-plane. The inner limit is defined by $\e\to0$, $\tilde r=r/\e$ fixed; the outer limit, by $\e\to0$, $r$ fixed. From the perspective of the inner limit, the outer limit sends all points to infinity ($\tilde r\to\infty$). From the perspective of the the outer limit, the inner limit sends all points to zero ($r\to 0$).}
\label{limits}
\end{figure}

Dual expansions are typically used to solve boundary value problems in which the solution exhibits rapid change in a very small region (or a finite number of such regions). The regions of rapid change are referred to as boundary layers. Frequently, this rapid change prevents a regular expansion from satisfying a given boundary condition, though a ``boundary layer" can sometimes arise away from any boundary. The usual means of solving these problems is to make use of two regular expansions: an inner expansion $f_{\text{in}}$ that is expected to be valid in the boundary layer, and an outer expansion $f_{\text{out}}$ that is expected to be valid outside of it. Suppose we have a one-dimensional problem with coordinate $r$, and that the boundary layer is at $r=r_b$ and has a thickness $\sim\delta(\e)$. Then the outer expansion is simply a regular series at fixed $r$, and the inner expansion is a regular series at fixed values of the rescaled coordinate $\tilde r \equiv (r-r_b)/\delta(\e)$; this can be written as $\trans_\e:r\mapsto\tilde r$. The inner expansion allows us to capture changes over the lengthscale $\delta(\e)$, since $\tilde r$ is of order unity when the original coordinate $r$ is of order $\delta(\e)$. Note that if we treat the problem on a two-dimensional plane with coordinates $(r,\e)$, then the inner and outer expansions can be visualized as expansions along flow lines defined by $r=$constant and $r/\e=$constant, as shown in Fig.~\ref{limits}.

For simplicity, suppose that $r_b=0$ and that boundary data is given at $r=0$ and $r=1$. In this case, the outer expansion typically fails to satisfy the boundary condition at $r=0$, but it can be made to satisfy the condition at $r=1$; conversely, the inner expansion can satisfy only the condition at $r=0$. This leaves each of the expansions underdetermined. The basic idea of dual expansions is to fully determine them by insisting that they agree in some region of mutual validity. Suppose that $f_{\rm in}$ is an $N$th-order asymptotic approximation of $\exact{f}$ in a region $D_{\text{in}}$, and $f_{\text{out}}$ is an $N$th-order asymptotic approximation in a region $D_{\rm out}$. Then by definition, $\exact{f}-f_{\text{in}}=o(\zeta_N(\e))$ in $D_{\text{in}}$ and $\exact{f}-f_{\text{out}}=o(\zeta_N(\e))$ in $D_{\text{out}}$. Subtracting the second equation from the first, we have the \emph{overlap matching condition}:
\begin{equation}
f_{\text{out}}-f_{\text{in}} = o(\zeta_N(\e)) \text{ in } D_{\text{out}}\cap D_{\text{in}}.
\end{equation}

Note that this condition relies on the existence of the \emph{overlap region} $D_{\text{out}}\cap D_{\text{in}}$. If that region is empty, then the condition is vacuous. And it may not be obvious that such a region ever exists, since the inner expansion trivially appears to be valid only for $\tilde r\sim 1$, and the outer expansion only for $r\sim 1$. However, if $f_{\text{out}}(r)$ is a uniform asymptotic approximation to $\exact{f}(r)$ on an interval $[a,1]$ for constant $a$, then it is also a uniform approximation on the extended interval $[\zeta_i(\e),1]$ for some $\zeta_i(\e)=o(1)$; similarly, if $f_{\text{in}}(\tilde r)$ is a uniform approximation to $\exact{f}(\tilde r)$ on $[0,b]$, then it is a uniform approximation on the extended interval $[0,1/\zeta_j(\e)]$ for some $\zeta_j(\e)=o(1)$ \cite{Lagerstrom, Eckhaus}. If we have access to the exact function $\exact{f}$, then we can explicitly determine the overlap of these extended regions. But in a typical application, without access to an exact solution, we must make use of the \emph{overlap hypothesis}, which states that the overlap region exists. In order to implement the overlap matching condition, one then simply assumes that the constructed asymptotic series of a given order are asymptotic approximations of the same order, and one then takes the overlap region to be the region in which $f_{\text{out}}-f_{\text{in}} = o(\zeta_N(\e))$.

In this paper, I will not make direct use of the overlap matching condition. Instead, I will use a second, simpler matching condition, which I will refer to as the \emph{coefficient-matching condition}:
\begin{equation}\label{matching_equation}
\expand^k_{\e}\trans^*\expand^m_{\e}\trans_*\exact{f} = \expand^k_{\e}\trans^*\expand^m_{\e}\trans_*\expand^m_{\e}\exact{f}.
\end{equation}
In this matching condition, we match results term-by-term in the expansions. On the left-hand side we have the inner expansion ($\expand^m_{\e}\trans_*\exact{f}$) written as a function of $r$ (via $\trans^*$) and then expanded in the outer limit; on the right-hand side, we have the outer expansion ($\expand^m_{\e}\exact{f}$) written as a function of $\tilde r$ (via $\trans_*$) and expanded in the inner limit, and then rewritten as a function of $r$ and re-expanded. (The right-hand side requires an extra expansion in order to remove terms that would appear as higher-order terms in the inner expansion. Refer to Appendix~\ref{matching_example} for an illustrative example.) We can write this schematically as
\begin{equation}
\expand_\e f_{\text{in}}(r)=\expand_r f_{\text{out}},
\end{equation}
meaning that when the inner expansion is re-expanded for small $\e$ at fixed $r$, and the outer expansion is re-expanded for small $r$ at fixed $\e$, the two results must agree term by term. We can then, for example, equate coefficients of $\e^n r^m$ on the left- and right-hand sides.

If we define the \emph{buffer region} by the inequalities $\e\ll r\ll 1$,\footnote{In the applications of matched asymptotic expansions in GR, the meanings that I have assigned to the terms ``overlap region" and ``buffer region" are often conflated, and the terms are often used interchangeably. For the sake of clarity, I distinguish between the two.} this equation states that the inner and outer expansions must agree term by term when they are both expanded in the buffer region. In other words, if the exact solution is expanded first for small $\e$ at fixed $r/\e$ (yielding an inner expansion), and then expanded at fixed $r$ (or in other words, for $r\gg\e$), it must agree, term by term, with the result of expanding first for small $\e$ at fixed $r$ and then expanding for $r\ll 1$.

From the perspective of the inner limit, the buffer region lies at asymptotic infinity ($\tilde r\to\infty$); from the perspective of the outer expansion, it lies asymptotically close to $r=0$. From this we can intuit a still simpler matching condition, which I will call the \emph{asymptotic matching condition}:
\begin{equation}
\lim_{\tilde r\to\infty}f\coeff{0}_{\text{in}}=\lim_{r\to0}f\coeff{0}_{\text{out}},
\end{equation}
where $f\coeff{0}_{\text{in}}$ and $f\coeff{0}_{\text{out}}$ are the leading-order terms in, respectively, the inner and outer expansions.

The three matching conditions I have discussed are obviously related. In fact, one can derive the asymptotic matching condition and (a condition similar to) the coefficient-matching condition from the overlap hypothesis. However, one should realize that the overlap hypothesis is merely sufficient to arrive at those two matching conditions: it is not necessary. Functions exist that do not satisfy the overlap hypothesis but nevertheless satisfy the coefficient-matching condition, for example \cite{Eckhaus}.

Once a matching condition has been used to fully determine the inner and outer expansions, one can construct a \emph{composite expansion} that is uniformly accurate on the full domain of the problem. This expansion consists of the sum of the inner and outer expansions, minus the terms that are common to both in the buffer region. Explicitly,
\begin{equation}\label{composite}
f_{\text{comp}} = \expand_\e^m\exact{f} + \trans^*\expand_\e^m\trans_*\exact{f} - \trans^*\expand_\e^m\trans_*\expand_\e^m\exact{f},
\end{equation}
which we can write schematically as
\begin{equation}
f_{\text{comp}} = f_{\text{out}}+f_{\text{in}}-\expand_r f_{\text{out}}.
\end{equation}
Note that this is a general asymptotic expansion of the form $\sum_n\zeta_n(\e)F\coeff{n}(r,r/\delta(\e))$.

The matching conditions presented here can be used in a variety of ways. Most traditionally, in the method of matched asymptotic expansions, the inner and outer solutions are determined as far as possible using the governing differential equation and boundary conditions, and then they are fully determined by imposing a matching condition. However, one can instead begin with the matching conditions to restrict the general form of one or the other of the expansions, or to develop a general expansion in the buffer region. In Appendix~\ref{matching_example}, I illustrate the method of matched asymptotic expansions with a simple example. For further information on dual expansions, see Refs.~\cite{Eckhaus, Lagerstrom}.

\subsection{Multiscale expansions}\label{multiple scales}
In some systems, rather than a rapid change occurring near a submanifold, rapid changes occur over the entire region of interest (in other words, the ``boundary layers" are dense in the region). When studying these systems, one cannot form a uniform approximation by making use of regular expansions in separate regions and then combining them. Instead, one must assume a general expansion from the start.

Suppose for simplicity that the rapid changes occur on a scale $\sim 1$ and the slow changes occur on a scale $\sim 1/\e$,\footnote{One could rescale the variables such that the rapid changes occur on the scale $\sim\e$ and the slow changes on the scale $\sim 1$, to accord with the description of a region dense with boundary layers.} and that we seek an approximation to $\exact{f(t,\e)}$ that is uniform on the time-interval $[0,1/\e]$. Then we proceed by introducing a fast time variable $\phi=\phi(t,\e)$ and a slow time variable $\t=\t(t,\e)$ satisfying $\pdiff{\phi}{t}=\omega(t,\e)$ and $\pdiff{\t}{t}=\e\tilde\omega(t,\e)$, where $\omega$ and $\tilde\omega$ are uniformly $O_s(1)$; changes in $\phi$ are of the same order as changes in $t$, while $\t$ changes appreciably only after $t$ changes by a very large amount. (In the simplest case, we have $\phi=t$ and $\t=\e t$.) We invert the transformation in order to write the frequencies as functions of the slow time alone: $\omega(\t,\e)$ and $\tilde\omega(\t,\e)$. I next note that while setting $\omega=1$ will lead to large errors on a timescale $1/\e$ (consider, for example, attempting to approximate $\cos(t+\e t)$ by $\cos t$), setting $\tilde\omega=1$ will lead to large errors only on extremely long timescales outside our range of interest. Hence, I will take the slow time to be given by $\t=\e t$. The remaining frequency, $\omega$, must be determined over the course of the calculation. To make such a goal feasible, I assume $\omega$ possesses a regular expansion $\sum_{n\ge0}\zeta_n(\e)\omega\coeff{\emph{n}}(\t)$.

I next assume that $\exact{f}(t,\e)$ can be written as a function $F(\phi,\t,\e)$, and that $F$ possesses a regular expansion: that is,
\begin{equation}\label{two-time expansion}
\exact{f}(t,\e)=F(\phi,\t,\e)=\sum_{n}\zeta_n(\e)F\coeff{\emph{n}}(\phi,\t).
\end{equation}
Suppose $\exact{f}$ is to satisfy some differential equation $\exact{D}[\exact{f}]=0$. After making the substitution $\exact{f}(t,\e)=F(\phi,\t,\e)$, we use the chain rule to convert derivatives with respect to $t$ into the sum of partial derivatives $\frac{d}{dt}=\omega(\t,\e)\frac{\partial}{\partial\phi} + \e\frac{\partial}{\partial\t}$. We then arrive at a partial differential equation in terms of $\phi$ and $\t$. Now, the essential step in a multiscale expansion consists of treating $\phi$ and $\t$ as independent variables at this point; that is, $F$ is taken to be a solution to the PDE for arbitrary values of $\phi$ and $\t$. Given this assumption, after substituting the expansions for $F$ and $\omega$, we can solve the equation by setting the coefficient of each $\zeta_n$ to zero. If I did not assume that $F$ solves the equation for arbitrary $\phi$ and $\t$, then the $\e$-dependence scattered throughout $\phi(t,\e)$ and $\t(t,\e)$ would prevent us from solving the equation in this manner.

\begin{figure}
\begin{center}
\includegraphics[scale=0.83]{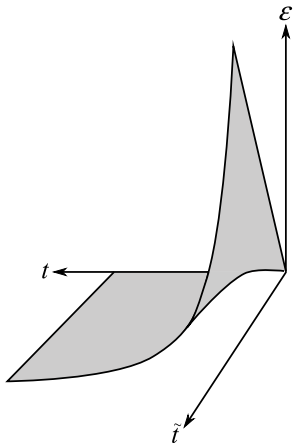}
\end{center}
\caption[The extended manifold of a multiscale expansion.]{In a multiscale expansion, we work on a manifold of larger dimension than that on which the original problem is posed. The solution is eventually evaluated on the submanifold defined by $\t=\e t$, shown here in grey.}
\label{submanifold}
\end{figure}

Treating $\phi$ and $\t$ as independent is equivalent to working on an enlarged manifold with coordinates $(\phi,\t,\e)$. The solution manifold on which $\exact{f}$ lives is a submanifold defined by $\phi=\phi(t,\e)$ and $\t=\t(t,\e)$. (See Fig.~\ref{submanifold} for an illustration in the simple case where $\phi=t$ and $\t=\e t$.) Determining $\omega$ can be viewed as a step in determining this submanifold; in fact, we can note that the transformation from the extrinsic coordinates $x^\alpha=(\phi,\t,\e)$ to the intrinsic coordinates $y^a=(t,\e)$ defines a set of basis vectors $e^\alpha_a$ on the submanifold, given by $e^\alpha_t=(\omega,\e,0)$ and $e^\alpha_\e=(\partial_\e\phi,t,1)$.

Because we are provided with sufficient boundary data for an ODE, rather than for a PDE, we must place some constraints on the function $F$. The most commonly imposed constraint is the \emph{non-secularity} condition, which says that integration constants must be chosen such that any secularly growing term vanishes. Other possible constraints include the demand that each coefficient $F\coeff{\emph{n}}(\phi,\t)$ is a periodic function of $\phi$, and the demand that each coefficient be unique. Obviously, one must apply such constraints judiciously and systematically.

Of course, this procedure relies on a host of assumptions. There is no guarantee that the exact solution $\exact{f}$ possesses an asymptotic expansion of the form \eqref{two-time expansion}; and even if it does, there is no guarantee that the terms in the expansion will necessarily solve the equation for arbitrary $\phi$ and $\t$. However, this method is extremely successful in practice. If one instead assumes a regular expansion, then the dependence of $\exact{f}$ on $\t$ will be expanded in powers of $\e t$. These powers of $t$ will eventually grow large, such that terms initially supposed to be high order become as large as the lower-order terms, preventing the expansion from providing a uniform approximation. In many cases, one can avoid this secular growth by using a rigorous method of averaging, which removes the rapid time dependence and recovers only the leading-order slow time-dependence. However, if one requires the fast time dependence as well, then the two-time method offers the most powerful means of doing so.

In Appendix~\ref{multiscale_example}, I illustrate the above ideas with an example. In that example, I demonstrate that even if the assumptions of the multiscale expansion fail at some order, the lower order solution can still yield a uniform asymptotic approximation; in addition, the failure of the assumptions is made manifest over the course of the calculation. For further information on multiscale expansions, see Ref.~\cite{Kevorkian_Cole}.

\subsection{Singular versus regular perturbation theory}
We should now take note of the essential differences between regular perturbation techniques and singular perturbation techniques. When a regular expansion of an exact solution $\exact{f}$ is substituted into a differential equation $\exact{D}[\exact{f}]=0$, the coefficients in the expansion are guaranteed to solve a hierarchy of differential equations, simply by setting the coefficients of each power of $\e$ to zero. Hence, when constructing a regular series solution to a differential equation, one can determine each term in the solution solely from the given differential equation (and its attendant boundary conditions). But a \emph{general} expansion of an exact solution is not guaranteed to satisfy any such hierarchy, because the coefficients in the expansions depend on $\e$. Hence, when constructing a general series solution to a differential equation, one must impose some extra conditions upon it---e.g., satisfying the overlap hypothesis in the method of matched asymptotic expansions, or satisfying a PDE rather than an ODE in the method of multiple scales---which are not guaranteed to be satisfied given only the form of the general expansion.

This means that proving general properties of solutions is much more difficult using singular perturbation theory. In regular perturbation theory, one can construct proofs of the form ``given an exact solution to such and such a boundary value problem, if it possesses a regular asymptotic expansion then that expansion has such and such behavior"; in singular perturbation theory, we must append further hypotheses to this statement. However, if we seek a very strong statement about the solution to a problem, we must in any case go beyond the form of such a proof: we must also prove that the exact solution actually \emph{does} possess the assumed expansion. This is a difficult feat regardless of whether the assumed expansion is regular or general. While it is usually easier in the case of regular expansions, techniques do exist for handling singular perturbation problems (see Ref.~\cite{Eckhaus} for examples). Furthermore, general expansions provide asymptotic solutions where regular series cannot, and they provide \emph{uniform} asymptotic solutions. Hence, in most cases of interest, their advantages far outweigh any disadvantages.

\section{Perturbation theory in General Relativity}\label{perturbations_in_GR}
In GR we typically do not begin with a predetermined manifold with predetermined boundary conditions that uniquely determine an exact solution. Instead, the manifold is (mostly) determined by the leading order ``background" solution to the Einstein equations. Within that manifold, we define boundary conditions that uniquely determine the perturbations. This somewhat complicates the problem, but it also makes the assumptions of singular perturbation theory more reasonable: since we do not seek an approximation to a \emph{unique} exact solution to a given boundary value problem, but only an approximation to \emph{some} exact solution to the EFE, it is eminently reasonable to impose the supplementary conditions required to construct general expansions.

\subsection{Regular perturbation theory}
Before describing singular perturbation theory in GR, I will briefly review regular perturbation theory. In its most geometric description, the formalism begins with a 5D manifold $\mathcal{N}$ carrying a 5D metric $\exact{g'}^{\mu\nu}$ of signature $(0,-,+,+,+)$ and a non-negative scalar field $\e:\mathcal{N}\to\mathbb{R}$. The manifold is foliated by 4D submanifolds $\man_\e$ defined by $d\e=0$, such that $\mathcal{N}\sim \man_{\e}\times\mathbb{R}$. When restricted to act on dual vectors tangent to $\man_\e$, $\exact{g'}^{\mu\nu}$ can be inverted, inducing a 4D Lorentzian metric $\exact{g}_\e$. Each member of the family of metrics $\lbrace\exact{g}_\e\rbrace$ is taken to be an exact solution of Einstein's equation at fixed $\e$. A regular expansion of the pair $(\man_\e,\exact{g}_\e)$ is an expansion around a known ``base" pair $(\man_0,g=\exact{g}_0)$. This expansion is performed by first defining a one-to-one relationship between points on $\man_0$ and $\man_\e$ via a diffeomorphism $\varphi_\e:\man_0\to\man_\e$, called an identification map. This map induces a flow on $\mathcal{N}$ with a tangent vector field $X$ that is non-vanishing and nowhere tangent to the submanifolds $\man_\e$. (Note that one could instead begin with a vector field and derive from it an identification map, but beginning with the identification map will be more useful in formalizing general expansions.)

In this context, the regular expansion $\exact{g}_\e=g+\sum_{n\ge 1}\e^n\hmn{}{\emph{n}}$ is given by an expansion along the flow induced by $X$:
\begin{align}
\phi_{\e}^*(\exact{g}_\e) &= e^{\e\!\Lie{X}}\exact{g}\big|_{\man_0},
\end{align}
where $\phi_{\e}^*\exact{g}_\e$ is the pull-back of $\exact{g}_\e$ onto the base manifold $\man_0$, and $\Lie{X}$ is the Lie derivative along the vector $X$. The background metric and the perturbations of it have clear geometrical interpretations: the background metric $g$ is the restriction of $\exact{g}_\e$ to the submanifold defined by $\e=0$; the first-order perturbation $\e \hmn{}{1}\equiv \e(\Lie{X}\exact{g})\big|_{\e=0}$ is the product of the ``distance" $\e$ along a flow line and the rate of change of $\exact{g}$ in the direction of the flow; and so on.

A choice of gauge corresponds to a choice of identification map $\map_\e$. A different choice, say $\psi_\e$, leads to a different tangent vector field $Y$, which in turn leads to a change
\begin{equation}
\psi_{\e}^*(\exact{g}_\e)-\phi_{\e}^*(\exact{g}_\e) = (e^{\e\!\Lie{Y}}-e^{\e\!\Lie{X}})\exact{g}\big|_{\man_0}.
\end{equation}
By expanding the exponentials, one finds that this induces changes $\hmn{}{\emph{n}}\to \hmn{}{\emph{n}}+\Delta \hmn{}{\emph{n}}$. At first and second order, the changes are given explicitly by
\begin{align}
\Delta \hmn{}{1} & = \Lie{\xi\dcoeff{1}}g, \label{gauge_trans1}\\
\Delta \hmn{}{2} & = \tfrac{1}{2}(\Lie{\xi\dcoeff{2}}+\Lie{\xi\dcoeff{1}}^2)g +\Lie{\xi_{(1)}}\hmn{}{1},\label{gauge_trans2}
\end{align}
where $\xi\dcoeff{1}\equiv(Y-X)\big|_{\e=0}$ and $\xi\dcoeff{2}\equiv[X,Y]\big|_{\e=0}$ are vector fields in the tangent bundle of $\man_0$. Note that $\xi\dcoeff{1}$ and $\xi\dcoeff{2}$ are linearly independent, so they can be chosen independently. In terms of coordinates, they correspond to the near-identity transformation
\begin{align}
x^\alpha \to x'^\alpha &= x^\alpha-\e\xi^\alpha_{(1)} +\tfrac{1}{2}\e^2\left(\xi^\alpha_{(1),\beta}\xi^\beta_{(1)}-\xi^\alpha_{(2)}\right) +\order{\e^3},
\end{align}
where the components on the right hand side are in the original coordinates $x^\alpha$. We say that the vectors $\xi\dcoeff{\emph{n}}$ are the generators of the gauge transformation. (See Ref.~\cite{Bruni_limits} for the precise meaning of this phrase.)

\subsection{Singular perturbation theory}
Although singular perturbation techniques have been utilized in many calculations in GR, the only formal description of them was provided by Kates \cite{Kates_structure}. I will review his description in this section, before extending it in the following sections. A singular perturbation problem is characterized by the limit $\e\to 0$ being singular: $\exact{g}_\e$ may not exist at $\e=0$, the topology or dimension of $\man_\e$ may change between $\e=0$ and $\e>0$, etc. This means that the 5D manifold $\mathcal{N}$ does not in general contain a ``base'' manifold $\man_0$; instead, it is given by $\mathcal{N}\sim\man_\e\times(0,\e^*]$. Hence, one cannot generically build an approximation on the limiting manifold by finding derivatives of the exact metric at $\e=0$. Instead, one works on a ``model manifold'' $\man_M$, on which one constructs a family of approximate solutions
\begin{equation}
g_M(\e)=g(\e)+\sum_{n\ge 1}\e^n h\coeff{n}(\e).
\end{equation}
The topology of the model manifold is taken to be compatible with the leading-order metric $g(\e)$. If there exists an identification map $\map_\e:\man_M\to\man_\e$, which maps a region $\mathcal{U}_M\subset\man_M$ to a region $\mathcal{U}_\e\subset\man_\e$, such that $g_M(\e)$ uniformly approximates $\map^*_\e \exact{g}_\e$ in the region $\mathcal{U}_M$ as $\e\to 0$, then $g_M(\e)$ is a uniform asymptotic approximation (as measured in some suitable norm) to the exact solution in the region $\mathcal{U}_\e$. Once again, the identification map induces a family of curves in the 5D manifold $\mathcal{N}$, but these curves will not, in general, continue smoothly to a base manifold $\man_0$.

As an example, consider a post-Newtonian expansion, which is singular~\cite{Futamase_Schutz, Futamase_particle1, Futamase_review, Rendall}. The Newtonian limit is given by $\e=v/c\to 0$, where $v$ is the supremum of the velocities in the system. In this limit, the light cones of the spacetime fold out into spatial surfaces, and the time-components of the metric blow up---alternatively, if we consider the inverse metric, we see that its time components vanish, such that it degenerates into a 3D spatial metric. Hence, we can infer that the manifold defined by $\e=0$ corresponds to the 3D spatial manifold of Newtonian theory.\footnote{The singular nature of the Newtonian limit is also signaled by the fact that hyperbolic wave equations become elliptic Poisson equations as the speed of gravity's propagation becomes infinite.} In this case, the model manifold and background metric are taken to be those of Minkowski spacetime.

In the next two subsections, I formulate dual expansions and multiscale expansions within this framework. The description of dual expansions follows that given by Kates \cite{Kates_structure}, which built on the work of D'Eath \cite{DEath, DEath_paper}; however, I more carefully formulate the matching conditions, specifically stressing the distinction between the strong matching condition (used in traditional singular perturbation theory) and a weak matching condition (often used in GR). My discussion of multiscale expansions is original to this work.

In the final subsection, I formulate the self-force problem as a free-boundary value problem. I then discuss means of solving the problem within the context of a self-consistent expansion with a fixed worldline.

\subsection{Dual expansions}\label{matching_conditions}
Though most of the description in this section carries over to a more general situation, I will restrict it to the pertinent case of a family of exact solutions $\exact{g}_\e$ containing a body of mass $\sim\e$, on a family of manifolds $\man_\e$. In this section, I only sketch the formalism of inner and outer limits for this system; in Secs.~\ref{fixed_worldline_formulation}--\ref{matching}, I provide further discussion of concrete applications of these limits.

Suppose that we are given two coordinate systems on $\man_\e$: a local coordinate system $X^\alpha=(T,R,\Theta^A)$ that is centered (in some approximate sense) on the small body, and a global coordinate system $x^\alpha$. For example, in an EMRI, the global coordinates might be the Boyer-Lindquist coordinates of the supermassive Kerr black hole (though we could consider the case in which both coordinate systems are centered on the small body); the local coordinates might be Schwarzschild-type coordinates for the small body. The local coordinates cover some region $D_I$ around (and possibly inside) the body, while the global coordinates cover a larger region $D_E$ outside the body. Assume, without loss of generality, that the two coordinate systems have overlapping domains, and that they are related by a map $\phi_\e:x^\alpha\mapsto X^\alpha$ in the region $D=D_I\cap D_E$.

A regular outer expansion $g_E(x,\e)=g(x)+\e\hmn{}{1}(x)+...$ is constructed by taking the limit $\e\to 0$ at fixed $x^\alpha$. In this limit, the body shrinks toward zero size as all other distances remain roughly constant. For simplicity, I assume that this limit continues to a base manifold $\man_0$. However, the limit certainly does not exist on a remnant curve $\gamma\coeff{0}$ corresponding to the ``position'' of the small body---for example, if one takes a regular limit of the Schwarzschild metric in Schwarzschild coordinates, then there is no limit defined at coordinate values corresponding to $r=0$. Hence, I take the model manifold in the outer expansion to be $\man_E=\man_0\cup\gamma\coeff{0}$, and I take the external background metric to be $g=\exact{g}_0$. Of course, this construction is not essential; the model manifold need not be defined by setting $\e=0$ in this way. But at the very least, for the outer expansion to be regular, we require $g=\lim_{\e\to0}\exact{g}$.

A regular inner expansion $g_I(\tilde X,\e)=g_B(\tilde X)+\e H\coeff{1}(\tilde X)+...$ is constructed by taking the limit $\e\to 0$ at fixed values of the scaled coordinates $\tilde X^\alpha = \trans(X^\alpha)= ((T-T_0)/\e,R/\e,\Theta^A)$. This limit is naturally singular: it follows flow lines that converge at a single point defined by $(T=T_0,R=0)$ in $\man_E$. Explicitly, since the metric written in these coordinates has the form $\exact{g}\sim \e^2g_{B\alpha\beta}d\tilde X^\alpha d\tilde X^\beta$, all distances vanish at $\e=0$. As discussed by D'Eath~\cite{DEath} (see also Ref.~\cite{Gralla_Wald}), to make the limit regular, one must use the conformally rescaled metric $\exact{\tilde g}_\e\equiv \e^{-2}\exact{g}_\e$. This rescaling effectively ``blows up" the distances in  spacetime, such that as $\e\to0$, the size of the small body remains constant while all other distances are sent to infinity; thus, the inner limit serves to ``zoom in" on a small region around the body. The background spacetime defined by $\e=0$ is then defined by the metric $g_B$ of the isolated small body, and the approximation is built on a model manifold $\man_I$ with the topology of that spacetime.\footnote{Note that $\man_I$ generically differs from $\man_E$. Consider, for example, the case of a small black hole orbiting a large black hole. The manifold $\man_I$ possesses a singularity at the ``position" of the small black hole but is otherwise smooth, while the manifold $\man_E$ possesses a singularity at the ``position" of the large black hole but possesses a smooth worldline where the small black hole should be.}

\begin{figure}[tb]
\begin{center}
\includegraphics[scale=0.83]{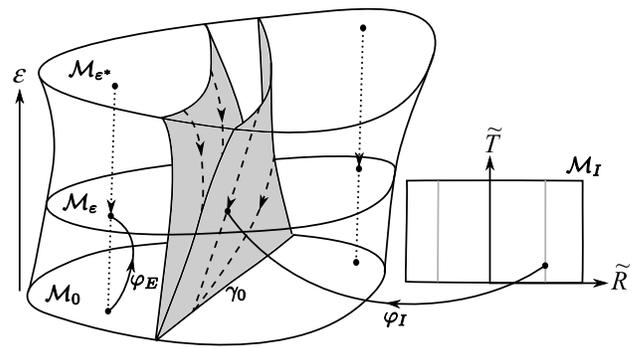}
\end{center}
\caption{Regular inner limit (dashed curves) and outer limit (dotted curves) on the manifold $\mathcal{N}\sim\man_\e\times(0,\e^*]$. The inner limit is generated by the map $\map_I$ from the manifold $\man_I$ on which the interior background metric of the isolated small body lives; these curves terminate at a point $(T=T_0,R=0)$ on $\gamma\coeff{0}$. The outer limit is generated by the map $\map_E$ from the manifold $\man_E$ on which the external background metric lives. The external manifold in this case is taken to be equal to $\man_0\cup\gamma\coeff{0}$. The gray region is a surface of constant $\tilde R$, which converges to the $\e$-independent worldline $\gamma\coeff{0}$.}
\label{regular_limit}
\end{figure}

The outer and inner expansions are related to the exact solution via identification maps $\map_E:\man_E\to\man_\e$ and $\map_I:\man_I\to\man_\e$, which respectively identify points on $\man_E$ and $\man_I$ with points on $\man_\e$. (See Fig.~\ref{regular_limit}.) These two identification maps induce a map  $\phi:\man_E\to\man_I$, given by $\phi=\map_I^{-1}\circ\map_E$, which has the identical coordinate description as the original transformation $\phi_\e$ between the global and local coordinates. Gauge transformations in the outer and inner expansions are generated by vector fields $\xi^\alpha(x)$ and $\tilde\xi^\alpha(\tilde X)$, which take their respective values in the tangent bundles of $\man_E$ and $\man_I$. Note that a gauge transformation in the outer expansion generically corresponds to a finite coordinate transformation in the inner expansion, due to the rescaling of the coordinates.

A uniform composite expansion is formed on a model manifold $\man\sim\man_\e$ by cutting out a portion of $\man_E$ and stitching part of $\man_I$ into the excised region. The local and global coordinates each cover a patch of $\man$, identified with the patches $D_I$ and $D_E$ on $\man_\e$ via the maps $\map_I$ and $\map_E$. The uniform metric is constructed on this manifold by adding together the inner and outer approximations in each coordinate system, then removing any ``double-counted'' terms that appear in both metrics.

How would one go about constructing such a uniform approximation if one did not have access to an exact solution? Just as in traditional perturbation theory, one would construct two separate asymptotic solutions to Einstein's equation, but now in possibly two different coordinate systems and on possibly two different manifolds. If one assumes that the two asymptotic solutions are approximations of a single exact solution, and if there exists an overlap region on the manifold $\man$ in which both approximations are valid to the same order, then they must agree in that overlap region. In this case, ``agreement" is defined by the existence of the unique map $\phi$ that relates the two expansions. As usual, we will not worry about a specific overlap region, but instead expand the two solutions in the buffer region.

However, before performing that expansion in the buffer region, one must write $g_I$ and $g_E$ in the same coordinate system. Let us choose this system to be the local coordinates $X^\alpha$. Then, adapting Eq.~\eqref{matching_equation}, the matching condition reads
\begin{equation}\label{strong_matching_condition}
\expand^k_\e\e^2\trans^*\expand^m_\e\e^{-2}\trans_*\exact{g}(X) = \expand^k_\e\e^2\trans^*\expand^m_\e\e^{-2}\trans_*\phi_*\expand^m_\e\exact{g}(x).
\end{equation}
On the left, we begin with the exact metric in the local coordinates $X^\alpha$. It is then expanded to $m$th-order in an inner expansion, by transforming into scaled coordinates $\tilde X^\alpha$ via $\trans$ (along with an appropriate conformal rescaling) and expanding. Next, it is expanded in the buffer region by re-expressing it in the unscaled local coordinates and expanding to $k$th order; this is equivalent to an expansion of the inner solution for $R\gg\e$. On the right-hand side of the equation, we begin with the exact metric in the global coordinates $x^\alpha$. It is expanded to $m$th order in those coordinates, yielding an outer expansion. It is then transformed to the starting point of the left-hand side, by transforming to the local coordinates via $\phi$, then to the scaled local coordinates via $\trans$, then re-expanding to $m$th order to yield an inner expansion. Finally, it is expanded in the buffer region by transforming back to the unscaled local coordinates and re-expanding. The content of this equation is that the expansion in the buffer region must be the same whether it is obtained by first performing an outer expansion or by first performing an inner expansion. Schematically, we can write
\begin{equation}
\expand_\e g_I(X) = \expand_R\phi_*g_E,
\end{equation}
which states that if the inner and outer expansions are written in the same coordinate system, then they must yield the same expansion in the buffer region.

But this is decidedly \emph{not} the matching condition that has been used in practice. Instead, what has been done in practice is the reverse: first, expand the two solutions in the buffer region, and only afterward find the coordinate transformation between them. This is accomplished by setting up a second local coordinate system $Y^{\alpha}=\phi_\gamma(x^\alpha)=(t,r,\theta^A)$ centered on a worldline $\gamma$ in $\man_E$; for example, these might be Fermi normal coordinates, and in the case of regular expansions, they would be centered on $\gamma\coeff{0}$. The outer expansion is then written in these local coordinates and expanded for small $r$, under the presumption that $r\sim R$. After performing this expansion (and the expansion of $g_I$ in the buffer region), one seeks a unique transformation $\phi_{\text{buf}}:Y^\alpha\mapsto X^\alpha$ that maps the buffer-region expansion of $g_E$ into the buffer-region expansion of $g_I$.\footnote{One can see that if everything is correct, the various transformations must be related as $\phi=\phi_{\text{buf}}\circ\phi_{\gamma}$.} Schematically, we can write
\begin{equation}\label{weak_matching_condition}
\expand_\e g_I(X) = \phi_{\text{buf}*}\expand_r \phi_{\gamma*}g_E.
\end{equation}
On the left, the inner expansion $g_I$ is expanded in the buffer region in the local unscaled coordinates $X^\alpha$. On the right, the outer expansion $g_E$ is transformed to the local coordinates $Y^\alpha$ via $\phi_\gamma$, then it is expanded in the buffer region (i.e., for small $r$). Hence, the two buffer-region expansions are written in two different coordinate systems: the inner expansion in the coordinates $X^\alpha$, and the outer expansion in the coordinates $Y^\alpha$. So, in order to make a comparison, as the final step on the right-hand side, the buffer-region expansion of $g_E$ is transformed to the coordinates $X^\alpha$ via $\phi_{\text{buf}*}$. In short, Eq.~\eqref{weak_matching_condition} states that if $g_I$ and $g_E$ are expanded in the buffer region, then the resulting expansions must be related by a coordinate transformation. 

I will call Eq.~\eqref{strong_matching_condition} the \emph{strong matching condition} and Eq.~\eqref{weak_matching_condition} the \emph{weak matching condition}. The weak condition follows from the strong condition, but not vice versa, and one can easily imagine situations in which the weak condition would be satisfied while the strong condition would not. In the weak matching condition, because the metric is already expanded for small $r$ before $\phi_{\text{buf}}$ is determined, $\phi_{\text{buf}}$ will itself be written as an expansion. Thus, the weak matching condition only requires an asymptotic approximation of $\phi_{\text{buf}}$ (or, equivalently, of $\phi=\phi_{\text{buf}}\circ\phi_{\gamma}$). Of course, one can only ever determine an asymptotic approximation---but in the strong matching condition, the approximation is for small $\e$, rather than for both small $\e$ and small $r$. This essentially reduces $\phi_{\text{buf}}$ to a gauge transformation in the buffer-region expansion defined by $R$ (or $r$) and $\e$ both being small. As mentioned above, a gauge transformation in the outer expansion corresponds to a finite coordinate transformation in the inner expansion, and vice versa. Hence, any choice of gauge on $\man_E$ must be compatible with the choice of background coordinates on $\man_I$ (and vice versa). The two matching conditions insist on this compatibility to differing extents.

One should note that though the description in this section makes use of two regular expansions, as in traditional matched asymptotic expansions, the same general description holds for two general expansions. The only differences are that $g\ne\lim_{\e\to 0}\exact{g}$ and that there is no need to conformally scale the metric to arrive at the inner expansion. This is particularly important if one wishes to allow the internal metric to vary on its ``natural" timescale $\tilde T\sim 1$ (i.e., the timescale determined by the mass of the small object). If the metric near the body varies on this timescale, then in the unscaled coordinate time $T=\e\tilde T$, the metric will have a functional dependence on the combination $T/\e$, which will be singular in the limit $\e\to0$. Thus, if both the expansions are to be regular, the internal metric can vary only on the external time $T$, corresponding to an internal slow evolution depending only on $\e\tilde T$. In other words, regularity requires that the internal solution varies quasistatically (see D'Eath's discussion~\cite{DEath}). Of course, for $\e>0$, one could construct a general inner expansion that is identical to the regular inner expansion by rescaling $R$ only, instead of both $T$ and $R$, and then simply assuming the inner expansion varies quasistatically; using this method, a global-in-time expansion can be constructed, and the metric is never conformally rescaled. Also, by using this method, one can remove the quasistatic assumption entirely.

Finally, before moving to the next singular perturbation technique, I will note that just as in traditional singular perturbation theory, there is a distinction between what I have called the overlap region and the buffer region. The buffer region corresponds simply to $\e\ll R\ll 1$. In order for us to express the outer solution in terms of the field $R$, the buffer region must lie within the region $D$, where both the local and global coordinate systems apply, but the size of the region is independent of the order of accuracy of the inner and outer solutions. As discussed in Refs.~\cite{Kates_motion, Thorne_Hartle, Gralla_Wald, my_paper,Futamase_review}, one can extract considerable information about the metric---and in particular, equations of motion for the small body---by working entirely within the buffer region, without ever constructing explicit inner and outer solutions or making use of an overlap hypothesis. This information is typically extracted by defining the mass and current moments of the body in the buffer region, which is possible because the buffer region lies at asymptotic infinity from the perspective of the inner expansion. Solving the Einstein equation then determines the evolution of these moments. In particular, an evolution equation for the body's mass dipole informs us of the motion of the body's center of mass relative to the chosen local coordinate system, providing an equation of motion for the body.

\subsection{Multiscale expansions}\label{multiple_scales_GR}
In the method of multiple scales, changes on both short and fast time scales occur throughout the spacetime. Thus, one cannot construct a uniform asymptotic approximation based on combining only two limit processes. If we consider a two-timescale expansion, with a fast time $t$ and a slow time $\t=\e t$, there are only two limits that can be easily envisioned: the slow-time limit $\e\to 0$ at fixed $\t$, which follows a congruence of curves in $\mathcal{N}$ that tend toward $t\to\infty$ as $\e\to 0$; or the fast-time limit $\e\to 0$ at fixed $t$, which follows a congruence of curves that tend toward $\t\to 0$. However, in a multiscale expansion, both quantities are to be kept fixed. As discussed above, this is accomplished by treating them as independent variables.

Consider the case of an expansion that holds fixed both a set of coordinates $x^\alpha$ and some scalar field $\zeta(x,\e)$ satisfying $\pdiff{\zeta}{x^\alpha}=o(1)$:
\begin{equation}
\exact{g}(x,\e)=g(x,\zeta)+\sum_{n\ge 1}\e^n \hmn{}{n}(x,\zeta).
\end{equation}
In the simplest case, $\zeta$ is equal to the product of $\e$ and one of the coordinates. When substituting this multiscale expansion into Einstein's equation, one would treat $\zeta$ and $x^\alpha$ as independent coordinates on an extended, 5D manifold $\widetilde{\man}_\e$; these 5D manifolds are stacked atop one another to form a 6D manifold $\widetilde{\mathcal{N}}\sim \man_\e\times\mathbb{R}^2$. The limit $\e\to 0$ is taken at fixed values of both $\zeta$ and $x$, and the actual solution is obtained by restricting the expansion to the submanifold defined by $\zeta=\zeta(x,\e)$.

As in traditional perturbation theory, one might require a fast-time variable $\phi$ that differs from the given coordinate time. Indeed, one might use any coordinates one likes on $\widetilde{\man}_\e$. To provide some flavor of the expansions, in this section I will define gauge transformations and sketch a multiscale expansion of the EFE for the simple case with coordinates $(x^\alpha,\zeta)$; I will occasionally provide details given the additional simplifying assumption $\partial_\mu\zeta=\e V_\mu(x,\zeta)$ for some $V_\mu=O_s(1)$.

Note that the gauge transformations in this expansion differ from those of a regular expansion. Gauge transformations are generated by transformations of the form
\begin{equation}
x^\alpha\to x'^\alpha = x^\alpha-\e\xi^\alpha(x,\zeta)+\order{\e^2},
\end{equation}
where $\xi=O_s(1)$. In order to determine the effect of this transformation, I expand the Lie derivative as
\begin{equation}
\Lie{\xi} = \Lie{\xi}\coeff{0}+\Lie{\xi}\coeff{1},
\end{equation}
where, e.g., for a vector $\xi(x,\zeta)$ and a tensor $T^\mu{}_\nu(x,\zeta)$,
\begin{align}
\Lie{\xi}\coeff{0}T^\mu{}_\nu & = \xi^{\rho}\frac{\partial T^\mu{}_\nu}{\partial x^\rho}-T^\rho{}_\nu\frac{\partial\xi^\mu}{\partial x^\rho}+T^\mu{}_\rho\frac{\partial\xi^\rho}{\partial x^\nu},\\
\Lie{\xi}\coeff{1}T^\mu{}_\nu & = \xi^{\rho}\frac{\partial T^\mu{}_\nu}{\partial\zeta}\frac{\partial\zeta}{\partial x^\rho} -T^\rho{}_\nu\frac{\partial\xi^\mu}{\partial\zeta}\frac{\partial\zeta}{\partial x^\rho} +T^\mu{}_\rho\frac{\partial\xi^\rho}{\partial\zeta}\frac{\partial\zeta}{\partial x^\nu}.
\end{align}
These definitions are independent of the behavior of $\zeta$. But in the particular case that $\partial_\mu\zeta=\e V_\mu(x,\zeta)$, the gauge transformation generated by a vector $\e\xi(x,\zeta)$ can be written as
\begin{align}
\Delta \hmn{}{1} & = \Lie{\xi}\coeff{0}g, \\
\Delta \hmn{}{2} & = \tfrac{1}{2}\Lie{\xi}\coeff{0}\Lie{\xi}\coeff{0}g +\Lie{\xi}\coeff{0}\hmn{}{1}+\Lie{\xi}\coeff{1}g.
\end{align}

Similarly, I expand the covariant derivative as
\begin{equation}
\del{\mu}V^{\nu}(x,\zeta) = \left(\del{\mu}\coeff{0}+\del{\mu}\coeff{1}\right)V^{\nu}(x,\zeta)
\end{equation}
where $\del{\mu}$ is compatible with $g(x,\zeta(x,\e))$, $\del{\mu}\coeff{0}$ is compatible with $g$ at fixed $\zeta$, and $\del{\mu}\coeff{1}$ is compatible with $g$ at fixed $x$. Explicitly,
\begin{align}
\del{\mu}\coeff{0}V^{\nu}(x,\zeta) &= \frac{\partial V^\nu}{\partial x^\mu}+\Gamma\coeff{0}{}^{\nu}_{\mu\rho}V^\rho,\\
\del{\mu}\coeff{1}V^{\nu}(x,\zeta) &= \frac{\partial V^\nu}{\partial\zeta}\frac{\partial\zeta}{\partial x^\mu}+\Gamma\coeff{1}{}^{\nu}_{\mu\rho}V^\rho,
\end{align}
where the Christoffel symbols are given by
\begin{align}
\Gamma\coeff{0}{}^{\alpha}_{\beta\gamma} &= \tfrac{1}{2}g^{\alpha\delta}\left(\frac{\partial g_{\delta\beta}}{\partial x^\gamma} +\frac{\partial g_{\delta\gamma}}{\partial x^\beta} -\frac{\partial g_{\beta\gamma}}{\partial x^\delta}\right)\\
\Gamma\coeff{1}{}^{\alpha}_{\beta\gamma} &=\tfrac{1}{2}g^{\alpha\delta}\left(\frac{\partial g_{\delta\beta}}{\partial\zeta}\frac{\partial\zeta}{\partial x^\gamma} +\frac{\partial g_{\delta\gamma}}{\partial\zeta}\frac{\partial\zeta}{\partial x^\beta} -\frac{\partial g_{\beta\gamma}}{\partial\zeta}\frac{\partial\zeta}{\partial x^\delta}\right)
\end{align}
The ``correction" $\del{\mu}\coeff{1}$ ensures that the total covariant derivative $\del{\mu}$ is compatible with the $\zeta$-dependence of $g$. Note that all three derivatives are metric compatible: $\nabla g=0$, $\nabla\coeff{0}g=0$, and $\nabla\coeff{1}g=0$.

By writing the Lie derivative in terms of the covariant derivative, we can express the gauge transformation generated by a vector field $\e\xi(x,\zeta)$ as
\begin{align}
\Delta \hmn{\alpha\beta}{1} & = 2\del{(\alpha}\coeff{0}\xi_{\beta)} \\
\Delta \hmn{\alpha\beta}{2} & = \xi^\gamma\del{\gamma}\coeff{0}\del{(\alpha}\coeff{0}\xi_{\beta)} +\del{(\gamma}\coeff{0}\xi_{\beta)}\del{\alpha}\coeff{0}\xi^\gamma +\del{(\alpha}\coeff{0}\xi_{\gamma)}\del{\beta}\coeff{0}\xi^\gamma \nonumber\\
&\quad +\xi^\gamma\del{\gamma}\coeff{0}\hmn{\alpha\beta}{1} +2\hmn{\gamma(\beta}{1}\del{\alpha)}\coeff{0}\xi^\gamma +2\del{(\alpha}\coeff{1}\xi_{\beta)}
\end{align}
assuming that $\partial_\mu\zeta=\e V_\mu(x,\zeta)$.

Note that because the background metric $g$ depends on $\zeta$, the Riemann tensor constructed from it can be expanded in powers of $\e$:
\begin{align}
R_{\mu\rho\nu\sigma}(x,\zeta) &= R\coeff{0}_{\mu\rho\nu\sigma}(x,\zeta) +\e R\coeff{1}_{\mu\rho\nu\sigma}(x,\zeta)\nonumber\\
&\quad +\e^2 R\coeff{2}_{\mu\rho\nu\sigma}(x,\zeta),
\end{align}
where $R\coeff{0}_{\mu\rho\nu\sigma}(x,\zeta)$ is constructed from $g$ and $\nabla\coeff{0}$, $R\coeff{1}_{\mu\rho\nu\sigma}(x,\zeta)$ contains one $\nabla\coeff{1}$ derivative, and $R\coeff{2}_{\mu\rho\nu\sigma}(x,\zeta)$ contains two $\nabla\coeff{1}$ derivatives. The $n$th-order perturbation of the Ricci tensor can be similarly expanded as $\delta^n R_{\mu\nu}[h]=\sum_{m=0}^2\e^m\delta^n R\coeff{n}_{\mu\nu}[h]$. This means that the vacuum Einstein equation $\exact{R}_{\mu\nu}=0$ becomes
\begin{align}
R\coeff{0}_{\mu\nu} &= 0, \\
\delta R\coeff{0}_{\mu\nu}[h\coeff{1}] & = R\coeff{1}_{\mu\nu}, \\
\delta R\coeff{0}_{\mu\nu}[h\coeff{2}] & = R\coeff{2}_{\mu\nu}-\delta R\coeff{1}_{\mu\nu}[h\coeff{1}]-\delta^2 R\coeff{0}_{\mu\nu}[h\coeff{1}], \\
&\ \ \vdots\nonumber
\end{align}

Similarly, the Bianchi identity on the background, $g^{\mu\nu}\del{\mu}G_{\nu\rho}[g]=0$, becomes
\begin{align}
\del{\mu}\coeff{0}\cdot G\coeff{0}_{\nu\rho} &= 0, \\
\del{\mu}\coeff{0}\cdot G\coeff{1}_{\nu\rho} &= -\del{\mu}\coeff{1}\cdot G\coeff{0}_{\nu\rho}, \\
\del{\mu}\coeff{0}\cdot G\coeff{2}_{\nu\rho} &= -\del{\mu}\coeff{1}\cdot G\coeff{1}_{\nu\rho}, \\
\del{\mu}\coeff{1}\cdot G\coeff{2}_{\nu\rho} &= 0,
\end{align}
where a dot indicates contraction over $\mu$ and $\nu$. And the Bianchi identitity on the full spacetime, $\exact{g}^{\mu\nu}{}^{\exact{g}}\del{\mu}\exact{G}_{\nu\rho}=0$, can be expanded schematically as
\begin{align}
\del{}\coeff{0}\cdot\delta G\coeff{0}[\hmn{}{1}] & = 0 \\
\del{}\coeff{0}\cdot\delta G\coeff{0}[\hmn{}{2}] & = -\left(\del{}\coeff{1}\cdot\delta G\coeff{0} +\del{}\coeff{0}\cdot\delta G\coeff{1}\right)[\hmn{}{1}] \nonumber\\
&\quad +\hmn{}{1}\del{}\coeff{0}\left(G\coeff{1} +\delta G\coeff{0}[\hmn{}{1}]\right) \nonumber\\
&\quad -\delta\Gamma\coeff{0}[\hmn{}{1}]\cdot\left(G\coeff{1} +\delta G\coeff{0}[\hmn{}{1}]\right)\nonumber\\
&\quad -\del{}\coeff{0}\cdot\delta^2G\coeff{0}[\hmn{}{1}]
\end{align}
where $\delta\Gamma\coeff{0}{}^{\alpha}_{\beta\gamma}[\hmn{}{1}] =\tfrac{1}{2}g^{\alpha\delta}(\del{\gamma}\coeff{0}\hmn{\delta\beta}{1} +\del{\beta}\coeff{0}\hmn{\delta\gamma}{1} -\del{\delta}\coeff{0}\hmn{\beta\gamma}{1})$, and the leading-order Einstein equation $R\coeff{0}_{\mu\nu} = 0$ and the Bianchi identity $g^{\mu\nu}\del{\mu}G_{\nu\rho}=0$ have already been imposed for compactness.

Note that the $\zeta$-dependence of $g$ allows the background to slowly react to the perturbation. Determining this reaction is the backreaction problem, which has been studied extensively in the past.

Recently, Hinderer and Flanagan \cite{Hinderer_Flanagan} have constructed a significantly more complicated two-timescale expansion tailored to EMRIs. In their method, all dynamical variables (i.e., the metric and the phase space variables of the worldline) are submitted to two-timescale expansions; this expansion captures both the fast dynamics of orbital motion and the slow dynamics of the particle's inspiral and the gravitational backreaction on the background spacetime. Since the metric and the worldline are related by the EFE, it is assumed that the metric can be written as a function of the phase space variables of the worldline. On each timeslice, the limit $\e\to0$ is then taken with the phase space variables held fixed. Specifically, the true worldline is specified by a set of action-angle variables $(J(\t,\e),\varphi(\t,\e))$ and a slow time variable $\t$. Expanding for $\e\to0$ with $\varphi$ and $\t$ held fixed results in a sequence of fast-time and slow-time equations. In the fast-time equations, wherein $\t$ (and therefore $J$) is treated as a constant, the metric is a function of $\varphi$ only; in other words, it is a functional of the geodesic that is instantaneously tangential to the true worldline. From this it follows that the leading-order fast-time equation yields a metric perturbation sourced by that geodesic, as in regular perturbation theory. However, that is only at fixed $\t$---the true worldline and metric perturbation emerge by allowing the variables to vary with $\t$, with a $\t$-dependence determined from the slow-time equations.

Although exceedingly useful for EMRIs, this procedure relies on the background metric being stationary at fixed $\t$, such that it has no fast time dependence, and on the geodesic motion in that background being integrable, such that the metric can be written in terms of the action-angle variables. In addition, it requires one to determine the slow evolution of the background metric, which has not yet been done.

\subsection{Self-consistent expansion}\label{fixed_worldline_formulation}
In Hinderer and Flanagan's formalism, the metric is written as a function of the phase space variables on the worldline, and then both the metric and those variables are submitted to a two-timescale expansion. The formalism I will now describe is a generalization of this: the metric is written as a functional of the worldline, and then the metric is expanded with that worldline held fixed. In order to motivate this approach, I will first provide a formulation of the exact problem to which we seek an approximate solution.

We wish to determine the mean motion of a small, spatially bounded matter distribution. (For the moment, I neglect the case of a black hole.) In principle, we have some matter field equations to go along with the EFE for this blob of matter. As governed by the field equations, the boundary of the blob traces out some surface in spacetime. In the interior of the boundary, the matter density is finite, and in the exterior it vanishes. To determine the motion of the body, we seek the equation for the generators of this boundary. This is a free-boundary value problem \cite{free_boundary_problems}, in which some boundary values are specified on a boundary that is free to move. In the context of bodies in GR, this problem has received some study  \cite{free_boundary_GR,free_boundary_GR2}, but it is still far from understood, and it must certainly be tackled numerically.

To make progress with an approximation scheme, I reformulate the problem. I surround the body by a tube $\Gamma$ embedded in the buffer region, such that for $\e\to0$, the radius of the tube vanishes. For the moment, consider $\Gamma$ to be defined by constant radius $R=\rad(\e)$ in the local coordinates $X^\alpha$. I assume that the body is fairly widely separated from all other matter sources, such that outside of $\Gamma$ there is a large vacuum region $\Omega$. I also assume that $\Gamma$ is in vacuum; since it lies in the buffer region around the small body, this means that I must restrict my approximation to a small body that is sufficiently compact to not fill the entire buffer region. Now, since the tube is close to the small body (relative to all external length scales), the metric on the tube is primarily determined by the small body's structure. In other words, the information about the body has now been transplanted into boundary conditions on the tube. Recall that the buffer region corresponds to $\tilde R\to\infty$. Hence, on the tube, we can construct a multipole expansion of the body's field, with the form $\sum\tilde R^{-n}$. I assume that the local coordinates $X^\alpha$ are mass-centered, such that the mass dipole term in this expansion vanishes. (See Ref.~\cite{STF_3} and references therein for discussion of multipole expansions in GR; see Refs.~\cite{STF_3, Thorne_Hartle} for discussion of mass-centered coordinates in the buffer region; see, e.g., Ref.~\cite{center_of_mass} for further discussion of definitions of center of mass.) This, then, is another free-boundary value problem: we must determine the equations of motion of the generators of the tube, given the boundary values of the metric on it, and in particular, given that the body lies at the ``center" of it.  With this formulation, we can also determine the motion of a black hole, rather than just a matter distribution.

Now suppose that I want to represent the motion of the body through the external spacetime $(g,\man_E)$, rather than through the exact spacetime. As we can see from Fig.~\ref{regular_limit}, this is easily accomplished by using the regular limit and taking the motion to be represented by the remnant worldline $\gamma\coeff{0}$. However, as mentioned in Sec.~\ref{introduction}, on long timescales this will provide a very poor representation of the motion. (See Ref.~\cite{my_paper} for a detailed discussion.)

Let us consider this from another direction. Assume that we were given the exact solution $\exact{g}_\e$ on $\man_\e$, along with the coordinate transformation $\phi_\e$ between the local coordinates $X^\alpha$ and the global coordinates $x^\alpha$ in the buffer region. At fixed $R=\rad(\e)=o(1)$, we could write this transformation as $x^\alpha=\phi^{-1}_\e(T,\rad,\Theta^A)$. In the limit of small $\e$, $\rad$ becomes small as well, meaning that this transformation can be expanded as $x=\phi^{-1}_\e(T,0,\Theta_0^A)+o(1)$, where $\Theta_0^A$ is an arbitrary choice of angles. This transformation thus defines a curve $z^\alpha(T,\e) \equiv \phi^{-1}_\e(T,0,\Theta_0^A)$ in the external manifold $\man_E$. Since the small body is centered ``at" $R=0$, this curve defines a meaningful long-term representative worldline $\gamma$. If we expand $\phi_\e(T,0,\Theta_0^A)$ for small $\e$, then it will not provide a uniform transformation between the inner and outer coordinates; it will contain secularly growing errors of the form $\e t$. So, instead, in order to construct a uniform asymptotic solution, when constructing the external approximation, one must hold $\gamma$ fixed. Determining $\gamma$ then amounts to determining the ``location" at which the (mass-centered) inner expansion is to be performed.

Since we will never be seeking $\phi$ directly, and in case the inverse of $\phi_\e$ does not exist at $R=0$, allow me to present the final reformulation of the problem. Define a tube $\Gamma_E[\gamma]\subset\man_E$ such that it is a surface of constant radius $r$ in Fermi normal coordinates centered on a worldline $\gamma\subset\man_E$. Using the map $\map_E$ from the regular expansion, this defines a tube $\Gamma=\phi_E(\Gamma_E)$. Now, the problem is the following: what equation of motion must $\gamma$ satisfy in order for $\Gamma$ to be mass-centered (in the sense that the mass dipole of the inner expansion vanishes when mapped to $\man_\e$ via $\map_I$)? Note that the worldline is a curve in the external manifold $\man_E$. It should not be thought of as a curve in the manifold $\man_\e$ on which the exact metric $\exact{g}_\e$ lives; in fact, if the small body is a black hole, then there is obviously no such curve. 

\begin{figure}[tb]
\begin{center}
\includegraphics[scale=0.83]{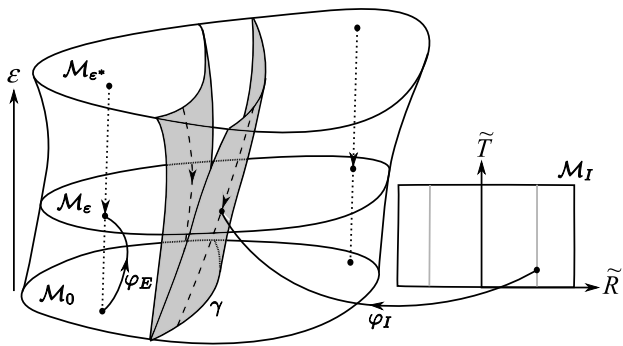}
\end{center}
\caption{Fixed-worldline expansion of a family of spacetimes. The dotted lines correspond to the outer limit, which lets the body shrink to zero size but keeps its motion fixed. The dashed lines correspond to the inner limit, which keeps the size of the body fixed. Here I display the singular inner limit, which does not rescale the inner time coordinate; hence, dashed lines originating at different times will terminate at different points on $\gamma$. The worldline lies in the manifold $\man_E=\man_0\cup\gamma\coeff{0}$, but it does not correspond to the remnant curve $\gamma\coeff{0}$ defined by the regular limit; instead, it is allowed to have $\e$-dependence, and it is determined by the particular value of $\e$ at which an approximate solution is sought.} 
\label{singular_limit}
\end{figure}

In order to determine the equation of motion of the worldline, I consider a family of metrics $g_E(x,\e;\gamma)$ parametrized by $\gamma$, such that when $\gamma$ is given by the correct equation of motion, we have $g_E(x,\e;\gamma(\e))=\map_E^*\exact{g}_\e(x)$. The metric in the outer limit is thus taken to be the general expansion
\begin{equation}\label{external ansatz}
\exact{g}(x,\e) =g_E(x,\e;\gamma)= g(x)+h(x,\e;\gamma),
\end{equation}
where
\begin{equation}
h(x,\e;\gamma)=\sum_{n=1}^{N_E}\e^n\hmn{E}{\emph{n}}(x;\gamma) +\order{\e^{N_E+1}}.
\end{equation}
Solving Einstein's equations will determine the worldline $\gamma$ for which the inner expansion is mass-centered. I will call this either a self-consistent or a fixed-worldline expansion. In it, the perturbations produced by the body are constructed about a fixed worldline determined by the particular value of $\e$ at which one seeks an approximation. Refer to Figs.~\ref{regular_limit} and \ref{singular_limit} for a graphical comparison between this expansion and a regular one.

In the remainder of this section, I present a sequence of perturbation equations that arise in this expansion scheme, along with a complementary sequence for the inner expansion. The equations were originally presented in Ref.~\cite{my_paper}. In that paper, I described a particular, lengthy method of solving the equations and deriving equations of motion. In Sec.~\ref{matching}, I will discuss an alternative approach using the method of matched asymptotic expansions.

My sequence of perturbation equations relies on a particular choice of gauge. I discuss the gauge freedom in the self-consistent expansion, and the effect of gauge transformations on the equation of motion, in Appendix~\ref{gauge}.
 
\subsubsection{Field equations in outer expansion}
To begin, I surround the body with a worldtube $\Gamma$, where $\Gamma$ is embedded in the buffer region, such that the field on it can be found from either the inner or outer expansion. I seek a solution in a vacuum region $\Omega$ outside of $\Gamma$; I further specify that $\Omega$ consists of the future domain of dependence of $\Gamma\cup\Sigma$, where $\Sigma$ is a spacelike initial-data surface.
 
Now, recall that in a multiscale expansion, the expanded equations are solved by assuming that they are valid for arbitrary values of the slow-time variable $\t$, not only on the true solution manifold defined by $\t=\e t$. Similarly, in the fixed-worldline expansion, one method of solving the expanded EFE will consist of assuming that it is valid for arbitrary worldlines; the true solution is found by choosing the true worldline. Solving the EFE with an arbitrary worldline seems to require reformulating it in a ``relaxed" form before expanding it, such that, for example, the linearized equation does not immediately determine $\gamma$ to be a geodesic. To accomplish this, I assume that the Lorenz gauge can be imposed everywhere in $\Omega$ on the entirety of $h$, such that $L_\mu[h]=0$, where $L_\mu$ is the operator defined by
\begin{equation}
L_\mu[h] =\left(g^\rho_\mu g^{\sigma\gamma}-\tfrac{1}{2}g^\gamma_\mu g^{\rho\sigma}\right)\!\del{\gamma}h_{\rho\sigma}.
\end{equation}
In Appendix~\ref{gauge}, I discuss the validity of this assumption.

With this choice of gauge, the vacuum Einstein equation $\exact{R}_{\mu\nu}=0$ is reduced to a weakly nonlinear wave equation that can be expanded and solved at fixed $\gamma$, leading to the sequence of wave equations
\begin{align}
E_{\mu\nu}[\hmn{E}{1}] &=0, \label{h_E1 eqn}\\
E_{\mu\nu}[\hmn{E}{2}] &=2\delta^2 R_{\mu\nu}[\hmn{E}{1}], \label{h_E2 eqn}\\
&\vdots\nonumber
\end{align}
where $E_{\mu\nu}$ is the relativistic wave operator defined by
\begin{equation}
E_{\mu\nu}[h] = \left(g^\rho_\mu g^\sigma_\nu\nabla^\gamma\del{\gamma} +2R_\mu{}^\rho{}_\nu{}^\sigma\right)\!h_{\rho\sigma}.
\end{equation}
More generally, we can write the $n$th-order equation as
\begin{equation}
E_{\mu\nu}\big[\hmn{E}{\emph{n}}\big]=S\coeff{n}_{\mu\nu}\big[\hmn{E}{1},...,\hmn{E}{n-1},\gamma\big],
\end{equation}
where the source term $S\coeff{\emph{n}}_{\mu\nu}$ consists of nonlinear terms in the expansion of the Ricci tensor. These equations can be solved for arbitrary $\Gamma$ (and hence for abritrary $\gamma$). The formal solution is given by
\begin{align}
\hmn{E\alpha\beta}{n} & = \frac{1}{4\pi}\!\oint\limits_{\partial\Omega}\!\!\Big(G_{\alpha\beta}{}^{\gamma'\delta'} \del{\mu'}\hmn{E\gamma'\delta'}{n}-\hmn{E\gamma'\delta'}{n}\del{\mu'} G_{\alpha\beta}{}^{\gamma'\delta'}\Big) dS^{\mu'}\nonumber\\
&\quad -\frac{1}{4\pi}\int_\Omega G_{\alpha\beta}{}^{\gamma'\delta'}S\coeff{n}_{\gamma'\delta'}dV',
\end{align}
where $G_{\alpha\beta\alpha'\beta'}$ is the retarded Green's function for $E_{\alpha\beta}$; I adopt the conventions of Ref.~\cite{Eric_review} for the Green's function. This formal solution requires boundary data on $\Gamma$. Since $\Gamma$ lies in the buffer region, the boundary data on it can be provided by the inner expansion, as discussed in Ref.~\cite{my_paper}.

In this scheme, the equation of motion is determined only once the perturbative EFE, rather than just the wave equation, is satisfied. One means of ensuring that the EFE is satisfied is to ensure that the gauge condition is satisfied. I assume that the acceleration of $\gamma$ possesses an expansion
\begin{equation}
a_\mu(t,\e)=\an{0}_\mu(t)+\e\an{1}_\mu(t;\gamma)+...\label{a expansion}.
\end{equation}
This is an expansion of a particular function of time on the worldline; it does not suggest a multiplicity of worldlines with differing accelerations. Substituting this expansion, along with that of $h_{\mu\nu}$, into the exact gauge condition $L_\mu[h]=0$ and solving with arbitrary $\gamma$, we arrive at the sequence of equations
\begin{align}
L\coeff{0}_\mu\big[\hmn{E}{1}\big] &=0, \label{gauge_expansion 1}\\
L\coeff{1}_\mu\big[\hmn{E}{1}\big] &= -L\coeff{0}_\mu\big[\hmn{E}{2}\big],
 \label{gauge_expansion 2}\\
&\ \ \vdots\nonumber
\end{align}
where $L\coeff{0}[f]\equiv L[f]\big|_{a=\an{0}}$, $L\coeff{1}[f]$ is linear in $\an{1}$, $L\coeff{2}[f]$ is linear in $\an{2}$ and quadratic in $\an{1}$, and so on. More generally, for $n>0$ the equations read
\begin{equation}\label{compact_gauge}
L\coeff{n}_\mu\big[\hmn{}{1}\big]=-\sum_{m=1}^n L_\mu\coeff{n-m}\big[\hmn{}{m+1}\big].
\end{equation}
In these expressions, $L[f]$ is first calculated on an arbitrary worldline, and then to find $L\coeff{\emph{n}}$, the expansion of the acceleration is inserted---while still holding $\gamma$ and $u^\mu$ fixed. These equations will determine successive terms $\an{n}_\mu$. At the end of the calculation, when $a^\mu$ has been determined to the desired accuracy, $h[\gamma]$ is evaluated for the particular worldline with acceleration $a_\mu=\an{0}_\mu+\e\an{1}_\mu+...$, just as at the end of a two-timescale expansion, the solution is evaluated for $\t=\e t$.

Beyond formalizing the self-consistent approach as a systematic approximation scheme, this method differs in one key respect from previous derivations of self-consistent equations of motion. Earlier derivations generally result in equations containing derivatives of the acceleration; in the case of an uncharged body in vacuum, this is seen in the gravitational antidamping term discovered by Havas \cite{damping} (as corrected by Havas and Goldberg \cite{damping2}). Such equations are unphysical, exhibiting, for example, runaway solutions. Traditionally, they have been made well behaved via an a posteriori ``reduction of order" \cite{Eric_review,Quinn_Wald}. However, my assumed expansion of the acceleration automatically yields a well-behaved, order-reduced equation of motion. Furthermore, the expansion of the acceleration was necessary to split the gauge condition (or, equivalently, the Bianchi identity) into a sequence of exactly solvable equations. Hence, we can see that a systematic approach, in which one requires exact solutions to the perturbation equations, eliminates the ill-behaved equations of motion that have plagued prior self-consistent derivations.

In particular, this method differs from the gauge-relaxation procedure that has been used historically in the gravitational self-force problem \cite{Mino_Sasaki_Tanaka,Quinn_Wald, Gralla_Wald, my_paper}. In that procedure, one constructs an approximate solution to the regular linearized Einstein equation by (i) solving the linear wave equation and (ii) ensuring the equation of motion enforces the relaxed gauge condition $L_\mu[\e\hmn{}{1}]=O(\e^2)$. The relaxed gauge condition accomplishes the same goal as my reformulation of the Einstein equation into relaxed form: it allows the body to move on an accelerated worldline rather than the geodesic worldline enforced by the regular linearized Einstein equation. However, this procedure is an a posteriori corrective measure, rather than part of a systematic expansion, and unlike Eq.~\eqref{gauge_expansion 1}, it results in an ill-behaved equation of motion requiring further a posteriori correction in the form of order-reduction.

Before proceeding to the inner expansion, I note that the first-order acceleration $\an{1}$ is determined by Eq.~\eqref{gauge_expansion 2}, which requires the second-order metric perturbation $\hmn{E}{2}$. Hence, if the gauge condition is to be used to determine the equation of motion, one must solve the second-order wave-equation in order to determine the first-order acceleration. However, in the method of matched asymptotic expansions, the inner expansion at any given order generically contains information that is of infinite order in the outer expansion. Therefore, if matching is used, it may be possible to derive an equation of motion using only the first-order term in the outer expansion. This will be the route explored in Sec.~\ref{matching}.

\subsubsection{Field equations in inner expansion}
For the inner expansion, I assume the existence of some local polar coordinates $X^\alpha=(T,R,\Theta^A)$, such that the metric can be expanded for $\e\to 0$ while holding fixed $\tilde R\equiv R/\e$, $\Theta^A$, and $T$. This leads to the ansatz
\begin{align}
\exact{g}(X,\e) & =g_I(T,\tilde R,\Theta^A,\e)\nonumber\\
&= g_B(T,\tilde R,\Theta^A)+ H(T,\tilde R,\Theta^A,\e),\label{internal ansatz}
\end{align}
where $H$ at fixed $(T,\tilde R,\Theta^A)$ is a perturbation beginning at order $\e$. The leading-order term $g_B(T,\tilde R,\Theta^A)$ at fixed $T$ is the metric of the small body if it were isolated. For example, if the body is a small Schwarzschild black hole of ADM mass $\e m(T)$, then in Schwarzschild coordinates $g_B(T,\tilde R,\Theta^A)$ is given by
\begin{align}
ds^2 &= -\left(1-2m(T)/\tilde R\right)dT^2 +\left(1-2m(T)/\tilde R\right)^{-1}\!\!\e^2d\tilde{R}^2 \nonumber\\
&\quad + \e^2\tilde{R}^2\left(d\Theta^2+\sin^2\Theta d\Phi^2\right).
\end{align}
Since the metric becomes one-dimensional at $\e=0$, the limit $\e\to0$ is singular. As mentioned in Sec.~\ref{matching_conditions}, the limit can be made regular by rescaling time as well, such that $\tilde T=(T-T_0)/\e$, and then rescaling the entire metric by a conformal factor $1/\e^2$. This is equivalent to using the above general expansion and assuming that the metric $g_B$ and its perturbations are quasistatic (evolving only on timescales $\sim 1$). Both are equivalent to assuming that the exact metric contains no high-frequency oscillations occurring on the body's natural timescale $\sim\e$. In other words, the body is assumed to be in equilibrium.

I seek a solution in a vacuum region outside the body. Given my assumptions, the vacuum EFE $\exact{G}=0$ can be expanded as
\begin{equation}
0=\exact{G} = G_I[g_B]+\delta G_I[H]+\delta^2 G_I[H]+...,
\end{equation}
where each term is further expanded as
\begin{align}
G_I[g_B] &= \e^{-2}\left(G\coeff{0}_I[g_B]+\e G\coeff{1}_I[g_B]+\e^2 G\coeff{2}_I[g_B]\right),\\
\delta^k G_I[g_B] &= \e^{-2}\!\!\left(\delta^kG\coeff{0}_I[H]+\e \delta^kG\coeff{1}_I[H]+\e^2 \delta^kG\coeff{2}_I[H]\right).
\end{align}
The overall factors of $\e^{-2}$ result from $\tilde R=R/\e$ and the fact that the Einstein tensor scales as the metric divided by two powers of length. The correction terms contain derivatives with respect to $T$, which are each suppressed by a factor of $\e$; specifically, $G_I\coeff{\emph{n}}$ and $\delta^kG_I\coeff{\emph{n}}$ consist of the terms in $G_I$ and $\delta^k G_I$ that contain $n$ derivatives with respect to $T$. Now, suppose $H$ possesses an expansion
\begin{equation}
H(T,\tilde R,\Theta^A,\e)=\sum_{n=1}^{N_I}\e^n H\coeff{\emph{n}}(T,\tilde R,\Theta^A).
\end{equation}
Substituting this expansion of $H$ into the above expansion of the EFE, and then solving order-by-order in powers of $\e$, leads to the sequence
\begin{align}
G\coeff{0}_I{}^{\mu\nu}[g_B] &= 0,\label{inner_eqn0}\\
\delta G\coeff{0}_I{}^{\mu\nu}[H\coeff{1}] &= -G\coeff{1}_I{}^{\mu\nu}[g_B],\label{inner_eqn1}\\
\delta G\coeff{0}_I{}^{\mu\nu}[H\coeff{2}] &= -\delta^2 G\coeff{0}_I{}^{\mu\nu}[H\coeff{1}] - \delta G\coeff{1}_I{}^{\mu\nu}[H\coeff{1}]\nonumber\\
&\quad-G\coeff{2}_I{}^{\mu\nu}[g_B],\label{inner_eqn2}\\
&\vdots\nonumber
\end{align}
Note that there is only one timescale here, so these equations automatically follow from the assumed form of the expansion of the metric; there is none of the potential failings of a two-timescale expansion. In Sec.~\ref{internal solution} and Appendix~\ref{perturbed_BH}, I discuss a particular solution to this sequence of equations.

\section{Calculation of the self-force from matched asymptotic expansions}\label{matching}
In this section, I consider the most intuitive means of solving the sequences of equations just presented: the method of matched asymptotic expansions. As outlined in the previous section, in this method the perturbation equations in the inner and outer expansions are solved independently, and then any free functions are identified by insisting that the two metrics agree in the buffer region around the body. Following the tradition of the field, in matching the two metrics I make use of the weak matching condition, rather than the strong condition.

My presentation of the matching procedure roughly follows that of Refs.~\cite{Eric_review,Eric_matching}, though most of my conclusions apply as well to the earlier calculation performed by Mino, Sasaki, and Tanaka \cite{Mino_Sasaki_Tanaka}. However, my goal is not simply to review those earlier calculations, but to pinpoint their underlying assumptions. First among these assumptions is a very strong restriction on the relationship between the inner and outer solutions: essentially, the two solutions must be assumed to differ only by generically ``small" coordinate transformations in the buffer region. This restriction is required because the weak matching condition, which has always been used in matched-expansion derivations of the self-force, is found to be \emph{too} weak to yield unique results. The required restriction amounts to introducing a ``refined" matching condition midway between the weak and strong conditions.

Second among the underlying assumptions is the restrictive choice of inner solution, which effectively already removes many of the integration constants that would normally be fixed by a matching procedure. As discussed in Sec.~\ref{traditional}, in traditional matched asymptotic expansions the leading-order inner and outer solutions are determined entirely by boundary conditions, while in the matched expansions used in the self-force problem, the leading-order solutions must be chosen based on some desired physical properties; only after the leading-order solutions are chosen can boundary conditions be imposed. In the self-force problem, the leading-order outer solution is taken to be some desired vacuum metric. For the EMRI problem, the desired metric is that of a Kerr black hole. Typically, for simplicity, the leading-order inner solution is taken to be that of a Schwarzschild black hole, though one could instead choose, for example, that of a neutron star~\cite{Love_numbers1,Love_numbers2}.

However, in derivations of the self-force, the inner and outer solutions have been even further restricted: the form of the perturbations have also been largely selected, rather than determined by matching. For example, the inner perturbations have been taken to be of a particular form presumed to correspond to the influence of tidal fields on the small black hole. And the outer perturbation has been taken to be that of a point particle. In this section, I will make use of these assumed forms for the inner and outer solutions. As shown in Refs.~\cite{DEath, Gralla_Wald, my_paper}, the assumption of a point particle perturbation can be removed, because the point particle solution follows directly from the assumed existence of an inner expansion. On the other hand, the assumed form of the tidally perturbed black hole metric has not, to my knowledge, been rigorously justified. Instead, I will point out the ways in which this metric restricts the generality of the inner solution.

My analysis begins with a discussion of the outer expansion. Section~\ref{internal solution} then describes the metric in the inner expansion. In Secs.~\ref{zeroth_order_matching}--\ref{first_order_matching}, I perform the matching procedure, focusing on the restrictions that must be imposed to yield a unique result. I conclude the section with a discussion of the method.

The calculations in this section make use of numerous computational techniques that are standard in the literature: near-coincidence expansions, Fermi and retarded coordinate systems, and curved-spacetime Green's functions, STF (symmetric trace-free) decompositions, and tensor-harmonic expansions. Refer to Ref.~\cite{Eric_review} for pedagogical reviews of the first three techniques. STF decompositions and tensor-harmonic expansions are briefly outlined in Appendices~\ref{STF tensors} and \ref{perturbed_BH}.

\subsection{Outer expansion}
I require an expansion of the background metric $g$ and the first-order external perturbation $\hmn{E}{1}$ in the buffer region. To find these expansions, I adopt Fermi coordinates $(t,x^a)$ centered on $\gamma$ and then expand in powers of the geodesic distance $r\equiv\sqrt{\delta_{ij}x^ix^j}$. The construction of the coordinate system is derived in Ref.~\cite{Eric_review}.

Uppercase Latin indices $I,J,K$ run from 0 to 3, corresponding to an orthonormal tetrad $e^\alpha_I$. Lowercase Latin indices $i,j,k$ run from 1 to 3 and correspond to the spatial part of either the coordinate basis or the orthonormal basis. I am interested only in components in the Cartesian-type coordinates $(t,x^a)$, but I will sometimes express these components in terms of $r$ and two angles $\theta^A$, which are defined in the usual way in terms of $x^a$. I also introduce the unit one-form $n_\alpha\equiv \partial_\alpha r$, which depends only on the angles $\theta^A$, and use the multi-index notation $n^L\equiv n^{i_1}...n^{i_\ell}\equiv n^{i_1...i_\ell}$. Angular brackets around indices denote the STF combination of the enclosed indices; a caret over a tensor denotes the STF part of that tensor. Finally, I define the coordinate one-forms $t_\alpha\equiv\partial_\alpha t$ and $x^a_\alpha\equiv \partial_\alpha x^a$.

In Fermi coordinates, the components of the background metric are given by
\begin{align}
g_{tt} &= -1-2ra_in^i-\tfrac{1}{3}r^2 a_i a^i-a_{\langle i}a_{j\rangle}\nhat^{ij}\nonumber\\
&\quad -r^2\etide_{ij}\nhat^{ij} +O(r^3), \label{Fermi background tt}\\
g_{ta} &= \tfrac{2}{3}r^2\epsilon_{aik}\btide^k_j\nhat^{ij}+O(r^3),\label{Fermi background ta}\\
g_{ab} &= \delta_{ab} -\tfrac{1}{9}r^2\delta_{ab}\etide_{ij}\nhat^{ij} -\tfrac{1}{9}r^2\etide_{ab}\nonumber\\
&\quad +\tfrac{2}{3}r^2\etide_{i\langle a}\nhat^i_{b\rangle}+O(r^3),\label{Fermi background ab}
\end{align}
where I have decomposed the components into irreducible STF pieces, and defined the tidal fields $\etide_{ab}\equiv R_{a0b0}$ and $\btide_{ab}\equiv\tfrac{1}{2}\epsilon_a{}^{cd}R_{0bcd}$. The tidal fields are functions on the worldline, and are therefore functions of $t$ only.

One should note that the coordinate transformation $x^\alpha(t,x^a)$ between Fermi coordinates and the global coordinates is $\e$-dependent, since Fermi coordinates are tethered to an $\e$-dependent worldline. If one were using a regular expansion, then this coordinate transformation would devolve into a background coordinate transformation to a Fermi coordinate system centered on a geodesic worldline, combined with a gauge transformation to account for the $\e$-dependence. But in the present general expansion, the transformation is purely a background transformation, because the $\e$-dependence in the transformation is reducible to the $\e$-dependence in the fixed worldline.

The transformation hence induces not only new $\e$-dependence into the perturbations $\hmn{E}{\emph{n}}$, but also $\e$-dependence in the background metric $g$. (Despite its $\e$-dependence, $g$ is the background metric of the outer expansion, and I will use it to raise and lower indices on $h$.) This new $\e$-dependence takes two forms: a functional dependence on $z^\alpha(t)=x^\alpha(t,x^a=0)$, the coordinate form of the worldline written in the global coordinates $x^\alpha$; and a dependence on the acceleration vector $a^\alpha(t)$ on that worldline. For example, the first type of dependence appears in the components of the Riemann tensor (or tidal fields) in Fermi coordinates, which are related to the components in the global coordinates via the relationship $R_{IJKL}(t)=R_{\alpha\beta\gamma\delta}(z^\mu(t))e^\alpha_I e^\beta_J e^\gamma_K e^\delta_L$. The second type of $\e$-dependence consists of factors of the acceleration $a^\mu(t)$, which has the assumed expansion $a_i(t,\e) = \an{0}_i(t)+\e\an{1}_i(t;\gamma)+\order{\e^2}$.

Hence, in the buffer region we can opt to work with the quantities $g$ and $h_E$, which are defined with $a$ fixed, or we can opt to re-expand these quantities by substituting into them the expansion of $a$. (In either case, we would still hold fixed the functional dependence on $z^\mu$.) Substituting the expansion of $a$ in Fermi coordinates yields the \emph{buffer-region expansions}
\begin{align}
g_{\mu\nu} & = g_{\mu\nu}\coeff{0}(t,x^a;\gamma)+\e g_{\mu\nu}\coeff{1}(t,x^a;\gamma)+\order{\e^2},\label{buffer_expansion g}\\
\hmn{E\alpha\beta}{\emph{n}} &= \hmn{\alpha\beta}{\emph{n}}(t,x^a;\gamma) +\order{\e}\label{buffer_expansion h},
\end{align}
where $g_{\mu\nu}\coeff{0}\equiv g_{\mu\nu}\big|_{a=\an{0}}$, $g_{\mu\nu}\coeff{1}=\an{1}_i \pdiff{g_{\mu\nu}}{a_i}\big|_{a=\an{0}}$, and $\hmn{\mu\nu}{\emph{n}}\equiv\hmn{E\mu\nu}{\emph{n}}\big|_{a=\an{0}}$. Because the inner expansion does not hold the acceleration fixed, for the sake of matching, I will use the buffer-region quantities $g\coeff{0}$, $g\coeff{1}$, and $\hmn{}{1}$.

In order to determine $\hmn{}{1}$, I rewrite the wave equation \eqref{h_E1 eqn} as
\begin{equation}
E_{\alpha\beta}[\hmn{E}{1}] = -16\pi (T_{\alpha\beta}-\tfrac{1}{2}g_{\alpha\beta}g^{\mu\nu}T_{\mu\nu}),
\end{equation}
where $T_{\alpha\beta}$ is the stress-energy tensor of a point particle, given by
\begin{equation}\label{stress_energy_point_particle}
T^{\mu\nu}(x) = \int_\gamma \frac{mu^\mu u^\nu}{\sqrt{|g|}} \delta^4(x-z(t))dt,
\end{equation}
where $u^\mu\equiv \diff{z^\mu}{t}$ is the four-velocity on $\gamma$, and $|g|$ denotes the absolute value of the determinant of $g_{\alpha\beta}$. Note that there is no contradiction between this equation and Eq.~\eqref{h_E1 eqn}, since the latter applies only in the vacuum region $\Omega$, where $T_{\alpha\beta}$ vanishes pointwise. The solution to this wave equation can be expressed in terms of an integral over the worldline $\gamma$. Near the worldline, the solution can then be expanded in powers of $r$. That calculation is presented in Appendix~\ref{point_particle_soln}. The result is the following:  
\begin{align}\label{external metric buffer expansion}
\hmn{Ett}{1} &= \frac{2m}{r}+\A{}{1,0}+3ma_in^i+r\left[4ma_ia^i+\A{i}{1,1}n^i\right.\nonumber\\
&\quad \left.+m\left(\tfrac{3}{4}a_{\langle i}a_{j\rangle} +\tfrac{5}{3}\etide_{ij}\right)\nhat^{ij}\right]+O(r^2),\\
\hmn{Eta}{1} &= \C{a}{1,0}+r\big(\B{}{1,1}n_a-2m\dot a_a+\C{ai}{1,1}n^i \nonumber\\
&\quad+\epsilon_{ai}{}^j\D{j}{1,1}n^i +\tfrac{2}{3}m\epsilon_{aij}\btide^j_k\nhat^{ik}\big)+O(r^2), \\
\hmn{Eab}{1} &= \frac{2m}{r}\delta_{ab}+(\K{}{1,0}-ma_in^i)\delta_{ab}+\H{ab}{1,0} \nonumber\\
&\quad+r\Big\lbrace\delta_{ab}\big[\tfrac{4}{3}ma_ia^i+\K{i}{1,1}n^i+\tfrac{3}{4}ma_{\langle i}a_{j\rangle}\nhat^{ij}\nonumber\\
&\quad-\tfrac{5}{9}m\etide_{ij}\nhat^{ij}\big] +\tfrac{4}{3}m\etide^i_{\langle a}\nhat_{b\rangle i} +4ma_{\langle a}a_{b\rangle}-\tfrac{38}{9}m\etide_{ab} \nonumber\\
&\quad+\H{abi}{1,1}n^i +\epsilon_i{}^j{}_{(a}\I{b)j}{1,1}n^i+\F{\langle a}{1,1}n^{}_{b\rangle}\Big\rbrace+O(r^2).
\end{align}
Here the uppercase hatted quantities are STF Cartesian tensors that are functions of time alone; they are named following the scheme of Eqs.~\eqref{generic_STF tt}--\eqref{generic_STF ab}. They are constructed from tail integrals, the acceleration, and $\etide$, and their exact form is specified in Table~\ref{STF wrt tail}. Together, they make up the Detweiler-Whiting regular field \cite{Detweiler_Whiting}, a solution to the homogeneous linearized Einstein equation. $\hmn{}{1}$ is given by setting $a_i=\an{0}_i$ in the above expressions.

The buffer-region expansion of the full metric in the outer limit can now be written as
\begin{equation}
g_{E\alpha\beta}=g\coeff{0}_{\alpha\beta}+\e g\coeff{1}_{\alpha\beta}+\e\hmn{\alpha\beta}{1}+\order{r^3,\e r^2,\e^2},
\end{equation}
where $g_{\alpha\beta}$ is given in Fermi coordinates in Eqs.~\eqref{Fermi background tt}--\eqref{Fermi background ab}.

\subsection{Inner expansion}\label{internal solution}
I assume that the internal solution is that of a perturbed Schwarzschild black hole, and I adopt retarded Eddington-Finkelstein coordinates $(U,X^a)$ adapted to that spacetime. The background metric $g_B$ is then given by
\begin{align}
g_{B} &= -f(U,\tilde R)dUdU -2\Omega_adUdX^a\nonumber\\
&\quad+(\delta_{ab}-\Omega_{ab})dX^a dX^b,
\end{align}
where $f(U,\tilde R)=1-\frac{2M(U)}{\tilde R}$, and $\widetilde\Omega_a\equiv X^a/R$ is a function of two angles $\Theta^A$. (Note that in this equation, I have written the metric in non-rescaled coordinates, but I have written the components of the metric in terms of the scaled coordinate $\tilde R$.) Here $M(U)$ is the Bondi mass of the spacetime, divided by the mass at $U=0$. The mass is allowed to depend on $U$ because $g_B$ is required only to solve Eq.~\eqref{inner_eqn0}, which contains no time-derivatives. Next, I expand the components of the metric perturbation $H$ as
\begin{align}
H_{\mu\nu}(U,\tilde R, \Theta^A,\e) &= \e H_{\mu\nu}\coeff{1}(U,\tilde R,\Theta^A)\nonumber\\
&\quad+\e^2 H_{\mu\nu}\coeff{2}(U,\tilde R,\Theta^A)+...
\end{align}
As a boundary condition on these perturbations, I require that they remain regular on the event horizon. In addition, I adopt the light cone gauge~\cite{light_cone_gauge}, defined in retarded polar coordinates by the condition $H^{(n)}_{UR}=H^{(n)}_{RR}=H^{(n)}_{RA}=0$. In this gauge, $U$ and $R$ maintain their geometrical meaning even in the perturbed spacetime: $U$ is constant on each outgoing light cone, and $R$ is an affine parameter on outgoing light rays. I assume that this gauge condition can always be imposed.

The first- and second-order perturbations, along with the time-dependence of $g_B$, must satisfy the vacuum Einstein equations \eqref{inner_eqn1}--\eqref{inner_eqn2}. In Appendix~\ref{perturbed_BH}, I show that $\diff{M}{U}=0$. This implies that Eq.~\eqref{inner_eqn1} becomes $\delta G\coeff{0}_I[H\coeff{1}]=0$, the linearized vacuum EFE for static perturbations. The solutions to this equation have been thoroughly studied~\cite{Zerilli, Regge_Wheeler, Wald_perturbations, Eric_perturbations}. Because of the spherical symmetry of the background spacetime, the equation can be most easily solved by decomposing $H\coeff{1}$ into spherical harmonics: the various harmonics decouple in the linearized Ricci tensor, such that they can be solved independently. In addition, for $\ell>0$, the harmonics can be decomposed into even- and odd-parity sectors, which also decouple. It is known that the gauge-invariant content of the monopole terms in the solution corresponds to a constant shift of the black hole's mass parameter; odd-parity dipole terms correspond to a shift to a nonzero, constant spin; and even-parity dipole perturbations correspond to a shift in center of mass, which can always be removed via a coordinate transformation. In addition, it is known that for all $\ell$, the solutions behave as $\sim\tilde R^\ell$ for $\tilde R\gg 1$.

In the derivation provided by Mino, Sasaki, and Tanaka \cite{Mino_Sasaki_Tanaka}, the monopole and dipole terms in $H\coeff{1}$ were set to zero, on the basis that they correspond to either pure gauge or to mere redefinitions of mass and angular momentum. However, this step is not justified, since the ``constant" shifts in the black hole's parameters are actually functions of time, with a time-dependence to be determined by the higher-order perturbation equations. Also, the fact that the even-parity dipole term corresponds to a shift in center of mass does not mean that it can be trivially ignored; this will be discussed further in the following sections. For $\ell>1$, Mino, Sasaki, and Tanaka took the terms to necessarily behave as $\tilde R^\ell$ in the buffer region. This means that $H\coeff{\emph{n}}$ cannot contain terms of $\ell>n$: since $\e^n\tilde R^\ell=\e^{n-\ell}R^\ell$, if $\ell>n$ then such a term would correspond to negative powers of $\e$ in the outer expansion. Hence, $H\coeff{1}$ can contain only monopole and dipole terms, and since these are set to zero, $H\coeff{1}$ itself must be zero. It then follows that Eq.~\eqref{inner_eqn2} becomes another linearized vacuum EFE for static perturbations, $\delta G\coeff{0}_I[H\coeff{2}]=0$. The solutions to this equation must be purely quadrupolar, since monopole and dipole terms are set to vanish and $H\coeff{2}$ cannot contain terms of $\ell>2$. However, even if the monopole and dipole terms are set to zero, this reasoning remains specious, because solutions with $\ell>n$ \emph{can} exist: though the asymptotically dominant terms behave as $\tilde R^\ell$, subdominant terms can grow less rapidly with $\ell$, as is shown explicitly in Appendix~\ref{perturbed_BH}. (However, these subdominant terms can be removed with a gauge transformation.)

In the derivation provided by Poisson Ref.~\cite{Eric_review,Eric_matching}, all of the above steps were taken, but the quadrupole terms were then further constrained. Rather than finding a general inner solution and then restricting it by imposing a matching condition, Poisson simplified the possible forms of the metric by first imposing a form of the asymptotic matching condition. (Refer back to Sec.~\ref{traditional} for the definition of this condition.) Specifically, he demanded that for $\tilde R\gg 1$, the metric must asymptotically approach that of a vacuum spacetime in retarded coordinates centered on a geodesic. This demand motivated the following ansatz for the internal metric in polar coordinates:
\begin{align}
g_{IUU} &= -f\left[1+\e^2\R^2e_1(\R)\etide^*(U)\right]+\order{\e^3},\\
g_{IUR} & = -1, \\
g_{IUA} & = R\!\left[\tfrac{2}{3}\e^2\R^2\!\left(e_2(\tilde R)\etide^*_A+b_2(\tilde R)\btide^*_A\right)\!+\order{\e^3}\right]\!, \\
g_{IRR} & = g_{IRA} = 0, \\
g_{IAB} & = R^2\Big[\Omega_{AB}-\tfrac{1}{3}\e^2\R^2\left(e_3(\R)\etide^*_{AB}+b_3(\R)\btide^*_{AB}\right)\nonumber\\
&\quad+\order{\e^3}\Big],
\end{align}
where $e_1$, $e_2$, $e_3$, $b_2$, and $b_3$ are undetermined functions constrained to approach 1 for $\tilde R\gg 1$, and the quantities $\ein^*$, $\bin^*_A$, etc., are constructed from tidal fields $\ein_{ab}$ and $\bin_{ab}$, as displayed in Eqs.~\eqref{Estar}--\eqref{BstarAB}. At $\R\to\infty$ (or $\e=0$), this metric is precisely the metric of a vacuum spacetime in retarded coordinates centered on a geodesic. Note that even with the constraint on the asymptotic behavior of the internal metric, the above ansatz is more restrictive than it need be: generally, the free functions of $\R$ could also be functions of $U$, with a $U$-dependence to be determined by the higher-order EFE.

One might wonder why the internal metric is constrained to approach that of a vacuum metric in coordinates centered on a geodesic, rather than being constrained to approach a vacuum metric in coordinates centered on an arbitrarily accelerating worldline, to agree with the form of the external background metric in retarded coordinates. The reason is that the terms linear in the acceleration in the external background metric are even-parity dipole terms, which have been set to zero to ensure that the coordinates are mass-centered. I will return to the relevance of these terms in the following sections, but I will note now that this assumed form already suggests that the body must be moving on a geodesic of some spacetime. That spacetime will turn out to be $g+h^R$, rather than $g$. See Ref.~\cite{Detweiler_review} for further discussion of this point.

Substituting the ansatz into the linearized EFE and imposing regularity at the event horizon determines the free functions. After transforming the resulting metric back into Cartesian-type coordinates, one finds
\begin{align}
g_{IUU} &= -f-\e^2 f^2\R^2\ein^*+O(\e^3),\\
g_{IUa} &= -\Omega_a+\tfrac{2}{3}\e^2\R^2 f(\ein^*_a+\bin^*_a)+O(\e^3),\\
g_{Iab} &= \delta_{ab}-\Omega_{ab}-\tfrac{1}{3}\e^2\R^2\left(1-\frac{2 M^2}{\R^2}\right)\ein^*_{ab} \nonumber\\
&\quad-\tfrac{1}{3}\e^2\R^2\bin^*_{ab}+O(\e^3),
\end{align}
where $\ein_a^*=\ein_A^*\Omega^A_a$, $\bin_a^*=\bin_A^*\Omega^A_a$, $\ein_{ab}^*=\ein_{AB}^*\Omega^A_a\Omega^B_b$, and $\bin_{ab}^*=\bin_{AB}^*\Omega^A_a\Omega^B_b$. Note that there is no a priori relationship between the mass $\e M$ of the internal spacetime and the mass $\e m$ of the point particle perturbation in the external spacetime. Similarly, although the inner solution was specifically constructed to asymptotically approach the form of an external metric in the buffer region, there is no priori relationship between $\ein_{ab}$ and $\etide_{ab}$ or between $\bin_{ab}$ and $\btide_{ab}$. These relationships are to be determined in the matching procedure.

To expand the metric in the buffer region, we rewrite $\tilde R$ as $R/\e$ and then re-expand in powers of $\e$; this corresponds to an expansion for $R\gg \e$. In order to agree with the external metric, which is constructed in Fermi coordinates and in the Lorenz gauge, we must also transform from retarded coordinates and the lightcone gauge into Fermi-like harmonic coordinates $(T,X^a)$; and the result must be decomposed into its irreducible STF pieces. That calculation is shown in Appendix~\ref{perturbed_BH}. The final result is
\begin{align}\label{internal metric buffer expansion}
g_{ITT} & = -1+\e\frac{2M}{R}+\tfrac{5}{3}\e MR\ein_{ij}\hat N^{ij}-R^2\ein_{ij}\hat N^{ij}\nonumber\\
&\quad +\order{\e^2,\e R^2, R^3}, \\
g_{ITa} & = 2\e MR\ein_{ai}N^i+\tfrac{2}{3}\e MR\epsilon_{aij}\bin^j_k\hat N^{ik}\nonumber\\
&\quad +\tfrac{2}{3}R^2\epsilon_{aik}\bin^k_j\hat N^{ij}+\order{\e^2,\e R^2, R^3}, \\
g_{Iab} & = \delta_{ab}\left(1+\e\frac{2M}{R}-\tfrac{5}{9}\e MR\ein_{ij}\hat N^{ij}-\tfrac{1}{9}R^2\ein_{ij}\hat N^{ij}\right) \nonumber\\
&\quad +\tfrac{64}{21}\e MR\ein_{i\langle a}\hat N_{b\rangle}{}^i-\tfrac{46}{45}\e MR\ein_{ab}-\tfrac{1}{9}R^2\ein_{ab} \nonumber\\
&\quad +\tfrac{2}{3}\e MR\ein_{ij}\hat N_{ab}{}^{ij}+\tfrac{2}{3}R^2\ein_{i\langle a}\hat N^i_{b\rangle}\nonumber\\
&\quad -\tfrac{4}{3}\e MR\epsilon_{jk(a}\bin_{b)}^kN^j+\order{\e^2,\e R^2, R^3},
\end{align}
where $N^i=X^i/R$. This is the metric that I will use in the matching procedure, even though, as pointed out above, it has already been heavily restricted.

In Appendix~\ref{perturbed_BH}, I derive the above metric without relying on an ansatz, enabling me to better characterize its generality.

\subsection{Zeroth-order matching}\label{zeroth_order_matching}
I now consider the relationship between the two metrics. Beginning with the zeroth-order weak matching condition, we have the metric in the outer expansion given by
\begin{align}
g_{Ett} &= -1-2r\an{0}_in^i-\tfrac{1}{3}r^2 \an{0}_i \an{0}{}^i-r^2\an{0}_{\langle i}\an{0}_{j\rangle}\nhat^{ij}\nonumber\\
&\quad -r^2\etide_{ij}\nhat^{ij} +O(\e, r^3),\\
g_{Eta} &= \tfrac{2}{3}r^2\epsilon_{aik}\btide^k_j\nhat^{ij}+O(\e, r^3),\\
g_{Eab} &= \delta_{ab} -\tfrac{1}{9}r^2\delta_{ab}\etide_{ij}\nhat^{ij} -\tfrac{1}{9}r^2\etide_{ab}+\tfrac{2}{3}r^2\etide_{i\langle a}\nhat^i_{b\rangle}\nonumber\\
&\quad+O(\e, r^3).
\end{align}
while the metric in the inner expansion is given by
\begin{align}
g_{ITT} &= -1-R^2\ein_{ij}\hat N^{ij} +O(\e, R^3),\label{gITT_buffer}\\
g_{ITa} &= \tfrac{2}{3}R^2\epsilon_{aik}\bin^k_j\hat N^{ij}+O(\e,R^3),\\
g_{Iab} &= \delta_{ab}-\tfrac{1}{9}R^2\delta_{ab}\ein_{ij}\hat N^{ij}-\tfrac{1}{9}R^2\ein_{ab}+\tfrac{2}{3}R^2\ein_{i\langle a}\hat N^i_{b\rangle}\nonumber\\
&\quad+O(\e,R^3)\label{gIab_buffer}.
\end{align}
It seems that we may immediately identify these two metrics and conclude that $T=t$, $X^a=x^a$, $\ein_{ab}=\etide_{ab}$, and most importantly, $\an{0}_\mu=0$. However, the matching condition does not require that these two metrics be identical, since they may be in different coordinate systems; the matching condition requires only that these two metrics be related by a diffeomorphism. But this condition places no restriction at all on the acceleration of the worldline: The form of the inner metric is that of an arbitrary background written in Fermi coordinates centered on a geodesic worldline. The form of the outer metric is that of a known background written in Fermi coordinates centered on a possibly accelerated worldline. Regardless of the value of the acceleration, if the geodesic is embedded in the external spacetime, then these two solutions are obviously related by a diffeomorphism, since the geodesic can be transformed to the accelerated worldline.

Evidently, some information has been lost here. I assumed from the beginning that the inner and outer expansions were performed ``around" the same worldline. In the inner expansion, the ``location" of the body is encoded into the coordinate system by the condition that the body's mass dipole vanishes in that coordinate system; in the outer expansion, the ``location" of the body is encoded in the worldline sourcing the perturbation. If we use the weak matching condition, in which we expand the metric before finding the coordinate transformation between the inner and outer expansions, then this information is lost.

However, one might wonder if this ambiguity might be removed by supplementing the weak matching condition with some other condition. One such condition appears obvious: the coordinate transformation between the inner and outer expansions in the buffer region must be ``small"--that is, it must vanish in the limit $\e\to0$. This removes the possibility of transforming from an arbitrary geodesic to an arbitrarily-accelerated worldline. In the buffer region, $r\to0$ as $\e\to0$, so this allows transformations that have no explicit $\e$-dependence, but which do have explicit $r$-dependence. I trust the reader to convince himself that under such a transformation, we must have $R=r$, $T=t$, the tidal fields appearing in the inner metric must be identical (up to $\order{\e}$ corrections) to those constructed from the Riemann tensor in the outer solution---and the leading-order term in the acceleration must vanish: $\an{0}=0$. The two coordinate systems may, of course, be related by rotations, but these are insignificant.

Hence, we can adopt a stronger, refined matching condition: the inner and outer expansions in the buffer region must be equal up to a unique small coordinate transformation. Unfortunately, this refined condition is still insufficient. The reason is that the inner expansion \emph{could} have included acceleration-type terms. In fact, we can always include such terms by transforming the metric into an accelerating frame. Suppose we begin with the Schwarzschild metric in Kerr-Schild form,
\begin{equation}\label{Kerr-Schild}
g_{B\mu\nu} = \eta_{\mu\nu} + \frac{\e M}{\bar R}\ell_\mu \ell_\nu,
\end{equation}
where $\eta=\text{diag}(-1,1,1,1)$ is the Minkowski metric, $\ell_\mu=\left(1,\frac{\bar X}{\bar R},\frac{\bar Z}{\bar R},\frac{\bar Z}{\bar R}\right)$ is a null vector, $\bar R=\displaystyle\sqrt{\bar X^2+\bar Y^2+\bar Z^2}$, and the (unscaled) coordinates are $(\bar T,\bar X,\bar Y,\bar Z)$. Now, by  using the flat-spacetime transformation from an inertial frame to an accelerated one, we can transform the metric to a new set of accelerated retarded coordinates $(U', R',\Theta'^A)$. For simplicity, assume that the acceleration is in the $\bar Z$-direction. Then the transformation is given by
\begin{align}
\bar T & = T_0(U')+R'(\cos\Theta'\sinh q(U')+\cosh q(U')),\\
\bar X & = R' \sin\Theta'\cos\Phi',\\
\bar Y & = R' \sin\Theta'\sin\Phi',\\
\bar Z & = Z_0(U')+R'(\cos\Theta'\cosh q(U')+\sinh q(U')),
\end{align}
where $T_0=\int\cosh q(U')dU'$, $Z_0=\int\sinh q(U')dU'$, and $q(U')=\int \alpha(U')dU'$, where $\alpha(U')$ is the magnitude of the acceleration. Under this transformation, $g_B$ maintains the form in Eq.~\eqref{Kerr-Schild}. $\eta_{\mu\nu}$ becomes the metric of flat spacetime in retarded coordinates, given by
\begin{align}
\eta &= -\left[(1+R'\alpha\cos\Theta')^2-R'^2 \alpha^2\right]dU'^2-2dU'dR'\nonumber\\
&\quad -2R'^2 \alpha\sin\Theta'dU'd\Theta'+ R'^2d\Omega'^2,
\end{align}
while $\ell_\mu$ takes on a more complicated (and unenlightening) form. Note that in flat spacetime, this transformation translates the spatial origin from $\bar Z=0$ to $\bar Z = Z_0(U')$. And in the spacetime of $g_B$, the same interpretation applies at large distances from the black hole---that is, in the buffer region. In other words, the new coordinates are not mass-centered: the center of mass is moving away from the origin.

Although the metric takes on an inconveniently complicated form in this non--mass-centered coordinate system, in principle one could use it in constructing the inner expansion $g_I$. If one does so, then when $g_I$ is expanded in the buffer region, it becomes $g_{I\mu\nu}=\eta_{\mu\nu}+O(\e,R^2)$, as we can infer immediately from the form of Eq.~\eqref{Kerr-Schild}. But in this expansion, $\eta_{\mu\nu}$ is the metric of flat spacetime centered on an accelerating worldline, not on a geodesic. Therefore, if we transform the metric to Fermi-type coordinates $(T,X^a)$, we arrive at
\begin{align}
g_{ITT} &= -1-2R\alpha\cos\Theta+O(\e,R^2),\\
g_{ITa} &= O(\e,R^2),\\
g_{Iab} &= \delta_{ab}+O(\e,R^2).
\end{align}
This metric agrees with the one in the outer expansion, regardless of the value of the acceleration. We may identify $\alpha\cos\Theta$ with $\an{0}_in^i$, and the matching procedure, even with the refined matching condition, provides no information whatsoever about the worldline.

We can readily see why the matching procedure has failed: we have not insisted on any relationship between the inner and outer expansions. In order for matching to be successful, we must insist that the ``position" of the black hole in the inner expansion can be identified with the position of the worldline in the outer expansion. To make this identification mathematically precise, I insist that the two expansions are to be expanded and matched in the buffer region only when the outer expansion is evaluated in a coordinate system centered on the worldline and the inner expansion is evaluated in a mass-centered coordinate system. If this condition is imposed, then the accelerating coordinate system $(U',R',\Theta^A)$ is inadmissible, since it is not mass-centered. Therefore, we can discount it and others like it---and we can once again, now more confidently, conclude that the acceleration of the worldline must vanish in the limit $\e\to0$. Such a condition serves to implicitly define the worldline, and it is necessary for the matching procedure to be well-defined and to yield unambiguous results.

Based on the above analysis of the zeroth-order matching procedure, I suggest the following matching condition: if the inner expansion is written in a mass-centered coordinate system and the outer expansion is written in a worldline-centered coordinate system, then the two expansions must be equal up to a small coordinate transformation when expanded in the limit of small (outer) radial coordinate distances. (Here ``outer" radial coordinate means a coordinate that is formally of order $1$ in the outer expansion and of order $1/\e$ in the inner expansion.) Making use of this condition allows us to determine the acceleration of the worldline at zeroth order. However, as we shall see in the next subsection, it requires still more restrictions.

\subsection{First-order matching}\label{first_order_matching}
Comparing the expression for the external solution with that for the internal solution, we find that the $1/r$ terms agree if and only if we make the identification $m=M$. In order for the other terms to be made to agree, there must exist a coordinate transformation, from the external coordinates to the internal coordinates, that induces a gauge transformation $g_E\to g_E+\e\delta g_E+\order{\e^2}$, where
\begin{align}
\delta g_{Ett} & = -\A{}{1,0}+r(2\an{1}_i-\A{i}{1,1})n^i+\order{r^2}, \label{required_trans}\\
\delta g_{Eta} & = -\C{a}{1,0}-\tfrac{1}{6}r\partial_t(\A{}{1,0}+3\K{}{1,0})n_a-r\C{ai}{1,1}n^i  \nonumber\\
&\quad-r\epsilon_{ai}{}^j\D{j}{1,1}n^i+2mr\etide_{ai}n^i+\order{r^2},\label{required_trans2}\\
\delta g_{Eab} & = -\delta_{ab}\K{}{1,0}-\H{ab}{1,0}-r\delta_{ab}\K{i}{1,1}n^i \nonumber\\
&\quad-\tfrac{3}{10}r\left(\K{\langle a}{1,1}-\A{\langle a}{1,1}+2\partial_t\C{\langle a}{1,1}\right)n_{b\rangle}\nonumber\\
&\quad-r\H{abi}{1,1}n^i-r\epsilon_i{}^j{}_{(a}\I{b)j}{1,1}n^i +\tfrac{12}{7}mr\etide_{i\langle a}\nhat_{b\rangle}{}^i \nonumber\\
&\quad +\tfrac{16}{5}mr\etide_{ab} +\tfrac{2}{3}mr\etide_{ij}\nhat_{ab}{}^{ij}\nonumber\\
&\quad -\tfrac{4}{3}mr\epsilon_{jk(a}\btide^k_{b)}n^j+\order{r^2}.\label{required_trans3}
\end{align}
I remind the reader that $a_i$ is to be set to $\an{0}_i=0$ in the explicit expressions for the uppercase script tensors.

This transformation is generated by a vector field satisfying $\delta g_{E\alpha\beta}=2\xi_{(\alpha;\beta)}$. I assume the field can be expanded as
\begin{equation}
\begin{array}{lr}
\xi_t = \sum_{n\ge 0}r^n\xi_t\coeff{\emph{n}}, & \xi_a = \sum_{n\ge 0}r^n\xi_a\coeff{\emph{n}},
\end{array}
\end{equation}
where the coefficients $\xi_t\coeff{n}$ and $\xi_a\coeff{n}$ are decomposed as
\begin{align}
\xi_t\coeff{\emph{n}} & = \sum_{\ell\ge 0}\Xi_L\coeff{\emph{n}}\nhat^L,\\
\xi_a\coeff{\emph{n}} & = \sum_{\ell\ge 1}\left(\Upsilon_{aL-1}\coeff{\emph{n}}\nhat_{L-1} +\epsilon_{ab}{}^c\nhat^{bL-1}\Lambda_{cL-1}\coeff{\emph{n}}\right) \nonumber\\
&\quad +\sum_{\ell\ge 0}\Psi_L\coeff{\emph{n}}\nhat_a{}^L.
\end{align}
The Cartesian tensors $\Xi_L$, $\Upsilon_L$, $\Lambda_L$, and $\Psi_L$ are STF in $L$, and they depend only on time.

Calculating $2\xi_{(\alpha;\beta)}$ from the above expansion is straightforward. Demanding that the result of this calculation agrees with Eqs.~\eqref{required_trans}--\eqref{required_trans3} at each order in $r$ then determines a sequence of equations for $\xi\coeff{\emph{n}}$. No $\order{1/r}$ terms appear in Eqs.~\eqref{required_trans}--\eqref{required_trans3}, so from the $\order{1/r}$ terms in $2\xi_{(\alpha;\beta)}$ we find that $\partial_a\xi_t\coeff{0}=0$ and $\partial_{(a}\xi_{b)}\coeff{0}=0$. From this we determine that $\xi_{\alpha}\coeff{0}$ must be independent of angle: $\xi_t\coeff{0} = \Xi\coeff{0}$ and
$\xi_a\coeff{0} = \Upsilon_a\coeff{0}$.

From the $\order{r^0}$ terms, we find
\begin{align}
\partial_t\Xi\coeff{0} &= -\tfrac{1}{2}\A{}{1,0},\\
(n_a+r\partial_a)\xi_t\coeff{1} &= -\partial_t\xi_a\coeff{0}-\C{a}{1,0},\\
(n_{(a}+r\partial_{(a})\xi_{b)}\coeff{1} &= -\tfrac{1}{2}\delta_{ab}\K{}{1,0}-\tfrac{1}{2}\H{ab}{1,0},
\end{align}
from the $tt$-, $ta$-, and $ab$-component, respectively. The first of these equations determines that $\Xi\coeff{0}=-\tfrac{1}{2}\int\!\A{}{1,0}dt$, the second determines that $\Xi_a\coeff{1}=-\C{a}{1,0}-\partial_t\Upsilon_a\coeff{0}$, and the last determines that $\Psi\coeff{1}=-\tfrac{1}{2}\K{}{1,0}$, $\Upsilon_{ab}\coeff{1}=-\tfrac{1}{2}\H{ab}{1,0}$, and $\Lambda_c\coeff{1}$ is arbitrary. All other terms in $\xi_\alpha\coeff{1}$ vanish.

Finally, from the $\order{r}$ terms, we find:
\begin{align}
\partial_t\xi_t\coeff{1} & = \etide^j_in^i\Upsilon_j\coeff{0}\! +\an{1}_in^i\! -\tfrac{1}{2}n^i\A{i}{1,1},\\
(2n_a+r\partial_a)\xi_t\coeff{2} & = -\partial_t\xi_a\coeff{1}+2\etide_{ai}n^i\Xi\coeff{0} \nonumber\\
&\quad -\tfrac{1}{6}\partial_t(\A{}{1,0}+3\K{}{1,0})n_a\nonumber\\
&\quad -\C{ai}{1,1}n^i-\epsilon_{ai}{}^j\D{j}{1,1}n^i-2m\etide_{ai}n^i \nonumber\\
&\quad -2R_{0i}{}^j{}_an^i\Upsilon_j\coeff{0},\\
2(2n_{(a}+r\partial_{(a})\xi_{b)}\coeff{2} & = \tfrac{4}{3}R_{0(ab)i}n^i\Xi\coeff{0} -\tfrac{4}{3}R^j{}_{(ab)i}n^i\Upsilon_j\coeff{0}\nonumber\\
&\quad -\delta_{ab}\K{i}{1,1}n^i+\tfrac{12}{7}m\etide_{i\langle a}\nhat_{b\rangle}{}^i\nonumber\\
&\quad +\tfrac{16}{5}m\etide_{ab}-\H{abi}{1,1}n^i \nonumber\\
&\quad -\tfrac{3}{10}\big(\K{\langle a}{1,1}\!\!-\A{\langle a}{1,1}\!+\!2\partial_t\C{\langle a}{1,1}\big)n_{b\rangle}\nonumber\\
&\quad +\tfrac{2}{3}m\etide_{ij}\nhat_{ab}{}^{ij} -\tfrac{4}{3}m\epsilon_{jk(a}\btide^k_{b)}n^j \nonumber\\
&\quad -\epsilon_i{}^j{}_{(a}\I{b)j}{1,1}n^i.
\end{align}
Again, these equations follow from the $tt$-, $ta$-, and $ab$-component, respectively. The first of them yields the equation of motion
\begin{equation}\label{matching equation of motion}
\partial_t^2\Upsilon_i\coeff{0}+\etide^j_i\Upsilon_j\coeff{0} = \tfrac{1}{2}\A{i}{1,1}-\partial_t\C{i}{1,0}-\an{1}_i,
\end{equation}
the second of them yields
\begin{align}
\Xi\coeff{2} & = -\tfrac{1}{12}\partial_t\A{}{1,0}, \\
\Xi\coeff{2}_{ab} & = \tfrac{5}{16}\partial_t\H{ab}{1,0}+\tfrac{5}{4}\etide_{ab}\Xi\coeff{0} -\tfrac{5}{8}\C{ab}{1,1}-\tfrac{5}{4}m\etide_{ab}\nonumber\\
&\quad +\tfrac{5}{4}\epsilon^j{}_{i\langle a}\btide_{b\rangle}^i\Upsilon_j\coeff{0},\\
\partial_t\Lambda_c\coeff{1} & = -\D{c}{1,1} +\tfrac{1}{2}\epsilon_c{}^{pq}\epsilon^{i}{}_{jp}\btide^j_q\Upsilon_i\coeff{0},
\end{align}
and the last of them yields (after some algebra)
\begin{align}
\Upsilon_a\coeff{2} & = -\tfrac{1}{2}\etide^j_a\Upsilon_j\coeff{0}+\tfrac{3}{16}\A{a}{1,1} -\tfrac{3}{8}\partial_t\C{a}{1,0},\\
\Upsilon_{ab}\coeff{2} & = \tfrac{6}{5}m\etide_{ab},\\
\Upsilon_{abc}\coeff{2} & = -\tfrac{1}{4}\H{abc}{1,1},\\
\Psi_a\coeff{2} & = -\tfrac{9}{20}(\K{a}{1,1}+\tfrac{1}{4}\A{a}{1,1} -\tfrac{1}{2}\partial_t\C{a}{1,0})+\tfrac{1}{2}\etide^j_a\Upsilon_j\coeff{0},\\
\Psi_{ab}\coeff{2} & = \tfrac{1}{3}m\etide_{ab},\\
\Lambda_{ab}\coeff{2} & = -\tfrac{1}{2}\I{ab}{1,1}-\tfrac{2}{3}m\btide_{ab} -\tfrac{2}{3}\btide_{ab}\Xi\coeff{0}+\tfrac{2}{3}\epsilon_i{}^j_{(a}\etide^i_{b)}\Upsilon_j\coeff{0}.
\end{align}
All other terms vanish.

In summary, the first three terms in the expansion of the gauge vector field are given by
\begin{align}
\xi_t\coeff{0} &= -\tfrac{1}{2}\int\!\!\A{}{1,0}dt,\\
\xi_a\coeff{0} &= \Upsilon_a\coeff{0},
\end{align}
where $\Upsilon_a\coeff{0}$ is a function of time satisfying the equation of motion \eqref{matching equation of motion},
\begin{align}
\xi_t\coeff{1} &= (\C{i}{1,0}-\partial_t\Upsilon_i\coeff{0})n^i,\\
\xi_a\coeff{1} &= \epsilon_a{}^{ij}n_i\left(\int\!\!\D{j}{1,0}dt +\tfrac{1}{2}\epsilon_j{}^{pq}\epsilon^\ell{}_{kp} \int\!\!\btide^k_q\Upsilon_\ell\coeff{0}dt\right) \nonumber\\
&\quad +\tfrac{1}{2}\K{}{1,0}n_a+\tfrac{1}{2}\H{ai}{1,0}n^i,
\end{align}
and
\begin{align}
\xi_t\coeff{2} &= \tfrac{5}{8}\big(-\tfrac{1}{2}\partial_t\H{ij}{1,0} +2\etide_{ij}\Xi\coeff{0}  +\C{ij}{1,1}+2m\etide_{ij} \nonumber\\
&\quad +2\epsilon^k{}_{p\langle i}\btide^c_{j\rangle}\Upsilon_k\coeff{0}\big)\nhat^{ij}+\tfrac{1}{12}\partial_t\A{}{1,0},\\
\xi_a\coeff{2} &= \left[\tfrac{1}{2}\Upsilon_j\coeff{0}\etide^j_i +\tfrac{9}{20}(\K{i}{1,1}+\tfrac{1}{4}\A{i}{1,1} -\tfrac{1}{2}\partial_t\C{i}{1,0})\right]\nhat_a^i  \nonumber\\
&\quad +\tfrac{1}{3}m\etide_{ij}\nhat_{ab}{}^{ij} -\tfrac{1}{2}(\Upsilon_j\coeff{0}\etide^j_a+\tfrac{3}{8}\A{a}{1,1} -\tfrac{3}{4}\partial_t\C{a}{1,0}) \nonumber\\
&\quad -\tfrac{6}{5}m\etide_{ai}n^i+\tfrac{1}{4}\H{abi}{1,1}n^i +\epsilon_{aij}\nhat^{ik}\Big(\tfrac{2}{3}\epsilon^{cd}{}_{(j} \etide_{k)c}\Upsilon_d\coeff{0} \nonumber\\
&\quad +\tfrac{1}{2}\I{jk}{1,1}+\tfrac{2}{3}m\btide_{jk} -\tfrac{1}{3}\btide_{jk}\int\!\!\A{}{1,0}dt\Big).
\end{align}
This is the most general transformation that succeeds in making the exterior solution identical to the interior solution, up to order $\e r$. It has one free function of time: $\Upsilon_a\coeff{0}$.

Despite the refinement of the matching condition formulated in the zeroth-order matching procedure, this coordinate transformation has failed to uniquely identify the acceleration of the worldline. Instead, it determines an equation for $\Upsilon_a\coeff{0}$, given by Eq.~\eqref{matching equation of motion}. Consider the meaning of this equation. In the internal solution, the mass dipole and all dipole perturbations have been set to zero, and an acceleration term in the buffer region corresponds to a dipole perturbation. Equation~\eqref{matching equation of motion} thus tells us that for any given acceleration $\an{1}_i$, we can perform a small, angle- and $r$-independent spatial translation (in the buffer region) that ensures the dipole perturbation vanishes.

In order to arrive at the correct equation, $\an{1}_i=\tfrac{1}{2}\A{i}{1,0}-\partial_t\C{i}{1,0}$, one must further restrict the matching condition. Recall that the coordinate transformation must be small, which implies that $\Upsilon_i\coeff{0}$ must remain of order unity. If the right-hand side of Eq.~\eqref{matching equation of motion} does not vanish, then $\Upsilon_i\coeff{0}$ will generically grow large; more precisely, on a timescale such as $\sim1/\e$, which becomes unbounded in the limit $\e\to 0$, $\Upsilon_i\coeff{0}$ will generically become larger than order unity. However, it will not \emph{necessarily} grow large (e.g., if the right-hand side of Eq.~\eqref{matching equation of motion} is purely oscillatory). Thus, we cannot conclude that the right-hand side must vanish based on the refined matching condition of the previous section. Instead, I propose a final version of the \emph{refined matching condition}: \emph{if the inner expansion is written in a mass-centered coordinate system and the outer expansion is written in a worldline-centered coordinate system, then the two expansions must be equal up to a} necessarily \emph{small coordinate transformation when the inner expansion is expanded in the limit of small (outer) radial coordinate distances.} In other words, the coordinate transformation must not only be small, but must necessarily remain so on all timescales of interest.

With this final refinement, we can conclude the following: if on an unbounded timescale, (i) the exact metric possesses inner and outer expansions, (ii) there exists a local coordinate system in which the metric in the inner expansion is given by that of a tidally perturbed black hole, up to errors of order $\e^3$, (iii) there exists a global coordinate system in which the metric in the outer expansion is that of the external background $g$ plus a point-particle solution to the wave equation~\eqref{h_E1 eqn}, up to errors of order $\e^2$, and (iii) the exact solution satisfies the refined matching condition presented above, then the worldline defining the point particle perturbation has an acceleration given by
\begin{equation}
\an{1}_i = \tfrac{1}{2}\A{i}{1,1}-\partial_t\C{i}{1,0},
\end{equation}
where $\A{i}{1,0}$ and $\C{i}{1,0}$ are obtained by setting $a_i=\an{0}_i=0$ in Table~\ref{STF wrt tail}. This is the MiSaTaQuWa equation.

It may be that some or all of these assumptions can be removed, even within the context of matched asymptotic expansions. For example, if the inner and outer expansions exist, then the solution to the wave equation~\eqref{h_E1 eqn} is necessarily that with a point particle source. Also, the inner metric could correspond to a body other than a black hole. If it were taken to be a tidally perturbed, otherwise spherically symmetric neutron star, for example, then the equation of motion would be unaffected: outside the star, the metric would be altered only by the presence of induced tidal moments, which scale as $\e^{2\ell+1}$ and hence would not appear in the first-order matching procedure \cite{Love_numbers1,Love_numbers2}. It is also possible that matching the inner and outer solutions to higher order in $r$ would show that $\Upsilon_a\coeff{0}$ must vanish, and that the refined matching condition is needlessly strong; however, there is no obvious indication of the order at which this would occur.

\subsection{Interpretation and commentary}\label{commentary}
Let us interpret the above calculation. First, note that a large part of the transformation consists of removing tail terms. This can be understood as follows: In the Fermi coordinates centered on the worldline in $g$, the spacetime appears to be that of a singular monopole perturbation $h^S$, plus a regular homogeneous perturbation $h^R$, plus the field of the smooth background metric $g$ expanded about some worldline. But in the coordinates $X^\alpha$, at a large distance $R\gg \e$ from the body, the spacetime appears to be simply a singular monopole perturbation atop some smooth background field that is expanded around a geodesic. Therefore, transforming between these coordinates can be understood as transforming from the Fermi coordinates of $g$ into the Fermi coordinates of $g+h^R$, where $h^R$ is the Detweiler-Whiting regular field. Reference~\cite{Detweiler_review} contains further discussion of this point.

For example, the angle- and $r$-independent monopole term $\A{}{1,0}=\tail_{tt}$ is removed because proper time must be measured in $g+h^R$, rather than in $g$. Similarly, the dipole terms in the perturbation are removed because the body is nonspinning and non-accelerating in $g+h^R$,  rather than in $g$. And if we proceeded to order $\e r^2$ in the matching procedure, we would find that the tidal fields appearing in the inner expansion are those of $g+h^R$, rather than those of $g$.

These general points are shared with the derivation in Refs.~\cite{Eric_matching,Eric_review}. Note, however, that my results differ considerably from those of Refs.~\cite{Mino_Sasaki_Tanaka,Eric_matching,Eric_review}. The first difference is that in those earlier calculations it was found that the tetrad on the worldline is not parallel-propagated in the external background spacetime. This followed from the fact that the spin dipole perturbation in the internal solution had been set to zero via a choice of gauge; effectively, the Fermi tetrad was required to rotate with the perturbed gravitational field, to set the total spin to zero. However, this is not necessary: transforming from the external, nonrotating frame, to the internal, rotating frame (or to a nonrotating frame in $g+h^R$), simply requires a gauge transformation (specifically generated by $\Lambda\coeff{1}_c$). There is no reason to require the external, background Fermi tetrad in $g$ to spin.

More importantly, my analysis has shown that the weak matching condition that is normally utilized is actually too weak to yield unique results. In order to arrive at an equation of motion, I have had to formulate a refined matching condition in a somewhat ad hoc manner. But even if that refinement is accepted, all of these derivations rely on another strong assumption: the form of the inner expansion, which fixes not only the background metric, but also the form the perturbations. By assuming that the metric in the inner expansion appears to that of a singular monopole perturbation atop some smooth background field that is expanded around a geodesic, the matching derivations of the self-force seem to implicitly assume a generalized equivalence principle: they assume that in vacuum, the black hole, as viewed from a distance, moves on a geodesic of some smooth spacetime. Given that such a geodesic exists, the matching procedure provides a means of determining \emph{which} smooth spacetime the geodesic lies in---but it does not prove the existence of the geodesic.

If the inner expansion is to be sufficiently general for the matching procedure to \emph{derive} the generalized equivalence principle, rather than assume it, then one must use a less restricted inner expansion. For example, at linear order, no acceleration-like dipole term (i.e. one behaving as $\sim r$ in the buffer region) can arise in the inner expansion without also introducing a mass dipole. However, in the inner expansion an acceleration term $\e r\an{1}$ corresponds to a second-order perturbation $\e^2\tilde R \an{1}$, so in order to maintain that no such term can arise in mass-centered coordinates, one must solve the second-order EFE in generality. In addition, if one uses the refined matching condition or some variant of it, one must prove that an acceleration term cannot arise from a generically small coordinate transformation in the buffer region. It is not immediately clear that at second order, being in mass-centered coordinates implies that any acceleration-like term must vanish.\footnote{Kinnersley's photon rocket is an example of an exact solution to the EFE with acceleration-like dipole terms sourced by radiation \cite{Kinnersley, Bonnor_rocket, Damour_rocket}.} Solving the EFE, in full generality, at second order also requires one to include potentially time-evolving shifts in the mass and spin of the black hole. The time-evolution, if any, of these parameters would be determined at second order. Indeed, a time-dependent correction to the mass was found in Ref.~\cite{my_paper}.

In that paper, I performed a second-order analysis, but of a different sort than the one just suggested. Instead of assuming that the small body is a black hole, and trying to solve the second-order EFE in the whole of that black hole's spacetime, I allowed the body to be arbitrarily structured, and I solved the second-order EFE in the outer expansion, and only in the buffer region. (See Refs.~\cite{Kates_motion, Gralla_Wald} for similar approaches at first order and in the case of a regular expansion, respectively.) Given the weaknesses of the method of matched asymptotic expansions, such a calculation provides a much firmer conclusion. However, if one is willing to accept the additional assumptions, matched asymptotic expansions provide a much simpler means of finding an equation of motion.

\section{Conclusion}\label{conclusion}
In this paper, I have discussed three types of general expansions: dual expansions, multiscale expansions, and a self-consistent expansion. Each of these can be used to systematically overcome the unbounded errors of a regular expansion, and at least two of them will likely be required in any successful analytical approach to the problem of motion for asymptotically small bodies. For example, whether a multiscale expansion or a self-consistent expansion is used to eliminate secular errors, it must be combined with an inner expansion to eliminate errors near the small body. And these expansions may need to be supplemented with a third expansion in the wave zone.

However, these expansions are not merely powerful tools for finding asymptotic solutions to differential equations: they have a rich underlying geometrical structure, which I have sketched, but which warrants further study. This underlying structure allows us to, for example, easily define and conceptualize the representative worldline of a black hole. Careful consideration of the underlying formalism can also reveal subtle yet important features of a calculation. For example, in moving from the realm of traditional perturbation theory to one with potentially multiple coordinate systems, two types of matching conditions arise, and the condition that has been implicitly used in previous derivations of the gravitational self-force is significantly weaker than the condition used in applied mathematics. In order to arrive at unique results with this matching condition, additional assumptions must be made, which weakens the conclusions of a matching calculation.

It is my hope that this paper, by highlighting points such as these, will spur further research into the foundations of singular perturbation techniques in GR, in order to promulgate their utility, illuminate their underlying structure, and put them on a more precise and deductive mathematical basis.

\begin{acknowledgments}
Thanks to Eric Poisson and Bob Wald for helpful feedback, and to the anonymous referee for a careful reading of the mathematical expressions. This work was supported by the Natural Sciences and Engineering Research Council of Canada.
\end{acknowledgments}

\appendix
\section{Illustrative example of matched asymptotic expansions}\label{matching_example}
To illustrate the procedure of matched asymptotic expansions, I consider the following boundary value problem:
\begin{equation}\label{matchDE}
\e\ddiff{\exact{f}}{r}+\diff{\exact{f}}{r}+\exact{f}=0, \qquad \exact{f}(0)=0,\ \exact{f}(1)=1.
\end{equation}
I assume that $\exact{f}$ possesses an outer expansion $f_{\text{out}}=f_{\text{out}}\coeff{0}(r)+\e f_{\text{out}}\coeff{1}(r)+...$. Substituting this expansion into Eq.~\eqref{matchDE} and equating coefficients of each power of $\e$ to zero, we find $\diff{f_{\text{out}}\coeff{0}}{r}+f_{\text{out}}\coeff{0}=0$ and $\diff{f_{\text{out}}\coeff{1}}{r}+f_{\text{out}}\coeff{1}=-\ddiff{f_{\text{out}}\coeff{0}}{r}$; the boundary conditions are $f_{\text{out}}\coeff{0}(0)=f_{\text{out}}\coeff{1}(0)=0$, $f_{\text{out}}\coeff{0}(1)=1$, and $f_{\text{out}}\coeff{1}(1)=0$. The general solution to the zeroth-order differential equation is $f_{\text{out}}\coeff{0}(r)=C\coeff{0}e^{-r}$. This solution can satisfy the boundary condition at $r=1$ (by setting $C\coeff{0}=e$), but it cannot satisfy the condition at $r=0$. Hence, we guess that there is a boundary layer at $r=0$, and we choose only to satisfy the boundary condition at $r=1$. Doing the same for $f_{\text{out}}\coeff{1}$, we find $f_{\text{out}}\coeff{1}=(1-r)e^{1-r}$, yielding the first-order outer expansion
\begin{equation}
f_{\text{out}}= e^{1-r}+\e(1-r)e^{1-r}+...
\end{equation}

Now, in order to construct our inner expansion, we require a choice of rescaled coordinate $\tilde r$. Suppose that we choose $\tilde r=r/\e^p$. Substituting this into Eq.~\eqref{matchDE} and taking the limit $\e\to 0$, we find that if $0<p<1$, then the leading-order differential equation is $\diff{f_{\text{in}}\coeff{0}}{\tilde r}=0$. If $p>1$, then the equation becomes $\ddiff{f_{\text{in}}\coeff{0}}{\tilde r}=0$. And if $p=1$, then it becomes $\ddiff{f_{\text{in}}\coeff{0}}{\tilde r}+\diff{f_{\text{in}}\coeff{0}}{\tilde r}=0$. This is called a distinguished limit (also known as a significant degeneration) of the equation, because it contains within it all the terms appearing in the other two limiting equations. We can intuit that a distinguished limit will yield an approximation with maximal information. And although there is no guarantee that a coordinate leading to a distinguished limit is the ideal choice, it has proven to be the most reliable one.

So, proceeding with the rescaled variable $\tilde r=r/\e$, I rewrite Eq.~\eqref{matchDE} as
\begin{equation}\label{matchDE2}
\ddiff{\exact{f}}{\tilde r}+\diff{\exact{f}}{\tilde r}+\e\exact{f}=0, \qquad \exact{f}(0)=0,
\end{equation}
where, technically, $\exact{f}$ stands in for $\trans_*\exact{f}$. I assume that $\exact{f}(\tilde r)$ possesses an inner expansion $f_{\text{in}}(\tilde r,\e)=f_{\text{in}}\coeff{0}(\tilde r)+\e f_{\text{in}}\coeff{1}(\tilde r)+...$. After substituting this into Eq.~\eqref{matchDE2} and solving order-by-order, we find
\begin{align}
f_{\text{in}} &= D\coeff{0}(1-e^{-\tilde r})\nonumber\\
&\quad+\e\left[D\coeff{1}(1-e^{-\tilde r})-D\coeff{0}\tilde r(1+e^{-\tilde r})\right]+...
\end{align}

We can now make use of one of the matching conditions to determine the integration constants $D\coeff{0}$ and $D\coeff{1}$. Since $\lim_{\tilde r\to\infty}f\coeff{0}_{\text{in}}(\tilde r)=D\coeff{0}$ and $\lim_{r\to0}f_{\text{out}}\coeff{0}=e$, the asymptotic matching condition implies $D\coeff{0}=e$. In order to determine $D\coeff{1}$, I next make use of the coefficient-matching condition. Rewriting $f_{\text{in}}$ as a function of $r$ and expanding to order $\e$, we find
\begin{equation}\label{fin buffer}
f_{\text{in}}(r)=e+\e D\coeff{1}-er+...
\end{equation}
Note that $e^{-r/\e}=o(\e^n)$ for all $n>0$, so it vanishes in this expansion. Dropping the ellipses, this yields the left-hand side of Eq.~\eqref{matching_equation}:
\begin{equation}
\expand^1_{\e}\trans^*\expand^1_{\e}\trans_*\exact{f}= e+\e D\coeff{1}-er.
\end{equation}
Next, expanding $f_{\text{out}}$ to linear order in $r$, we find
\begin{equation}
f_{\text{out}} = e(1-r)+\e(1-2r)e+...
\end{equation}
This expansion contains an order-$\e r$ term, which is smaller than any term in Eq.~\eqref{fin buffer}; such a term would be matched by a term from $f\coeff{2}_{\text{in}}$, and so we can neglect it here. The extra expansion on the right-hand side of Eq.~\eqref{matching_equation} serves to remove such terms, and so we have
\begin{equation}
\expand^1_{\e}\trans^*\expand^1_{\e}\trans_*\expand^1_{\e}\exact{f} = e(1-r)+e\e.
\end{equation}
Hence, the matching condition now determines that $D\coeff{1}=e$, and we have fully determined the inner expansion:
\begin{equation}
f_{\text{in}}= e(1-e^{-\tilde r})+\e e\left[(1-e^{-\tilde r})-\tilde r(1+e^{-\tilde r})\right]+...
\end{equation}

Using these results, we can construct the uniformly accurate composite expansion
\begin{align}
f_{\text{comp}} &= \left\lbrace e^{1-r}+\e(1-r)e^{1-r}\right\rbrace+\Big\lbrace e(1-e^{-r/e}) \nonumber\\
&\quad +\e e\left[(1-e^{-r/e})-r/\e(1+e^{-r/\e})\right]\Big\rbrace \nonumber\\
&\quad - \left\lbrace e(1-r)+e\e \right \rbrace \nonumber\\
& = e^{1-r}-(1+r)e^{1-r/\e}\nonumber\\
&\quad +\e\left[(1-r)e^{1-r}-e^{1-r/\e}\right],
\end{align}
where the first equality should be compared to Eq.~\eqref{composite}.

\begin{figure}
\begin{center}
\includegraphics[width=2.7 in, height = 2.7 in]{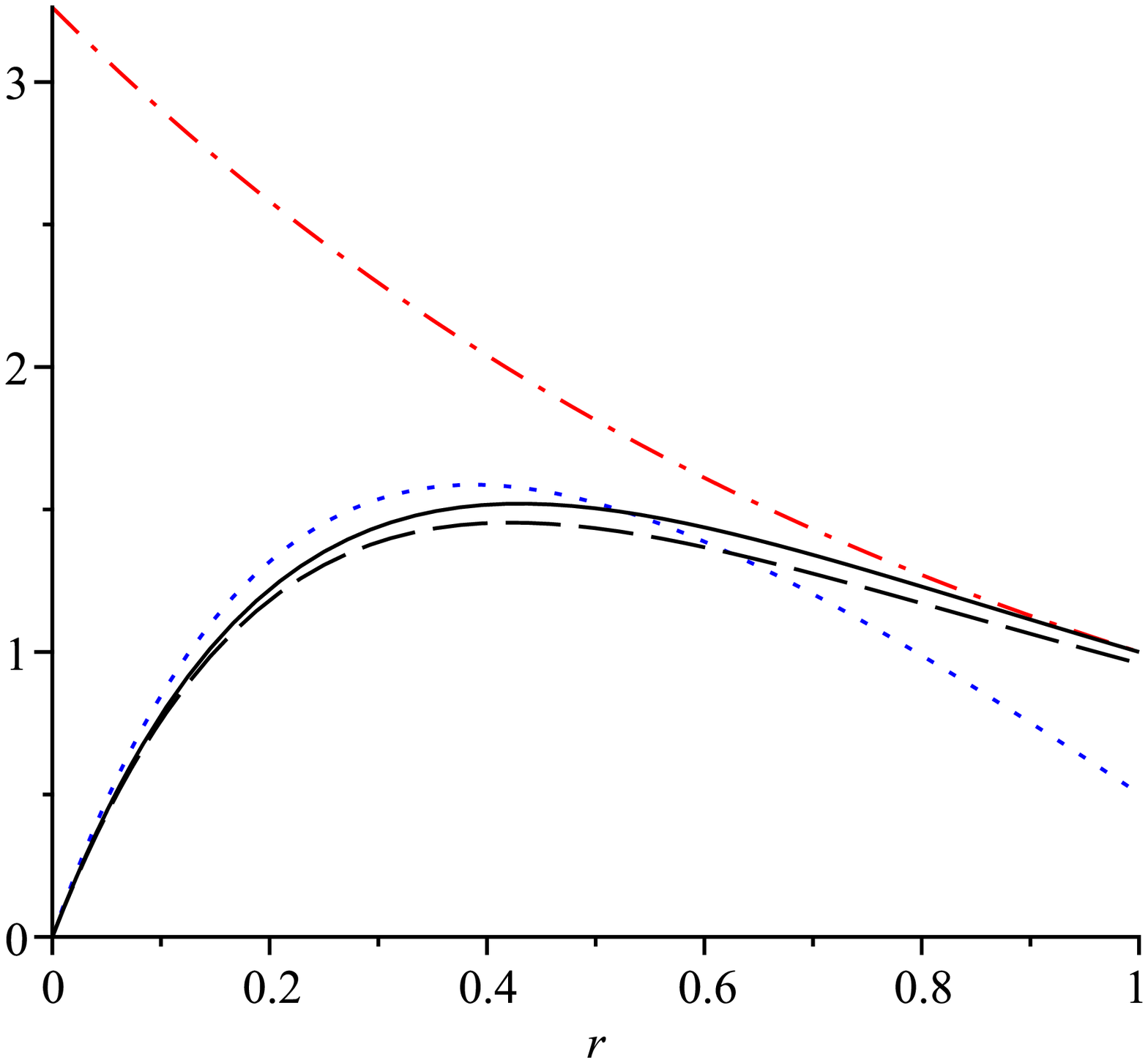}
\includegraphics[width=2.7 in, height = 2.7 in]{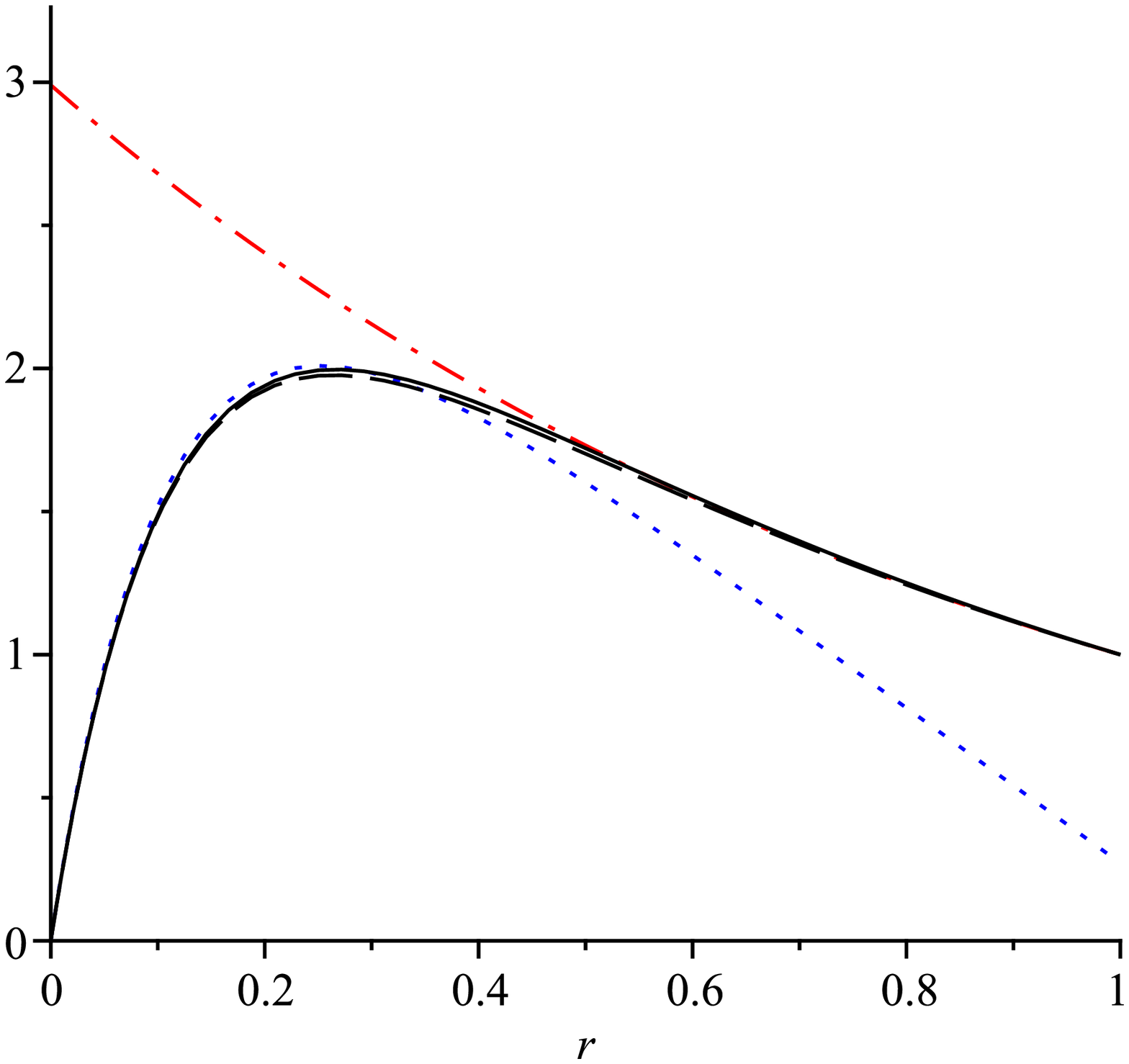}
\end{center}
\caption[An example of matched asymptotic expansions]{Comparisons of the exact solution $\exact{f}$ (the solid black curve), the inner solution $f_{\text{in}}$ (dotted blue), the outer solution $f_{\text{out}}$ (dot-dashed red), and the composite solution $f_{\text{comp}}$ (dashed black). The upper plot displays the solutions for $\e=0.2$; the lower, for $\e=0.1$.}
\label{matching_example_figure}
\end{figure}

In this case, we can compare our results to the exact solution to Eq.~\eqref{matchDE}, which is given by
\begin{equation}
\exact{f}=\frac{\exp\left[-(1-\sqrt{1-4\e})\frac{r}{2\e}\right]-\exp\left[-(1+\sqrt{1-4\e})\frac{r}{2\e}\right]}{\exp\left[-(1-\sqrt{1-4\e})\frac{1}{2\e}\right]-\exp\left[-(1+\sqrt{1-4\e})\frac{1}{2\e}\right]}.
\end{equation}
Note that this function does not exist at $\e=0$. Hence, the regular series $f_{\text{out}}$ is not a Taylor series expansion of $\exact{f}$. However, the limit $\lim_{\e\to0}\exact{f}$ does exist, and $f_{\text{out}}$ is given by $\sum\frac{\e^n}{n!}\lim_{\e\to 0}\frac{\partial^n\exact{f}}{\partial\e^n}$.
One can straightforwardly check that $f_{\text{in}}$ (when written as a function of $r$) is a uniform, first-order asymptotic approximation on an extended domain $D_{\text{in}}=\lbrace r\,:\, 0\le r\ll 1\rbrace$, and  $f_{\text{out}}$ is a uniform, first-order approximation on $D_{\text{out}}=\lbrace r\,:\,|\e\ln\e|\ll r\le 1\rbrace$. So at first order, the overlap region exists, and it is given by $|\e\ln\e|\ll r\ll 1$ (which is notably smaller than the buffer region). One can also verify that $f_{\text{comp}}$ is a uniform approximation on the whole interval $[0,1]$. Figure~\ref{matching_example_figure} shows a graphical comparison of the exact, inner, outer, and composite solutions.

\section{Illustrative example of multiscale expansions}\label{multiscale_example}
In this appendix, I present an illustrative example of multiscale expansions, along with a resultant discussion of their utility. I consider the following differential equation, adapted from the text by Kevorkian and Cole~\cite{Kevorkian_Cole}
\begin{equation}\label{DE1}
\ddiff{\exact{f}}{t}+2\e\diff{\exact{f}}{t}+\exact{f}=0, \qquad \exact{f}(0,\e)=0,\ \diff{\exact{f}}{t}(0,\e)=1.
\end{equation}
Suppose we wish to solve this problem using a regular power series $\exact{f}(t,\e)=\sum_{n\ge 0}\e^nf\coeff{n}(t)$. After substituting this series and equating powers of $\e$, we arrive at the sequence of equations
\begin{equation}
\begin{array}{lll}
\displaystyle\ddiff{f\coeff{0}}{t}+f\coeff{0} & = 0, & \quad f\coeff{0}(0)=0,\ \displaystyle\diff{f\coeff{0}}{t}(0)=1, \vspace*{5.5 pt}\\ 
\displaystyle\ddiff{f\coeff{1}}{t}+f\coeff{1} & = -2\displaystyle\diff{f\coeff{0}}{t}, & \quad f\coeff{1}(0)=0,\ \displaystyle\diff{f\coeff{1}}{t}(0)=0.
\end{array}
\end{equation}
The solutions to these equations are easily found to be $f\coeff{0}(t)=\sin t$ and $f\coeff{1}(t)=-t\sin t$, so we have 
\begin{equation}\label{regular}
\exact{f}(t,\e)=\sin t-\e t\sin t+...
\end{equation}
Based on the unbounded growth of this solution, we surmise that it fails to uniformly approximate the exact solution on any unbounded interval $[0,1/\e^p]$, $p>0$.

To improve on this solution, I adopt the following assumption: there exists a function $F(t,\tilde t,\e)$ satisfying the equality $F(t,\tilde t=\e t, \e)=\exact{f}(t,\e)$. Substituting this into Eq.~\eqref{DE1} and making use of the chain rule $\frac{d}{dt}=\frac{\partial}{\partial t}+\e\frac{\partial}{\partial \t}$, we arrive at
\begin{equation}\label{DE2}
\pddiff{F}{t}+F+2\e\left(\frac{\partial^2F}{\partial\tilde t\partial t}+\pdiff{F}{t}\right)+\e^2\left(\pddiff{F}{\tilde t}+2\pdiff{F}{\tilde t}\right)=0.
\end{equation}
Now, the fundamental idea in a multiscale expansion is that the function $F$ satisfies this equation not just when $\tilde t=\e t$, but also when $\tilde t$ is treated as an independent coordinate. This means that if one assumes a regular expansion $F(t,\tilde t,\e)=\sum_{n\geq 0}\e^n F\coeff{\emph{n}}(t,\tilde t)$, then the coefficient of each power of $\e$ in Eq.~\eqref{DE2} must vanish; if we insisted on solving the equation only at $\tilde t=\e t$, then the $\e$-dependence embedded in $\tilde t$ would prevent us from concluding that the equation must be satisfied order-by-order in this way. (Of course, we could always solve the equation by setting the coefficient of each power of $\e$ to zero, but we could not deduce that each coefficient \emph{must} vanish.)

So, following this procedure, we arrive at a new sequence of equations, 
\begin{align}
\pddiff{F\coeff{0}}{t}+F\coeff{0} &= 0, \\
\pddiff{F\coeff{1}}{t}+F\coeff{1} &= -2\pdiff{F\coeff{0}}{t}-2\frac{\partial^2 F\coeff{0}}{\partial\tilde t\partial t},\label{F1}\\
\pddiff{F\coeff{2}}{t}+F\coeff{2} &= -2\pdiff{F\coeff{1}}{t}-2\frac{\partial^2 F\coeff{1}}{\partial\tilde t\partial t}-\pddiff{F\coeff{0}}{\tilde t}\nonumber\\
&\quad-2\pdiff{F\coeff{0}}{\tilde t},\label{F2}
\end{align}
subject to the initial conditions $F\coeff{\emph{n}}(0,0)=0$ for $n\ge 0$, $\pdiff{F\coeff{0}}{t}(0,0)=1$, and $\pdiff{F\coeff{\emph{n}}}{t}(0,0) =-\pdiff{F\coeff{\emph{n}-1}}{\tilde t}(0,0)$ for $n>0$. The solution to the first equation is
\begin{equation}
F\coeff{0}=A\coeff{0}(\tilde t)\sin t+B\coeff{0}(\tilde t)\cos t,
\end{equation}
where the initial conditions on $F\coeff{0}$ do not fully determine the slow evolution of $A\coeff{0}$ and $B\coeff{0}$, but only impose $A\coeff{0}(0)=1$ and $B\coeff{0}(0)=0$. The general solution to the second equation is
\begin{align}
F\coeff{1} &= A\coeff{1}(\tilde t)\sin t+B\coeff{1}(\tilde t)\cos t\nonumber\\
&\quad-\left(A\coeff{0}(\tilde t)+\pdiff{A\coeff{0}}{\tilde t}(\tilde t)\right)(\cos t+2t\sin t)\nonumber\\
&\quad-\left(B\coeff{0}(\tilde t)+\pdiff{B\coeff{0}}{\tilde t}(\tilde t)\right)t\cos t.
\end{align}
I now make a final assumption, called the \emph{no-secularity condition}: the ratio of successive terms, $F\coeff{\emph{n}+1}/F\coeff{\emph{n}}$, must be bounded. This means that the terms $t\sin t$ and $t\cos t$ are inadmissible, and so their coefficients must vanish. In this case, we have $\diff{A\coeff{0}}{\tilde t}+A\coeff{0}=0$ and $\diff{B\coeff{0}}{\tilde t}+B\coeff{0}=0$, subject to $A\coeff{0}(0)=1$ and $B\coeff{0}(0)=0$. Solving these equations, we find $A\coeff{0}=e^{-\tilde t}$ and $B\coeff{0}=0$. Hence, we have now fully determine $F\coeff{0}$ to be
\begin{equation}
F\coeff{0}=e^{-\tilde t}\sin t.
\end{equation}
And we have
\begin{equation}
F\coeff{1} = A\coeff{1}(\tilde t)\sin t+B\coeff{1}(\tilde t)\cos t,
\end{equation}
where the initial conditions on $F\coeff{1}$ imply that $A\coeff{1}(0)=B\coeff{1}(0)=0$.

If we ceased our work here, there would be no signal that our assumed expansion cannot, in fact, satisfy the non-secularity condition. Following the same procedure for Eq.~\eqref{F2} as we did for Eq.~\eqref{F1}, we find that in order to avoid secular growth in $F\coeff{2}$, the functions $A\coeff{1}$ and $B\coeff{1}$ must satisfy the equations $\diff{A\coeff{1}}{\tilde t}+A\coeff{1}+\tfrac{1}{2}e^{-\tilde t}=0$ and $\diff{B\coeff{1}}{\tilde t}+B\coeff{1}-e^{-\tilde t}=0$, along with the initial conditions $A\coeff{1}(0)=B\coeff{1}(0)=0$. The solutions to these equations are the secularly growing functions $A\coeff{1}=\tilde t e^{-\tilde t}$ and $B\coeff{1}=-\tfrac{1}{2}\tilde t e^{-\tilde t}$. Thus, in order to avoid secular growth in $F\coeff{2}$, we must introduce secular growth into $F\coeff{1}$. In other words, the expansion has failed.

In this case, we can determine the precise reason for the failure. The exact solution to the original ODE is
\begin{equation}
\exact{f}(t,\e) = \frac{e^{-\e t}}{\sqrt{1-\e^2}}\sin(t\sqrt{1-\e^2}).
\end{equation}
If we expand this in a regular power series, we arrive at $\exact{f}(t,\e)=\sin t-\e t\sin t+...$, agreeing with the regular expansion given in Eq.~\eqref{regular}. But we find by inspection that $\exact{f}(t,\e)$ cannot be written as $F(t,\tilde t,\e)$ in such a way that a regular expansion of $F$ satisfies the no-secularity condition. While $e^{-\e t}$ can be written as $e^{-\tilde t}$ to remove secular growth, an expansion of $\sin(t\sqrt{1-\e^2})$ will violate the condition. However, we \emph{can} write $\exact{f}(t,\e)=\tilde F(\phi,\tilde t,\e)$, where $\phi=\Omega(\e)t$, $\Omega(\e)=\sqrt{1-\e^2}$, and $\tilde F$ is given by
\begin{equation}
\tilde F(\phi,\tilde t,\e) = \frac{e^{-\tilde t}}{\Omega(\e)}\sin\phi.
\end{equation}
This function possesses the regular expansion $\tilde F(\phi,\tilde t,\e)=(1+\tfrac{1}{2}\e)e^{-\tilde t}\sin\phi+o(\e)$, which, when expressed in terms of $t$, is a uniform approximation to $\exact{f}(t,\e)$. One might wonder if we could have discovered this expansion without access to the exact solution. The answer, fortunately, is that we could have: substituting $\exact{f}=\tilde F=\sum\e^n \tilde F\coeff{\emph{n}}(\phi,\tilde t)$ and $\Omega(\e)=\sum_{n\ge 0}\e^n\Omega\coeff{\emph{n}}$ into Eq.~\eqref{DE1} and then solving for arbitrary $\phi$ and $\tilde t$ yields a sequence of equations that determine the $\tilde F\coeff{\emph{n}}$ and $\Omega\coeff{\emph{n}}$ \cite{Kevorkian_Cole}.

There are several points to note from this example. First, while an expansion method might appear to be working, it might still fail at higher order. Second, although we cannot be guaranteed that this failure will reveal itself in the course of our perturbation calculation, that will typically be the case, as it was here. Third, even though my assumptions about $F$ proved to be false, and even though $F\coeff{0}+\e F\coeff{1}$ fails to provide a uniform first-order approximation to $\exact{f}$, the term $F\coeff{0}$ alone, the only term in $F$ that was fully determined without any obvious contradiction, \emph{does} provide a uniform zeroth-order approximation. (This can easily be checked by calculating the supremum norm of $|\exact{f}-F\coeff{0}|$.)

\section{Gauge transformations in the self-consistent expansion}\label{gauge}
In the self-consistent expansion of Sec.~\ref{fixed_worldline_formulation}, the outer expansion is defined not only by holding $x^\alpha$ fixed, but also by demanding that the mass dipole of the body vanishes when calculated in coordinates centered on $\gamma$. If we perform a gauge transformation generated by a vector $\xi\coeff{1}{}^\alpha(x;\gamma)$, then the mass dipole will no longer vanish in those coordinates. Hence, a new worldline $\gamma'$ must be constructed, such that in coordinates centered on that new worldline, the mass dipole vanishes. In other words, in the outer expansion we have the usual gauge freedom of regular perturbation theory, so long as the worldline is appropriately transformed as well: $(h,\gamma)\to(h',\gamma')$. The transformation law for the worldline is well known \cite{self_force_gauge}; it will be worked out again presently. The only new feature of this gauge freedom is that $\e$-dependence can be incorporated into the transformation, because the gauge vectors can be functionals of the old worldline; this allows, for example, a ``first-order" gauge vector that is constructed from the tail integral $\tail[\gamma]$. However, to maintain the form of the expansion, we must also insist that $\xi\coeff{n}=O_s(1)$. Of course, in addition to this gauge freedom, one can still perform global, $\e$-independent background coordinate transformations.

I first justify, to some extent, the assumption that the Lorenz gauge condition can be imposed on the entirety of $h$. If we begin with the metric in an arbitrary gauge, then the gauge vectors $\e\xi\dcoeff{1}[\gamma]$, $\e^2\xi\dcoeff{2}[\gamma]$, etc., induce the transformation
\begin{align}
h\to h'&= h+\Delta h\nonumber\\
&=h+\e\Lie{\xi\dcoeff{1}}g+\tfrac{1}{2}\e^2(\Lie{\xi\dcoeff{2}}+\Lie{\xi\dcoeff{1}}^2g)\nonumber\\
&\quad+\e^2\Lie{\xi\dcoeff{1}}\hmn{}{1}+\ldots
\end{align}
If $h'$ is to satisfy the gauge condition $L_\mu[h']$, then $\xi$ must satisfy $L_\mu[\Delta h]=-L_\mu[h]$. After a trivial calculation, this equation becomes
\begin{align}
\sum_{n>0}\frac{\e^n}{n!}\Box\xi\dcoeff{\emph{n}}^\alpha &= -\e L^{\alpha}\big[\hmn{}{1}\big]-\e^2 L^{\alpha}\big[\hmn{}{2}\big]\nonumber\\
&\quad-\e^2 L^\alpha\big[\tfrac{1}{2}\Lie{\xi\dcoeff{1}}^2g+\Lie{\xi\dcoeff{1}}\hmn{}{1}\big]+\order{\e^3}.
\end{align}
Assuming that this equation is solved for arbitrary $\gamma$, we can equate coefficients of powers of $\e$, leading to a sequence of wave equations of the form
\begin{equation}
\Box\xi\dcoeff{\emph{n}}^\alpha = W^{\alpha}\dcoeff{\emph{n}},
\end{equation}
where $W^{\alpha}\dcoeff{\emph{n}}$ is a functional of $\xi\dcoeff{1},...,\xi\dcoeff{\emph{n}-1}$ and $\hmn{}{1},...,\hmn{}{\emph{n}}$. I seek a solution in the region $\Omega$ described in Sec.~\ref{fixed_worldline_formulation}. The formal solution reads
\begin{align}
\xi\dcoeff{n}^\alpha &=-\frac{1}{4\pi}\int_\Omega G^\alpha{}_{\alpha'}W^{\alpha'}\dcoeff{\emph{n}}dV'\nonumber\\
&\quad+\frac{1}{4\pi}\!\oint\limits_{\partial\Omega}\!\!\Big(G^{\alpha}{}_{\gamma'}\del{\mu'}\xi\dcoeff{n}^{\gamma'}-\xi\dcoeff{n}^{\gamma'}\del{\mu'} G^{\alpha}{}_{\gamma'}\Big) dS^{\mu'},
\end{align}
where $G_{\alpha\alpha'}$ is the retarded Green's function for the vector wave-equation. From this we see that the Lorenz gauge condition can be adopted to any desired order of accuracy, given the existence of self-consistent data on the worldtube $\Gamma$ of asymptotically small radius. I leave the question of that data's existence to future work. 

I now turn to the question of how the worldline transforms under a gauge transformation. I begin with the equation of motion derived in Ref.~\cite{my_paper}, written in the Lorenz gauge in Fermi coordinates centered on $\gamma$: 
\begin{align}\label{accelerations}
\partial_t^2 M_a+\etide_{ai}M^i &= -m\an{1}_a - \btide_{ai}S^i \nonumber\\
&\quad+ m\left[\tfrac{1}{2}\A{a}{1,1}-\partial_t\C{a}{1,0}\right]_{a^\mu=0},
\end{align}
where $M^i$ is the mass dipole of the small body, $S^i$ is its spin, and $\A{a}{1,1}$ and $\C{a}{1,0}$ are irreducible pieces of $\hmn{}{1}$, specifically
\begin{align}
\A{a}{1,1} &= \frac{3}{4\pi}\int n_a\hmn{Ett}{1,1}d\Omega,\\
\C{a}{1,0} &= \hmn{Eta}{1,0}.
\end{align}
They are given in terms of tail integrals in Table~\ref{STF wrt tail}.

The worldline is defined to be that of the body if $M^i$ vanishes for all time. Given that $M^i(t=0)=\partial_tM^i(t=0)=0$, this is possible if and only if the right-hand side of Eq.~\eqref{accelerations} vanishes. If, for simplicity, I neglect the Papapetrou spin term, then the first-order acceleration of $\gamma$ must be
\begin{align}
\an{1}_a &= \lim_{r\to0}\left(\frac{3}{4\pi}\int\frac{n_a}{2r}\hmn{Ett}{1}d\Omega -\partial_t\hmn{Eta}{1}\right)\\
&= \lim_{r\to0}\frac{1}{4\pi}\int\left(\tfrac{1}{2}\partial_a\hmn{Ett}{1} -\partial_t\hmn{Eta}{1}\right)d\Omega\\
&= \lim_{r\to0}\frac{3}{4\pi}\int\left(\tfrac{1}{2}\partial_i\hmn{Ett}{1} -\partial_t\hmn{Eti}{1}\right)n^i_a d\Omega \label{reg force},
\end{align}
where it is understood that explicit appearances of the acceleration are to be set to zero on the right-hand side. The first equality follows directly from Eq.~\eqref{accelerations} and the definitions of $\A{a}{1,1}$ and $\C{a}{1,0}$. The second and third equalities follow from the STF decomposition of $\hmn{}{1}$ and the integral identities \eqref{n_integral}--\eqref{nhat_integral}. The form of the force in the second line is the method of regularization used by Quinn and Wald~\cite{Quinn_Wald}; the form in the third line is used to derive a gauge-invariant equation of motion, as was was first noted by Gralla~\cite{Gralla_gauge}.

Now, suppose that we had not chosen a worldline for which the mass dipole vanishes, but instead had chosen some ``nearby" worldline. Then Eq.~\eqref{accelerations} provides the relationship between the acceleration of that worldline, the mass dipole relative to it, and the first-order metric perturbations (I again neglect spin for simplicity). The mass dipole is given by $M_i=\frac{3}{8\pi}\lim_{r\to 0}\int r^2\hmn{tt}{2}n_i d\Omega$, which has the covariant form
\begin{equation}
M_{\alpha'} = \frac{3}{8\pi}\lim_{r\to0}\int\! g^\alpha_{\alpha'}n_\alpha r^2\hmn{\mu\nu}{2}u^\mu u^\nu d\Omega,
\end{equation}
where a primed index corresponds to a point on the worldline. Note that the parallel propagator does not interfere with the angle-averaging, because in Fermi coordinates, $g^\alpha_{\beta'}=\delta^\alpha_\beta+O(\e,r^2)$. One can also rewrite the first-order-metric-perturbation terms in Eq.~\eqref{accelerations} using the form given in Eq.~\eqref{reg force}. We then have Eq.~\eqref{accelerations} in the covariant form
\begin{align}\label{gauge-invariant form}
&\frac{3}{8\pi}\lim_{r\to0}\int\! g^\alpha_{\alpha'}\! \left(\!g_{\alpha\beta}\frac{D^2}{d\tau^2}+\etide_{\alpha\beta}\!\right)\!n^\beta r^2\hmn{\mu\nu}{2}u^\mu u^\nu d\Omega\big|_{a=\an{0}} \nonumber\\
&=-\frac{3m}{8\pi}\lim_{r\to0}\int\!g^\alpha_{\alpha'}\left(2\hmn{\beta\mu;\nu}{1}-\hmn{\mu\nu;\beta}{1}\right)u^\mu u^\nu n_\alpha^\beta d\Omega\big|_{a=\an{0}}\nonumber\\
&\quad -m\an{1}_{\alpha'}.
\end{align}

Now consider a gauge transformation generated by $\e\xi\coeff{1}[\gamma]+\tfrac{1}{2}\e^2\xi\coeff{2}[\gamma]+...$, where $\xi\coeff{1}$ is bounded as $r\to0$, and $\xi\coeff{2}$ diverges as $1/r$. More specifically, I assume the expansions $\xi\coeff{1}=\xi\coeff{1,0}(t,\theta^A)+O(r)$ and $\xi\coeff{2}=\frac{1}{r}\xi\coeff{2,-1}(t,\theta^A)+O(1)$.\footnote{The dependence on $\gamma$ appears in the form of dependence on proper time $t$. Each term could in addition depend on the acceleration, but such dependence would not affect the result.} This transformation preserves the presumed form of the outer expansion, both in powers of $\e$ and in powers of $r$. According to Eqs.~\eqref{gauge_trans1}--\eqref{gauge_trans2}, the metric perturbations transform as 
\begin{align}
\hmn{\mu\nu}{1} &\to \hmn{\mu\nu}{1}+2\xi\coeff{1}_{(\mu;\nu)},\\
\hmn{\mu\nu}{2} &\to \hmn{\mu\nu}{2}+\xi\coeff{2}_{(\mu;\nu)}+\hmn{\mu\nu;\rho}{1}\xi\coeff{1}{}^\rho + 2\hmn{\rho(\mu}{1}\xi\coeff{1}{}^\rho{}_{;\nu)} \nonumber\\
&\quad+\xi\coeff{1}{}^\rho\xi\coeff{1}_{(\mu;\nu)\rho}+\xi\coeff{1}{}^\rho{}_{;\mu}\xi\coeff{1}_{\rho;\nu}+\xi\coeff{1}{}^\rho{}_{;(\mu}\xi\coeff{1}_{\nu);\rho}.
\end{align}
Using the results for $\hmn{}{1}$, the effect of this transformation on $\hmn{tt}{2}$ is given by
\begin{equation}
\hmn{tt}{2}\to\hmn{tt}{2}-\frac{2m}{r^2}n^i\xi\coeff{1}_i+O(r^{-1}).
\end{equation}
The order-$1/r^2$ term arises from $\hmn{\mu\nu;\rho}{1}\xi\coeff{1}{}^\rho$ in the gauge transformation. On the right-hand side of Eq.~\eqref{gauge-invariant form}, the metric-perturbation terms transform as
\begin{align}
(2\hmn{\beta\mu;\nu}{1}-\hmn{\mu\nu;\beta}{1})u^\mu u^\nu n^\beta &\to (2\hmn{\beta\mu;\nu}{1}-\hmn{\mu\nu;\beta}{1})u^\mu u^\nu n^\beta\nonumber\\
&\quad +2n_\beta\left(g_\beta^\gamma\frac{D^2}{d\tau^2}+\etide_\beta^\gamma\right)\xi\coeff{1}_\gamma.
\end{align}
The only remaining term in the equation is $m\an{1}_\alpha$. If we extend the acceleration off the worldline in any smooth manner, then it defines a vector field that transforms as $a^\alpha\to a^\alpha+\e\Lie{\xi\coeff{1}}a^\alpha+...$. Since $\an{0}=0$, this means that $\an{1}\to\an{1}$---it is invariant under a gauge transformation.

From these results, we find that the left- and right-hand sides of Eq.~\eqref{gauge-invariant form} transform in the same way:
\begin{align}\label{gauge_effect}
{\rm LHS/RHS} &\to {\rm LHS/RHS} \nonumber\\
&\quad- \frac{3}{4\pi}\lim_{r\to0}\int g^\alpha_{\alpha'} n_\alpha^\beta\left(g_\beta^\gamma\frac{D^2}{d\tau^2}+\etide_\beta^\gamma\right)\xi\coeff{1}_\gamma d\Omega.
\end{align}
Therefore, Eq.~\eqref{gauge-invariant form} provides a gauge-invariant relationship between the acceleration of a chosen fixed worldline, the mass dipole of the body relative to that worldline, and the first-order metric perturbations. So suppose that we begin in the Lorenz gauge, and we choose the fixed worldline $\gamma$ such that the mass dipole vanishes relative to it. Then in some other gauge, the mass dipole will no longer vanish relative to $\gamma$, and we must adopt a different, nearby fixed worldline $\gamma'$. If the mass dipole is to vanish relative to $\gamma'$, then the acceleration of that new worldline must be given by $a_\alpha=\e\an{1}_\alpha+o(\e)$, where
\begin{equation}
\an{1}_{\alpha'}=-\frac{3}{8\pi}\lim_{r\to0}\int\! g^\alpha_{\alpha'}(2\hmn{\beta\mu;\nu}{1}-\hmn{\mu\nu;\beta}{1})u^\mu u^\nu n_\alpha^\beta d\Omega.\big|_{a=\an{0}}.
\end{equation}
Hence, this is a covariant and gauge-invariant form of the first-order acceleration. (By that I mean the \emph{equation} is valid in any gauge, not that the value of the acceleration is the same in every gauge; under a gauge transformation, a new fixed worldline is adopted, and the value of the acceleration on the new worldline is related to that on the old worldline according to Eq.~\eqref{gauge_effect}.) An argument of this form was first presented by Gralla~\cite{Gralla_gauge} for the case of a regular expansion of the worldline; it is now extended to the case of a fixed-worldline expansion.

\section{STF multipole decompositions}\label{STF tensors}
This appendix briefly reviews the use of STF decompositions and collects several useful formulas. Refer to Ref.~\cite{STF_1,STF_2,STF_3} for thorough reviews. All formulas in this section are either taken directly from Refs.~\cite{STF_1} and \cite{STF_2} or are easily derivable from formulas therein.

Any Cartesian tensor field depending on two angles $\theta^A$ spanning a sphere can be expanded in a unique decomposition in terms of symmetric trace-free tensors. Such a decomposition is equivalent to a decomposition in terms of tensorial harmonics, but it is sometimes more convenient. It begins with the fact that the angular dependence of a Cartesian tensor $T_{S}(\theta^A)$ can be expanded in a series of the form
\begin{equation}\label{nhat_expansion}
T_S(\theta^A)=\sum_{\ell\geq 0}T_{S\langle L\rangle}\nhat^L,
\end{equation}
where $S$ and $L$ denote multi-indices $S=i_1...i_s$ and $L=j_1...j_\ell$, angular brackets denote an STF combination of indices, $n^a$ is a Cartesian unit vector, $n^L\equiv n^{j_1}\ldots n^{j_\ell}$, and $\nhat^L\equiv n^{\langle L\rangle}$. This is entirely equivalent to an expansion in spherical harmonics. Each coefficient $T_{S\langle L\rangle}$ can be found from the formula
\begin{equation}
T_{S\langle L\rangle} = \frac{(2\ell+1)!!}{4\pi\ell!}\int T_S(\theta^A)\nhat_L d\Omega,
\end{equation}
where $!!$ is a double factorial (defined by $x!!=x(x-2)...1$). These coefficients can then be decomposed into irreducible STF tensors. For example, for $s=1$, we have
\begin{equation}\label{decomposition_1}
T_{a\langle L\rangle} = \hat T^{(+1)}_{aL}+\epsilon^j{}_{a\langle i_\ell}\hat T^{(0)}_{L-1\rangle j}+\delta_{a\langle i_\ell}\hat T^{(-1)}_{L-1\rangle},
\end{equation}
where the $\hat T^{(n)}$'s are STF tensors given by
\begin{align}
\hat T^{(+1)}_{L+1} & \equiv T_{\langle L+1\rangle}, \\
\hat T^{(0)}_{L} & \equiv \frac{\ell}{\ell+1}T_{pq\langle L-1}\epsilon_{i_\ell\rangle}{}^{pq}, \\
\hat T^{(-1)}_{L-1} & \equiv \frac{2\ell-1}{2\ell+1}T^j{}_{jL-1}.
\end{align} 
Similarly, for a symmetric tensor $T_S$ with $s=2$, we have
\begin{align}\label{decomposition_2}
T_{ab\langle L\rangle} & = \mathop{\STF}_L\mathop{\STF}_{ab}\Big( \epsilon^p{}_{ai_\ell}\hat T^{(+1)}_{bpL-1} + \delta_{ai_\ell}\hat T^{(0)}_{b L-1} \nonumber\\
&\quad +\delta_{a i_\ell}\epsilon^p{}_{bi_{\ell-1}}\hat T^{(-1)}_{pL-2} +\delta_{ai_\ell}\delta_{bi_{\ell-1}}\hat T^{(-2)}_{L-2}\Big) \nonumber\\
&\quad +\hat T^{(+2)}_{abL}+\delta_{ab}\hat K_L,
\end{align}
where
\begin{align}
\hat T^{(+2)}_{L+2} & \equiv T_{\langle L+2\rangle}, \\
\hat T^{(+1)}_{L+1} & \equiv \frac{2\ell}{\ell+2}\mathop{\STF}_{L+1}(T_{\langle pi_\ell\rangle qL-1}\epsilon_{i_{\ell+1}}{}^{pq}), \\
\hat T^{(0)}_L & \equiv \frac{6\ell(2\ell-1)}{(\ell+1)(2\ell+3)}\mathop{\STF}_L(T_{\langle ji_\ell\rangle}{}^j{}_{L-1}), \\
\hat T^{(-1)}_{L-1} & \equiv \frac{2(\ell-1)(2\ell-1)}{(\ell+1)(2\ell+1)}\mathop{\STF}_{L-1}(T_{\langle jp\rangle q}{}^j{}_{L-2}\epsilon_{i_{\ell-1}}{}^{pq}), \\
\hat T^{(-2)}_{L-2} & \equiv \frac{2\ell-3}{2\ell+1}T_{\langle jk\rangle}{}^{jk}{}_{L-2} \\
\hat K_L & \equiv \tfrac{1}{3}T^j{}_{jL}.
\end{align}
These decompositions are equivalent to the formulas for addition of angular momenta, $J=S+L$, which results in terms with angular momentum $\ell-s\leq j\leq \ell+s$; the superscript labels $(\pm n)$ in these formulas indicate by how much each term's angular momentum differs from $\ell$.

By substituting Eqs.~\eqref{decomposition_1} and \eqref{decomposition_2} into Eq.~\eqref{nhat_expansion}, we find that a scalar, a Cartesian 3-vector, and the symmetric part of a rank-2 Cartesian 3-tensor can be decomposed as, respectively,
\begin{align}
T(\theta^A) &= \sum_{\ell\ge0}\hat A_L\nhat^L, \label{generic_STF tt}\\
T_a(\theta^A) &= \sum_{\ell\ge1}\left[\hat C_{aL-1}\nhat^{L-1} + \epsilon^i{}_{aj}\hat D_{iL-1}\nhat^{jL-1}\right]\nonumber\\
&\quad+\sum_{\ell\ge0}\hat B_L\nhat_{aL},\label{generic_STF ta}\\
T_{(ab)}(\theta^A) &= \delta_{ab}\sum_{\ell\ge0}\hat K_L\nhat^L+\sum_{\ell\ge0}\hat E_L\nhat_{ab}{}^L \nonumber\\
&\quad+\sum_{\ell\ge1}\left[\hat F_{L-1\langle a}\nhat^{}_{b\rangle}{}^{L-1} +\epsilon^{ij}{}_{(a}\nhat_{b)i}{}^{L-1}\hat G_{jL-1}\right] \nonumber\\
&\quad+\sum_{\ell\ge2}\left[\hat H_{abL-2}\nhat^{L-2}+\epsilon^{ij}{}_{(a}\hat I_{b)jL-2}\nhat_{i}{}^{L-2}\right].\label{generic_STF ab}
\end{align}
Each term in these decompositions is algebraically independent of all the other terms.

The unit vector $n_i$ satisfies the following integral identities:
\begin{align}
\int\nhat_Ld\Omega &=0 {\rm\ if\ } \ell>0, \label{n_integral}\\
\int n_L d\Omega & = 0 {\rm\ if\ } \ell {\rm\ is\ odd}, \\
\int n_L d\Omega & = 4\pi\frac{\delta_{\lbrace i_1 i_2}...\delta_{i_{\ell-1}i_{\ell\rbrace}}}{(\ell+1)!!} {\rm\ if\ } \ell {\rm\ is\ even}, \label{nhat_integral}
\end{align}
where the curly braces indicate the smallest set  of permutations of indices that make the result symmetric. For example, $\delta_{\lbrace ab}n_{c\rbrace}=\delta_{ab}n_c+\delta_{bc}n_a+\delta_{ca}n_b$.

\section{Linear perturbation due to a point particle}\label{point_particle_soln}
In this appendix, I present the solution to the wave equation with a point particle source. I also present the Detweiler-Whiting decomposition \cite{Detweiler_Whiting} of the solution into its singular and regular pieces.

\subsection{Solution to the wave equation}
The solution to the wave equation is\footnote{This is not the standard form presented in, e.g., Ref.~\cite{Eric_review}; however, one can easily show that it is equivalent to that form by using the Green's-function identities in Appendix~A of Ref.~\cite{my_paper}.}
\begin{equation}
\hmn{E\alpha\beta}{1} = 2m\int_\gamma G_{\alpha\beta\alpha'\beta'}(2u^{\alpha'}u^{\beta'}+g^{\alpha'\beta'})dt'.
\end{equation}
I seek an expansion of this equation in Fermi normal coordinates, in the case that $x$ is near to a point on $\gamma$. The domain of integration can be split into two: the points in the convex normal neighbourhood $\mathcal{N}(x)$---that is, the points that are connected to $x$ by a unique geodesic---and the points in the complement of $\mathcal{N}(x)$. In the convex normal neighbourhood, the Green's function can be decomposed as 
\begin{equation}
G_{\alpha\beta\alpha'\beta'}=U_{\alpha\beta\alpha'\beta'}\delta(\sigma(x,x'))+V_{\alpha\beta\alpha'\beta'}\theta(-\sigma(x,x')).
\end{equation}
After performing a change of variables using $dt'=\frac{d\sigma}{\sigma_{\alpha'}u^{\alpha'}}$ to evaluate the delta function, the metric perturbation becomes
\begin{equation}
\hmn{E\alpha\beta}{1} =\frac{2m}{r_{\text{ret}}}U_{\alpha\beta\alpha'\beta'}(2u^{\alpha'}u^{\beta'}+g^{\alpha'\beta'}) +\tail_{\alpha\beta}(u).
\end{equation}
where primed indices now refer to the point on $\gamma$ connected to $x$ by a null geodesic; this point is given by $z^\alpha(u)$, where $u$ is the retarded time; $r_{\text{ret}}$ is the retarded distance between $x$ and $z^\alpha(u)$, given by $\sigma_{\alpha'}u^{\alpha'}$; and the tail integral is given by
\begin{align}
\tail_{\alpha\beta}(u) & = 2m\int_{t^<}^{u} V_{\alpha\beta\alpha'\beta'}(2u^{\alpha'}u^{\beta'}+g^{\alpha'\beta'})dt'\nonumber\\
&\quad + 2m\int^{t^<}_{-\infty} G_{\alpha\beta\alpha'\beta'}(2u^{\alpha'}u^{\beta'}+g^{\alpha'\beta'})dt'\nonumber\\
&= 2m\int_{-\infty}^{u^-} G_{\alpha\beta\alpha'\beta'}(2u^{\alpha'}u^{\beta'}+g^{\alpha'\beta'})dt',
\end{align}
where $t^<$ is the first time at which the worldline enters $\mathcal{N}(x)$. The two formulas for the tail are equivalent because the upper limit of integration $u^-=u-0^+$ falls short of the past light cone of $x$, avoiding the divergent behavior of the Green's function there. 
 
The first term in $\hmn{E}{1}$, sometimes called the ``direct term", can be expanded in powers of $r$ using the following: the near-coincidence expansion $U_{\alpha\beta\alpha'\beta'}=g^{\alpha'}_\alpha g^{\beta'}_\beta(1+\order{r^3})$; the transformation between $r_{\text{ret}}$ and the Fermi radial distance $r$, given by
\begin{align}
\retr &= r\big[1+\tfrac{1}{2}ra_in^i-\tfrac{1}{8}r^2a_ia_jn^{ij}-\tfrac{1}{8}r^2\dot a_{\bar\alpha}u^{\bar\alpha}-\tfrac{1}{3}r^2\dot a_i n^i \nonumber\\
&\quad +\tfrac{1}{6}r^2\etide_{ij}n^{ij}+\order{r^3}\big];
\end{align}
and the coordinate expansion of the parallel-propagators, obtained from the formula $g^{\alpha'}_\alpha=b^{\alpha'}_I b^I_\alpha$, where the retarded tetrad $b^\alpha_I$ is given in terms of the Fermi tetrad in Eqs.~(226)--(227) of Ref.~\cite{Eric_review}, and the coordinate expansion of the Fermi tetrad is given in Eqs.~(123)--(124) of the same reference. The tail integral can be similarly expanded as follows: noting that $u=t-r+\order{r^2}$, we can expand $\tail(u)$ about $t$ as $\tail(t)-r\partial_t \tail(t)+...$; each term can then be expanded using the near-coincidence expansions $V_{\alpha\beta}^{\alpha''\beta''}=g^{\gamma''}_{(\alpha} g^{\delta''}_{\beta)}R^{\alpha''}{}_{\gamma''}{}^{\beta''}{}_{\delta''}+\order{r}$ and $\tail_{\alpha\beta}(t)=g^{\bar\alpha}_\alpha g^{\bar\beta}_\beta(\tail_{\bar\alpha\bar\beta}+r\tail_{\bar\alpha\bar\beta i}n^i)+\order{r^2}$, where barred indices correspond to the point $\bar x=z(t)$, connected to $x$ by a spatial geodesic perpendicular to $\gamma$, and $\tail_{\bar\alpha\bar\beta\bar\gamma}$ is given by
\begin{equation}
\tail_{\bar\alpha\bar\beta\bar\gamma}=2m\int_{-\infty}^{t^-} \del{\bar\gamma}G_{\bar\alpha\bar\beta\alpha'\beta'}(2u^{\alpha'}u^{\beta'}+g^{\alpha'\beta'})dt'.
\end{equation}
This yields the expansion
\begin{equation}
\tail_{\alpha\beta}(u)=g^{\bar\alpha}_\alpha g^{\bar\beta}_\beta(\tail_{\bar\alpha\bar\beta}+r\tail_{\bar\alpha\bar\beta i}n^i-4mr\etide_{\bar\alpha\bar\beta}) +\order{r^2}.
\end{equation}
As with the direct part, the final coordinate expansion is found by substituting $g^{\bar\alpha}_\alpha=e^{\bar\alpha}_I e^I_\alpha$.

Combining the expansions of the direct and tail parts of the perturbation, we arrive at the expansion in Fermi coordinates:
\begin{align}
\hmn{Ett}{1} & = \frac{2m}{r}(1+\tfrac{3}{2}ra_in^i +\tfrac{3}{8}r^2a_ia_jn^{ij} -\tfrac{15}{8}r^2\dot a_{\bar\alpha}u^{\bar\alpha} \nonumber\\
&\quad +\tfrac{1}{3}r^2\dot a_i n^i +\tfrac{5}{6}r^2\etide_{ij}n^{ij}) +(1+2ra_in^i)\tail_{00} \nonumber\\
&\quad +r\tail_{00i}n^i+\order{r^2},\\
\hmn{Eta}{1} & = 4ma_a-\tfrac{2}{3}mrR_{0iaj}n^{ij}+2mr\etide_{ai}n^i-2mr\dot a_a \nonumber\\
&\quad +(1+ra_in^i)\tail_{0a}+r\tail_{0ai}n^i+\order{r^2}, \\
\hmn{Eab}{1} & = \frac{2m}{r}(1-\tfrac{1}{2}ra_in^i +\tfrac{3}{8}r^2a_ia_jn^{ij} +\tfrac{1}{8}r^2\dot a_{\bar\alpha}u^{\bar\alpha} \nonumber\\
&\quad +\tfrac{1}{3}r^2\dot a_i n^i -\tfrac{1}{6}r^2\etide_{ij}n^{ij})\delta_{ab}+4mra_aa_b \nonumber\\
&\quad -\tfrac{2}{3}mrR_{aibj}n^{ij}-4mr\etide_{ab}+\tail_{ab} +r\tail_{abi}n^i \nonumber\\
&\quad +\order{r^2}.
\end{align}
As the final step, each of these terms is decomposed into irreducible STF pieces using the formulas \eqref{nhat_expansion}, \eqref{decomposition_1}, and \eqref{decomposition_2}, to yield
\begin{align}
\hmn{Ett}{1} &= \frac{2m}{r}+\A{}{1,0}+3ma_in^i+r\big[4ma_ia^i+\A{i}{1,1}n^i \nonumber\\
&\quad +m\left(\tfrac{3}{4}a_{\langle i}a_{j\rangle} +\tfrac{5}{3}\etide_{ij}\right)\nhat^{ij}\big]+O(r^2),\\
\hmn{Eta}{1} &= \C{a}{1,0}+r\big(\B{}{1,1}n_a-2m\dot a_a+\C{ai}{1,1}n^i \nonumber\\
&\quad +\epsilon_{ai}{}^j\D{j}{1,1}n^i +\tfrac{2}{3}m\epsilon_{aij}\btide^j_k\nhat^{ik}\big)+O(r^2), \\
\hmn{Eab}{1} &= \frac{2m}{r}\delta_{ab}+(\K{}{1,0}-ma_in^i)\delta_{ab}+\H{ab}{1,0} \nonumber\\
&\quad +r\Big\lbrace\delta_{ab}\big[\tfrac{4}{3}ma_ia^i+\K{i}{1,1}n^i +\tfrac{3}{4}ma_{\langle i}a_{j\rangle}\nhat^{ij}\nonumber\\
&\quad -\tfrac{5}{9}m\etide_{ij}\nhat^{ij}\big] +\tfrac{4}{3}m\etide^i_{\langle a}\nhat_{b\rangle i} +4ma_{\langle a}a_{b\rangle} \nonumber\\
&\quad -\tfrac{38}{9}m\etide_{ab}+\H{abi}{1,1}n^i +\epsilon_i{}^j{}_{(a}\I{b)j}{1,1}n^i\nonumber\\
&\quad +\F{\langle a}{1,1}n^{}_{b\rangle}\Big\rbrace+O(r^2),
\end{align}
where the uppercase script tensors are specified in Table~\ref{STF wrt tail}. The naming convention for those tensors follows that in Eqs.~\eqref{generic_STF tt}--\eqref{generic_STF ab}. 

\begin{table}[tb]
\caption[The regular field in terms of $\etide_{ab}$ and $\tail_{\alpha\beta}$]{Symmetric trace-free tensors in the first-order metric perturbation in the buffer region, written in terms of the electric-type tidal field $\etide_{ab}$, the acceleration $a_i$, and the tail of the perturbation.}  
\begin{ruledtabular}
\begin{tabular}{l}
$\begin{array}{ll}
\A{}{1,0} &= \tail_{00}\\
\C{a}{1,0} &= \tail_{0a}+ma_a\\
\K{}{1,0} &= \tfrac{1}{3}\delta^{ab}\tail_{ab}\\
\H{ab}{1,0} &= \tail_{\langle ab\rangle}\\
\A{a}{1,1} &= \tail_{00a}+2\tail_{00}a_a+\tfrac{2}{3}m\dot a_a\\
\B{}{1,1} &= \tfrac{1}{3}\tail_{0ij}\delta^{ij}+\tfrac{1}{3}\tail_{0i}a^i\\
\C{ab}{1,1} &= \tail_{0\langle ab\rangle}+2m\etide_{ab}+\tail_{0\langle a}a_{b\rangle} \\
\D{a}{1,1} &= \tfrac{1}{2}\epsilon_a{}^{bc}(\tail_{0bc}+\tail_{0b}a_c)\\
\K{a}{1,1} &= \frac{1}{3}\delta^{bc}\tail_{bca}+\tfrac{2}{3}m\dot a_a\\
\H{abc}{1,1} &= \tail_{\langle abc\rangle}\\
\F{a}{1,1} &= \tfrac{3}{5}\delta^{ij}\tail_{\langle ia\rangle j}\\
\I{ab}{1,1} &= \tfrac{2}{3}\displaystyle{\mathop{\STF}_{ab}} \left(\epsilon_b{}^{ij}\tail_{\langle ai\rangle j}\right)
\end{array}$
\end{tabular}
\end{ruledtabular}
\label{STF wrt tail}
\end{table}  

\subsection{Singular and regular pieces}
The Detweiler-Whiting singular field is given by
\begin{equation}
h^S_{\alpha\beta} = 2m\int G^S_{\alpha\beta\alpha'\beta'}(2u^{\alpha'}u^{\beta'}+g^{\alpha'\beta'})dt',
\end{equation}
where $G^S$ is the singular Green's function. Using the Hadamard decomposition
\begin{equation}
G^S_{\alpha\beta\alpha'\beta'}=\tfrac{1}{2}U_{\alpha\beta\alpha'\beta'}\delta(\sigma) -\tfrac{1}{2}V_{\alpha\beta\alpha'\beta'}\theta(\sigma),\label{GS_Hadamard}
\end{equation}
we can write this as
\begin{align}
h^S_{\alpha\beta} &= \frac{m}{\retr}U_{\alpha\beta\alpha'\beta'}(2u^{\alpha'}u^{\beta'}+g^{\alpha'\beta'})\nonumber\\
&\quad+ \frac{m}{\advr}U_{\alpha\beta\alpha''\beta''}(2u^{\alpha''}u^{\beta''}+g^{\alpha''\beta''}) \nonumber\\
&\quad-2m\int^v_u V_{\alpha\beta\bar\alpha\bar\beta}(u^{\bar\alpha}u^{\bar\beta}+\tfrac{1}{2}g^{\bar\alpha\bar\beta})d\bar t \label{hS}
\end{align}
where primed indices now refer to the retarded point $x'=z(u)$, where $u$ is the retarded time; double-primed indices refer to the advanced point $x''=z(v)$, where $v$ is advanced time; $\advr$ is the advanced distance between $x$ and $z^\alpha(v)$, given by $-\sigma_{\alpha''}u^{\alpha''}$; barred indices refer to points in the segment of the worldline between $z(u)$ and $z(v)$. The first term in Eq.~\eqref{hS} can be read off from the calculation of the retarded field. The other terms are expanded using the identities $v=u+2r+O(r^2)$ and $\advr=\retr(1+\tfrac{2}{3}r^2\dot a_in^i)$; see Ref.~\cite{Eric_review} for details (though the expansion therein is for $h^S_{\alpha\beta;\gamma}$, rather than $h^S_{\alpha\beta}$). The final result is
\begin{align}
h^S_{tt} &= \frac{2m}{r}+3ma_in^i+mr\big[4a_ia^i+\tfrac{3}{4}a_{\langle i}a_{j\rangle}\nhat^{ij}\nonumber\\
&\quad +\tfrac{5}{3}\etide_{ij}\nhat^{ij}\big]+O(r^2),\\
h^S_{ta} &=r\big(-2m\dot a_a +\tfrac{2}{3}m\epsilon_{aij}\btide^j_k\nhat^{ik}\big)+O(r^2), \\
h^S_{ab} &= \frac{2m}{r}\delta_{ab}-ma_in^i\delta_{ab} +r\Big\lbrace\delta_{ab}\big[\tfrac{4}{3}ma_ia^i\nonumber\\
&\quad+\left(\tfrac{3}{4}ma_{\langle i}a_{j\rangle}-\tfrac{5}{9}m\etide_{ij}\right)\nhat^{ij}\big] +\tfrac{4}{3}m\etide^i_{\langle a}\nhat_{b\rangle i} \nonumber\\
&\quad  +4ma_{\langle a}a_{b\rangle}-\tfrac{38}{9}m\etide_{ab} \Big\rbrace+O(r^2).
\end{align}

The regular field could be calculated from the regular Green's function. But it is more straightforwardly calculated using $h^R_{\alpha\beta}=\hmn{\alpha\beta}{1}-h^S_{\alpha\beta}$. The result is
\begin{align}
h^R_{tt} &= \A{}{1,0}+r\A{i}{1,1}n^i+\order{r^2}, \\
h^R_{ta} &= \C{a}{1,0} +r\Big(\B{}{1,1}n_a+\C{ai}{1,1}n^i +\epsilon_{ai}{}^j\D{j}{1,1}n^i\Big) \nonumber\\
&\quad +\order{r^2},\\
h^R_{ab} &= \delta_{ab}\K{}{1,0}+\H{ab}{1,0}+r\Big(\delta_{ab}\K{i}{1,1}n^i+\H{abi}{1,1}n^i \nonumber\\
&\quad+\epsilon_i{}^j{}_{(a}\I{b)j}{1,1}n^i+\F{\langle a}{1,1}n^{}_{b\rangle}\Big) +\order{r^2}.
\end{align}

\section{Metric of a tidally perturbed black hole}\label{perturbed_BH}
In this appendix, I present some general results for perturbations of Schwarzschild in a light cone gauge. Over the course of the calculation, I highlight the restrictions that must be imposed on the metric in order to arrive at the usual result for a tidally perturbed black hole. In the final section of the appendix, I present that metric, along with its expansion in the buffer region. The notation and definitions in the first section mostly follows that of Ref.~\cite{Eric_perturbations}.

\subsection{Metric expansion, perturbation equations, and gauge condition}
The exact metric $\exact{g}$ is expanded as $g_I(\tilde X,\e)=g_B(\tilde X)+H(\tilde X,\e)$, where $H(\tilde X,\e)=\e H\coeff{1}(\tilde X)+\e^2 H\coeff{2}(\tilde X)+...$, and $\tilde X^\alpha=(U,\tilde R,\Theta^A)$ are (scaled) retarded Eddington-Finkelstein coordinates adapted to the background metric $g_B$, where $\tilde R\equiv R/\e$. As described in Sec.~\ref{fixed_worldline_formulation}, the terms in the inner expansion of $\exact{g}$ must satisfy the sequence of equations \eqref{inner_eqn0}--\eqref{inner_eqn2}, which I rewrite here:
\begin{align}
G\coeff{0}_I{}^{\mu\nu}[g_B] &= 0,\label{inner_eqn0alt}\\
\delta G\coeff{0}_I{}^{\mu\nu}[H\coeff{1}] &= -G\coeff{1}_I{}^{\mu\nu}[g_B],\label{inner_eqn1alt}\\
\delta G\coeff{0}_I{}^{\mu\nu}[H\coeff{2}] &= -\delta^2 G\coeff{0}_I{}^{\mu\nu}[H\coeff{1}] - \delta G\coeff{1}_I{}^{\mu\nu}[H\coeff{1}]\nonumber\\
&\quad -G\coeff{2}_I{}^{\mu\nu}[g_B],\label{inner_eqn2alt}\\
&\vdots\nonumber
\end{align}
In these equations, $G_I\coeff{\emph{n}}$ and $\delta^kG_I\coeff{\emph{n}}$ consist of the terms in $G_I$ and $\delta^k G_I$ that contain $n$ derivatives with respect to $U$. The first equation, \eqref{inner_eqn0alt}, is the ordinary Einstein equation for $g_B$, except that all derivatives with respect to $U$ are removed. As the solution to this equation, I take the Schwarzschild metric
\begin{equation}
g_B = -f(U,\tilde R) dU^2 -2dUdR+R^2d\Omega^2, 
\end{equation}
where $f=1-\frac{2M(U)}{\tilde R}$, and $M(U)$ is the Bondi mass of $g_B$ at time $U$ divided by the initial mass. The dependence on $U$ can not be determined at this stage, because time-derivatives appear only in the higher-order equations.

Rather than fully solving the perturbation equations \eqref{inner_eqn1alt} and \eqref{inner_eqn2alt}, I will solve only certain parts of them, in order to pinpoint several key points about the general solution. First, I adopt the light cone gauge. This gauge choice consists of setting $H^{(n)}_{UR}=H^{(n)}_{RR}=H^{(n)}_{RA}=0$, which preserves the geometrical meaning of the retarded coordinates in the perturbed spacetime: $U$ remains constant on each outgoing light cone, and $R$ remains an affine parameter on outgoing light rays. Second, as a boundary condition, I insist that the perturbations must be regular on the event horizon.

Because of the spherical symmetry of the background, it is convenient to expand the perturbations in tensorial harmonics:
\begin{align}
H^{(n)}_{\mathsf{ab}} &= \sum_{\ell m}P^{(n)\ell m}_{\mathsf{ab}}Y^{\ell m},\\
H^{(n)}_{\mathsf{a}A} &= R\sum_{\ell m}\left(J_{\mathsf{a}}^{(n)\ell m}Y^{\ell m}_A+H_{\mathsf{a}}^{\ell m}X^{\ell m}_A\right),\\
H^{(n)}_{AB} &= R^2\sum_{\ell m}\Big(K^{(n)\ell m}\Omega_{AB}Y^{\ell m}+G^{(n)\ell m}Y^{\ell m}_{AB}\nonumber\\
&\quad +H_2^{(n)\ell m}X^{\ell m}_{AB}\Big), 
\end{align} 
where I have split the coordinates into the two sets $X^{\mathsf{a}}=(U,R)$ and $\Theta^A$, the various harmonics will be defined below, and the coefficients of the harmonics are functions of $U$ and $\tilde R$. In the context of this expansion, the light cone gauge is imposed by setting $P^{(n)\ell m}_{UR}=P^{(n)\ell m}_{RR}=P^{(n)\ell m}_{RA}=0$.

I define the various harmonics as follows: The scalar functions $Y^{\ell m}(\Theta^A)$ are the usual orthonormal spherical harmonics, which satisfy $[\Omega^{AB}D_AD_B+\ell(\ell+1)]Y^{\ell m}=0$, where $\Omega_{AB}$ is the metric of a unit 2-sphere, and $D_A$ is the covariant derivative compatible with $\Omega_{AB}$. The even-parity vector harmonics $Y^{\ell m}_A$ and odd-parity vector harmonics $X^{\ell m}_A$ are defined as
\begin{equation}
Y^{\ell m}_A\equiv D_AY^{\ell m},\quad X^{\ell m}_A\equiv-\epsilon_A{}^BD_BY^{\ell m},
\end{equation}
where $\epsilon_{AB}$ is the Levi-Civita tensor on the unit two-sphere. The even-parity and odd-parity tensor harmonics $Y^{\ell m}_{AB}$ and $X^{\ell m}_{AB}$ are defined as
\begin{align}
Y^{\ell m}_{AB} &= \left[D_A D_B+\tfrac{1}{2}\ell(\ell+1)\Omega_{AB}\right]Y^{\ell m},\\
X^{\ell m}_{AB} &= -\tfrac{1}{2}\left(\epsilon_A{}^CD_B+\epsilon_B{}^CD_A\right)D_CY^{\ell m}.
\end{align}
In this appendix, I will forgo any discussion of the odd-parity terms, since the even-parity terms are sufficient for my purpose of delineating the types of restrictions required to arrive at the usual form of a tidally perturbed metric.

In order to determine the effect of a gauge transformation, I write an even-parity gauge vector $\Xi\coeff{1}_\alpha=(\Xi\coeff{1}_{\mathsf{a}},\Xi\coeff{1}_A)$ as
\begin{equation}
\Xi\coeff{1}_{\mathsf{a}} = \sum_{\ell m}\xi^{(1)\ell m}_{\mathsf{a}}Y^{\ell m},\quad \Xi\coeff{1}_A = R\sum_{\ell m}\xi^{(1)\ell m}Y_A^{\ell m}.
\end{equation}
This vector has the following first-order effects:
\begin{align}
\Delta P\coeff{1}_{UU} &= -\frac{2M}{\R^2}\xi\coeff{1}_U+\frac{2Mf}{\R^2}\xi\coeff{1}_R,\label{gauge1}\\
\Delta P\coeff{1}_{UR} &= -\pdiff{}{\R}\xi\coeff{1}_U+\frac{2M}{\R^2}\xi\coeff{1}_R,\\
\Delta P\coeff{1}_{RR} &= -2\pdiff{}{\R}\xi\coeff{1}_R,\\
\Delta J_U\coeff{1} &= -\frac{1}{\R}\xi\coeff{1}_U,\\
\Delta J_R\coeff{1} &= -\pdiff{}{\R}\xi\coeff{1}-\frac{1}{\R}\xi\coeff{1}_R+\frac{1}{\R}\xi\coeff{1},\\
\Delta K\coeff{1} &= -\frac{2f}{\R}\xi\coeff{1}_R+\frac{2}{\R}\xi\coeff{1}_U+\frac{\ell(\ell+1)}{\R}\xi\coeff{1},\\
\Delta G\coeff{1} &= -\frac{2}{\R}\xi\coeff{1},\label{gauge2}
\end{align}
where for simplicity I have omitted the harmonic labels $\ell m$. I neglect $U$-derivatives in the transformation, since they are second-order effects in the present scheme.

Now, due to the spherical symmetry of the background metric, each mode in the harmonic expansion decouples from all the others in $\delta G_I$. Likewise, the even- and odd-parity sectors decouple. I write the even-parity terms in $\delta G\coeff{0}_I$ as
\begin{align}
Q^{\mathsf{ab}}_{\ell m} &= \int \delta G\coeff{0}_I{}^{\mathsf{ab}}Y^{\ell m} d\Omega,\\
Q^{\mathsf{a}}_{\ell m} &= \frac{2R^2}{\ell(\ell+1)}\int\delta G\coeff{0}_I{}^{\mathsf{a}A} Y_A^{\ell m}d\Omega,\\
Q^\flat_{\ell m} &= R^2\int\delta G\coeff{0}_I{}^{AB}\Omega_{AB}Y^{\ell m}d\Omega,\\
Q^\sharp_{\ell m} &= \frac{4R^4}{(\ell-1)\ell(\ell+1)(\ell+2)}\int \delta G\coeff{0}_I{}^{AB} Y_{AB}^{\ell m}d\Omega.
\end{align}
Explicitly, these gauge-invariant quantities are given by~\cite{Eric_perturbations}
{\allowdisplaybreaks \begin{align}
Q^{UU} & = -\pddiff{}{\tilde R}\tilde K-\frac{2}{\tilde R}\pdiff{}{\tilde R}\tilde K+\frac{f}{\tilde R}\pdiff{}{\tilde R}\tilde P_{RR}-\frac{2}{\tilde R}\pdiff{}{\tilde R}\tilde P_{UR} \nonumber\\
&\quad+\frac{\ell(\ell+1)\tilde R+4M}{2\tilde R^3}\tilde P_{RR},\\
Q^{UR} & = -\frac{\R-M}{\R^2}\pdiff{}{\R}\tilde K+\frac{1}{\R}\pdiff{}{\R}\tilde P_{UU}+\frac{1}{\R^2}\tilde P_{UU}\nonumber\\
&\quad -\frac{\ell(\ell+1)+4}{2\R^2}\tilde P_{UR}+\frac{f}{\R^2}\tilde P_{RR}+\frac{\mu}{2\R^2}\tilde K,\\
Q^{RR} & = \frac{(\R-M)f}{\R^2}\pdiff{}{\R}\tilde K -\frac{f}{\R}\pdiff{}{\R}\tilde P_{UU}+\frac{\mu\R+4M}{2\R^3}\tilde P_{UU}\nonumber\\
&\quad +\frac{2f}{\R^2}\tilde P_{UR}-\frac{f^2}{\R^2}\tilde P_{RR}-\frac{\mu f}{2\R^2}\tilde K,\\
Q^U & = -\pdiff{}{\tilde R}\tilde P_{UR}+\pdiff{}{\tilde R}\tilde K+\frac{2}{\tilde R}\tilde P_{UR}-\frac{\tilde R-M}{\tilde R^2}\tilde P_{RR},\\
Q^R & = \pdiff{}{\tilde R}\tilde P_{UU}-f\pdiff{}{\tilde R}\tilde K-\frac{2(\tilde R-M)}{\tilde R^2}\tilde P_{UR}\nonumber\\
&\quad +\frac{(\tilde R-M)f}{\tilde R^2}\tilde P_{RR},\\
Q^\flat & = -\pddiff{}{\R}\tilde P_{UU}+f\pddiff{}{\R}\tilde K -\frac{2}{\R}\pdiff{}{\R}\tilde P_{UU}\nonumber\\
&\quad+\frac{2(\R-M)}{\R^2}\pdiff{}{\R}\tilde P_{UR}-\frac{(\R-M)f}{\R^2}\pdiff{}{\R}\tilde P_{RR}\nonumber\\
&\quad+\frac{2(\R-M)}{\R^2}\pdiff{}{\R}\tilde K+\frac{\ell(\ell+1)}{\R^2}\tilde P_{UR}\nonumber\\
&\quad-\frac{\ell(\ell+1)\R^2-2\mu M\tilde R-4M^2}{2\R^4}\tilde P_{RR},\\
Q^\sharp & = 2\tilde P_{UR}-f\tilde P_{RR},
\end{align}}
where $\mu\equiv \ell(\ell+1)-2$, and I have introduced the gauge-invariant combinations
\begin{align}
\tilde P_{UU} &= P_{UU}-\frac{2M}{\R}J_U+\frac{2Mf}{\R}J_R-Mf\pdiff{}{\R}G,\\
\tilde P_{UR} &= P_{UR}-\pdiff{}{\R}(\R J_U)+\frac{2M}{\R}J_R-M\pdiff{}{\R}G,\\
\tilde P_{RR} &= P_{RR}-2\pdiff{}{\R}(\R J_R)+\R^2\pddiff{}{\R}G+2\R\pdiff{}{\R}G,\\
\tilde K &= K+2J_U-2fJ_R+\R f\pdiff{}{\R}G+\tfrac{1}{2}\ell(\ell+1)G.
\end{align}
I have omitted the indices $(n)\ell m$ for simplicity.

\subsection{First-order solution}
The first-order equation reads $\delta G\coeff{0}_I{}^{\alpha\beta}[H\coeff{1}]=-G\coeff{1}_I[g_B]$. The source term in this equation has a single non-vanishing component,
\begin{equation}
G\coeff{1}_I{}^{RR}[g_B] = \frac{2}{\R^2}\diff{M}{U}.
\end{equation}
Thus, in the notation introduced above, the $RR$, $\ell=0$ equation reads $Q^{RR}_{(0)00} = 4\sqrt{\pi}\R^{-2}\diff{M}{U}$, while all the other equations are source-free. For $\ell\geq 2$, the equations can be solved for arbitrary $\ell$. However, because various quantities are defined only for $\ell\geq 2$, the equations for the low multipoles $\ell=0$ and $\ell=1$ must be dealt with individually. I will write undetermined functions of $U$ as $A^{(n)\ell m}_k(U)$.

I begin by solving the $\ell=0$ equations. For $\ell=0$, the quantities $J_{\mathsf{a}}$, $G$, $Q^{\mathsf{a}}$, and $Q^\sharp$ are undefined. So the only equations are $Q^{UU}_{(0)00}[H\coeff{1}]=Q^{UR}_{(0)00}[H\coeff{1}]=Q^\flat_{(0)00}[H\coeff{1}]=0$ and $Q^{RR}_{(0)00}[H\coeff{1}] = 4\sqrt{\pi}\R^{-2}\diff{M}{U}$, in which $J_{\mathsf{a}}$ and $G$ are set to zero. Given the gauge condition, the only functions appearing in these equations are $K^{(1)00}$ and $P^{(1)00}_{UU}$. I first solve $Q^{UU}_{(0)00}[H\coeff{1}]=0$, which reads explicitly $-\pddiff{K^{(1)00}}{\tilde R}-\frac{2}{\tilde R}\pdiff{K^{(1)00}}{\tilde R}=0$. The solution to this equation is
\begin{equation}\label{K100}
K^{(1)00} = A^{(1)00}_1+\frac{1}{\R}A^{(1)00}_2.
\end{equation}
Substituting this into $Q^{UR}_{(0)00}[H\coeff{1}]=0$ yields an equation for $P^{(1)00}_{UU}$ that can be readily solved to find
\begin{equation}\label{P100}
P^{(1)00}_{UU} = A^{(1)00}_1-\frac{M}{\R^2}A^{(1)00}_2+\frac{1}{\tilde R}A^{(1)00}_3.
\end{equation}
Substituting these results into $Q^\flat_{(0)00}[H\coeff{1}]$ and $Q^{RR}_{(0)00}[H\coeff{1}]$, we find that both quantities are identically zero. Hence, from the equation $Q^{RR}_{(0)00}[H\coeff{1}] = 4\sqrt{\pi}\R^{-2}\diff{M}{U}$ we can conclude
\begin{equation}
\diff{M}{U} = 0;
\end{equation}
that is, the mass of the internal background is constant. The functions $A^{(1)00}_k$, $k=1,2,3$, can be be determined only by solving the second-order EFE.

Next, I proceed to the $\ell=1$ equations. For $\ell=1$, the quantities $G$ and $Q^\sharp$ are undefined, and the field equations read $Q^{\mathsf{ab}}_{(0)1m}[H\coeff{1}]=Q^{\mathsf{a}}_{(0)1m}[H\coeff{1}]=Q^\flat_{(0)1m}[H\coeff{1}]=0$, in which $G$ is set to zero. Solving $Q^{UU}_{(0)1m}=0$ yields
\begin{equation}
K^{(1)1m} = A^{(1)1m}_1+\frac{1}{\R}A^{(1)1m}_2.
\end{equation}
Substituting this into $Q^U_{(0)1m}=0$ and solving then yields
\begin{equation}
J_U^{(1)1m} = -\frac{1}{2\R}A^{(1)1m}_2+\frac{1}{\R^2}A^{(1)1m}_3+\R A^{(1)1m}_4.
\end{equation}
Substituting these results into $Q^R_{(0)1m}=0$ and solving then yields  
\begin{equation}
P_{UU}^{(1)1m} = -\frac{M}{\R^2}A^{(1)1m}_2+\frac{1}{\R^2}A^{(1)1m}_3-2\R A^{(1)1m}_4+A^{(1)1m}_5.
\end{equation}
Two of the remaining equations, $Q^{UR}_{(0)1m}=0=Q^{RR}_{(0)1m}$ fixes several of the free functions in these solutions: $A^{(1)1m}_4=0$ and $A^{(1)1m}_5=0$. The final equation, $Q^\flat_{(0)1m}=0$, yields no further information. Putting these results together, we find
\begin{align}
P\coeff{1}_{UU}{}^{1m} & = -\frac{M}{\R^2}A^{(1)1m}_2+\frac{1}{\R^2}A^{(1)1m}_3,\label{P11m}\\
J\coeff{1}_U{}^{1m} & = -\frac{1}{2\R}A^{(1)1m}_2+\frac{1}{\R^2}A^{(1)1m}_3,\\
K\coeff{1}{}^{1m} & = A^{(1)1m}_1+\frac{1}{\R}A^{(1)1m}_2.\label{K11m}
\end{align}

Finally, I proceed to the $\ell\geq 2$ equations. As with $\ell=0,1$, the equation $Q^{UU}_{(0)\ell m}=0$ can be immediately solved to find
\begin{equation}
K^{(1)\ell m} = A^{(1)\ell m}_1+\frac{1}{\R}A^{(1)\ell m}_2.
\end{equation}
Using this result and $Q^{R}_{(0)\ell m}=0$, we can express $P^{(1)\ell m}_{UU}$ in terms of $G^{(1)\ell m}$ and $J^{(1)\ell m}_U$; next, we can use $Q^U_{(0)\ell m}=0$ to express $G^{(1)\ell m}$ in terms of $J^{(1)\ell m}_U$; finally, substituting these results into $Q^\sharp_{(0)\ell m}=0$, we can solve for $J^{(1)\ell m}_U$ to find
\begin{align}
J^{(1)\ell m}_U &= -\frac{1}{2\R}A^{(1)\ell m}_2 +\frac{4M+\mu\R}{\R^2}A^{(1)\ell m}_4\nonumber\\
&+A^{(1)\ell m}_5\R^\ell (-f)^{\ell-1}\ {}_2F_1\!\!\left(2-\ell, 1-\ell;-2\ell;-\frac{2M}{\R f}\right) \nonumber\\
&+\frac{A^{(1)\ell m}_6}{\R^{\ell+1}(-f)^{2+\ell}}\ {}_2F_1\!\!\left(2+\ell, 3+\ell;2+2\ell;-\frac{2M}{\R f}\right),\label{Jlm}
\end{align}
where ${}_2F_1$ is a hypergeometric function. The term next to $A^{(1)\ell m}_6$ diverges at the unperturbed event horizon, $\R=2M$, violating the boundary condition. Therefore, we have $A^{(1)\ell m}_6=0$.\footnote{The term next to $A^{(1)\ell m}_6$ corresponds to an induced multipole moment. The fact that  it must vanish agrees with the no-hair theorem.} Next, I make my first restriction of the solution: if $J^{(1)\ell m}_U$ is expressed in terms of $R=\e\R$, then the term next to $A^{(1)\ell m}_5$ behaves as $\e^{-\ell}$. Since $H\coeff{1}$ is accompanied by a factor of $\e$ in the metric, this term behaves as $\e^{1-\ell}$. If I assume that $R\sim r$, where $r$ is, for example, the Fermi radial coordinate centered on the body's worldline in the external spacetime, then these negative powers of $\e$ would also appear in the outer expansion. By assumption, no negative powers do appear in the outer expansion; therefore, I set $A^{(1)\ell m}_5=0$. So $J^{(1)\ell m}_U$ simplifies to 
\begin{equation}
J^{(1)\ell m}_U = -\frac{1}{2\R}A^{(1)\ell m}_2 +\frac{4M+\mu\R}{\R^2}A^{(1)\ell m}_4.
\end{equation}

Recall that we had expressed $P^{(1)\ell m}_{UU}$ and $G^{(1)\ell m}$ in terms of $J^{(1)\ell m}_U$. With $J^{(1)\ell m}_U$ determined, we now have
\begin{align}
P^{(1)\ell m}_{UU} &= -\frac{M}{\R^2}\left(A^{(1)\ell m}_2-2\ell(\ell+1)A^{(1)\ell m}_4\right)+A^{(1)\ell m}_3,\\
G^{(1)\ell m} &= -\frac{4}{\R}A^{(1)\ell m}_4 + A^{(1)\ell m}_7.
\end{align}
Substituting these results into $Q^{UR}_{(0)\ell m}=0$ and then $Q^{RR}_{(0)\ell m}=0$, we determine
\begin{equation}
A^{(1)\ell m}_7 = -\frac{2A^{(1)\ell m}_1}{\ell(\ell+1)},\quad A^{(1)\ell m}_3=0.
\end{equation}
The sole remaining equation, $Q^\flat_{(1)\ell m}=0$, yields no new information. Hence, the first-order calculation is now complete. We have found that $\diff{M}{U}=0$; the $\ell=0$ modes in $H\coeff{1}$ are given by Eqs.~\eqref{K100} and \eqref{P100}; the $\ell=1$ modes are given by Eqs.~\eqref{P11m}--\eqref{K11m}; and the $\ell\geq2$ modes are given by
\begin{align}
P^{(1)\ell m}_{UU} &= -\frac{M}{\R^2}\left(A^{(1)\ell m}_2-2\ell(\ell+1)A^{(1)\ell m}_4\right),\\
J^{(1)\ell m}_U &= -\frac{1}{2\R}A^{(1)\ell m}_2 +\frac{4M+\mu\R}{\R^2}A^{(1)\ell m}_4,\\
K^{(1)\ell m} &= A^{(1)\ell m}_1+\frac{1}{\R}A^{(1)\ell m}_2,\\
G^{(1)\ell m} &= -\frac{4}{\R}A^{(1)\ell m}_4-\frac{2A^{(1)\ell m}_1}{\ell(\ell+1)}.
\end{align}

These results can be simplified by a refinement of the lightcone gauge. For $\ell=0$, the function $A^{(1)00}_2$ can be removed via
\begin{align}
\xi_R^{(1)00} & = \xi^{(1)00}_R(U),\\
\xi_U^{(1)00} & = f\xi^{(1)00}_R - \tfrac{1}{2}A^{(1)00}_2,
\end{align} 
leaving
\begin{align}
P^{(1)00}_{UU} &= A^{(1)00}_1+\frac{1}{\tilde R}A^{(1)00}_3,\\
K^{(1)00} &= A^{(1)00}_1.
\end{align}
Although this does not exhaust the residual freedom within the lightcone gauge, since $\xi_R(U)$ is arbitrary, the remaining freedom cannot be used to remove either $A^{(1)00}_1$ or $A^{(1)00}_3$. However, if we were to transform out of the lightcone gauge, $A^{(1)00}_1$ could be removed, leaving only a $1/\R$ mass monopole term, corresponding to a time-dependent correction to the mass.

For $\ell\geq1$, we can removed \emph{all} $A^{(1)\ell m}_k$ via
\begin{align} 
\xi^{(1)1 m}_U &= -\tfrac{1}{2}A^{(1)1m}_2+\frac{1}{\R}A^{(1)1m}_3,\\
\xi^{(1)1 m}_R &= -\frac{1}{2M}A^{(1)1m}_3,\\
\xi^{(1)1 m} &= -\tfrac{1}{2}\R A^{(1)1m}_2-\frac{1}{2M}-\tfrac{1}{2}A^{(1)1m}_3,
\end{align}
and
\begin{align}
\xi^{(1)\ell m}_U & = -\tfrac{1}{2}A^{(1)\ell m}_2+\left(\mu+\frac{4M}{\R}\right)A^{(1)\ell m}_4,\\
\xi^{(1)\ell m}_R & = -2A^{(1)\ell m}_4,\\
\xi^{(1)\ell m} & = -2A^{(1)\ell m}_4-\frac{\R A^{(1)\ell m}_1}{\ell(\ell+1)}.
\end{align}
This exhausts the residual freedom in the gauge condition.

In order to arrive at the usual form of a tidally perturbed metric, we must go beyond these gauge refinements by setting the entirety of $H\coeff{1}$ to zero. That is, we must choose $A^{(1)00}_1=A^{(1)00}_3=0$. If the odd-parity calculation had been performed, we would find that $\ell=1$ terms, corresponding to time-dependent spin terms, must be set to zero. (Recall also that I have restricted the possible perturbations by disallowing terms that would contain negative powers of $\e$ in the unscaled coordinates.) Hence, we can conclude that to arrive at the usual metric of a tidally perturbed black hole, we must restrict the perturbation by setting numerous functions to zero, without any evident justification.

\subsection{Second-order solution}
With $H\coeff{1}$ set to zero, and with $M$ determined to be a constant, the second-order EFE becomes the homogeneous, linear equation
\begin{equation}
\delta G\coeff{0}_I{}^{\alpha\beta}[H\coeff{2}]=0.
\end{equation}
This equation is solved in the same manner as the first-order equation. The calculation of the low multipoles $\ell=0$ and $\ell=1$ proceeds just as at first order, yielding
\begin{align}
P^{(2)00}_{UU} &= A^{(2)00}_1-\frac{M}{\R}A^{(2)00}_2+\frac{1}{\tilde R}A^{(2)00}_3,\\
K^{(2)00} &= A^{(2)00}_1+\frac{1}{\R}A^{(2)00}_2,
\end{align}
and
\begin{align}
P_{UU}^{(2)1m} & = -\frac{M}{\R^2}A^{(2)1m}_2+\frac{1}{\R^2}A^{(2)1m}_3,\\
J_U^{(2)1m} & = -\frac{1}{2\R}A^{(2)1m}_2+\frac{1}{\R^2}A^{(2)1m}_3,\\
K^{(2)1m} & = A^{(2)1m}_1+\frac{1}{\R}A^{(2)1m}_2.
\end{align}

For $\ell\geq 2$, the calculation proceeds just as at first order, up until the point marked by Eq.~\eqref{Jlm}. When the term that diverges on the event horizon is removed, the analogue of that equation reads
\begin{align}
&J^{(2)\ell m}_U = -\frac{1}{2\R}A^{(2)\ell m}_2 +\frac{4M+\mu\R}{\R^2}A^{(2)\ell m}_4 \nonumber\\
& +A^{(2)\ell m}_5\R^\ell (-f)^{\ell-1}\ {}_2F_1\!\!\left(2-\ell, 1-\ell;-2\ell;-\frac{2M}{\R f}\right).
\end{align}
At first order, this solution was simplified by setting the coefficient of $\R^\ell$ to zero, because it led to negative powers of $\e$ when written in terms of $R$. However, at second order, this term will be multiplied by $\e^2$ in the metric, so it will scale as $\e^{2-\ell}$. Hence, the term is acceptable for $\ell=2$, but must be set to zero for $\ell>2$. I will hence deal with these two cases separately.

For $\ell=2$, we have
\begin{equation}
{}_2F_1\!\!\left(2-\ell, 1-\ell;-2\ell;-\frac{2M}{\R f}\right) = 1,
\end{equation}
and so 
\begin{align}
J^{(2)2m}_U &= -\frac{1}{2\R}A^{(2)2m}_2 +\frac{4M+4\R}{\R^2}A^{(2)2m}_4\nonumber\\
&\quad-A^{(2)2m}_5\R^2f.
\end{align}
As at first order, this determines $P^{(2)2m}_{UU}$ and $G^{(2)2m}$:
\begin{align}
P^{(2)\ell m}_{UU} &= -\frac{M}{\R^2}\left(A^{(2)2 m}_2-12A^{(2)2m}_4\right)+A^{(2)2m}_3\nonumber\\
&\quad+3\R(\R-4M)A^{(2)2m}_5,\\
G^{(2)\ell m} &= -\frac{4}{\R}A^{(2)2 m}_4 + \R^2 A^{(2)2m}_5+A^{(2)2 m}_7.
\end{align}
As at first order, we determine $A^{(2)2 m}_3$ and $A^{(2)2 m}_7$ from $Q^{UR}_{(0)\ell m}=0$ and then $Q^{RR}_{(0)\ell m}=0$, which yield
\begin{align}
A^{(2)2 m}_7 &= -\tfrac{1}{3}A^{(2)2m}_1-2M^2A^{(2)2m}_5,\\
A^{(2)2m}_3 &= 12M^2A^{(2)2m}_5.
\end{align}
Putting these results together, we have the solution
\begin{align}
P_{UU}^{(2)2 m} & = 3\R^2f^2 A_5^{(2)2m}-\frac{M}{\R^2}A_2^{(2)2 m}+\frac{12M}{\R^2}A_4^{(2)2 m},\\
J_U^{(2)2 m} & = -\frac{1}{2\R}A_2^{(2)2m}+\frac{4}{\R}\left(1+\frac{M}{\R}\right)A_4^{(2)2 m}\nonumber\\
&\quad -\R^2fA_5^{(2)2 m},\\
K^{(2)2 m} & = A^{(2)2 m}_1+\frac{1}{\R}A^{(2)2m}_2,\\
G^{(2)2 m} & = -\tfrac{1}{3}A_1^{(2)2m}+\R^2\left(1-\frac{2M^2}{\R^2}\right)A_5^{(2)2 m}\nonumber\\
&\quad-\frac{4}{\R}A_4^{(2)2 m}.
\end{align}

For $\ell>2$, the calculation is proceeds just as at first order, leading to the solution
\begin{align}
P_{UU}^{(2)\ell m} & = -\frac{M}{\R^2}A_2^{(2)\ell m}+\frac{2M}{\R^2}\ell(\ell+1)A_4^{(2)\ell m},\\
J_U^{(2)\ell m} & = -\frac{1}{2\R}A_2^{(2)\ell m}+\frac{\mu\R+4M}{\R^2}A_4^{(2)\ell m},\\
K^{(2)\ell m} & = A^{(2)\ell m}_1+\frac{1}{\R}A^{(2)\ell m}_2,\\
G^{(2)\ell m} & = -\frac{2}{\ell(\ell+1)}A_1^{(2)\ell m}-\frac{4}{\R}A_4^{(2)\ell m}.
\end{align}

Just as at first order, with an appropriate gauge refinement, we can remove the functions $A^{(2)00}_2$, $A^{(2)1 m}_k$, $A_1^{(2)\ell m}$, $A_2^{(2)\ell m}$, and $A_4^{(2)\ell m}$, where $\ell\geq2$, thereby exhausting the freedom within the lightcone gauge. In order to arrive at the usual form of the tidally perturbed black hole metric, we must then set $A_1^{(2)00}=A_3^{(2)00}=0$. (And I again remind the reader that I have disallowed terms that would possess a negative power of $\e$ when expressed in unscaled coordinates.) This leaves only one undetermined function: $A_5^{(2)\ell m}$. Although I do not show the odd-parity calculation, it yields an analogous result:  to yield the usual form of the metric, all but one of the possible undetermined functions must be set to zero, leaving a single quadrupole term. After imposing these restrictions, the only non-vanishing components of the metric perturbation are
\begin{align}
H\coeff{2}_{UU} &= \sum_m 3\R^2f^2 A_5^{(2)2m}Y^{2m},\\
H\coeff{2}_{UA} &= -R\sum_m\left(\R^2f A_5^{(2)2 m}Y_A^{2m}+\R^2 f B^{(2)2m}X^{2m}_A\right),\\
H\coeff{2}_{AB} &= R^2\sum_m\bigg[\R^2\left(1-\frac{2M^2}{\R^2}\right)A_5^{(2)2 m}Y^{2m}_{AB}\nonumber\\
&\quad +\R^2 B^{(2)2m}X^{2m}_{AB}\bigg].
\end{align}

\subsection{Tidally perturbed black hole metric and its expansion in the buffer region}
After the metric perturbations have been restricted as described in the preceding two sections, the inner expansion has the following form:
\begin{align}
g_{IUU} &= -f+\e^2\sum_m 3\R^2f^2 A_5^{(2)2m}Y^{2m}+O(\e^3),\\
g_{IUA} &= -R\Bigg[\e^2\R^2f\sum_mA_5^{(2)2 m}Y_A^{2m}\nonumber\\
&\quad +\e^2\R^2f\sum_mB^{(2)2m}X^{2m}_A+O(\e^3)\Bigg],\\
g_{IAB} &= R^2\Bigg[\Omega_{AB}+\e^2\R^2\left(1-\frac{2M^2}{\R^2}\right)\sum_mA_5^{(2)2 m}Y^{2m}_{AB}\nonumber\\
&\quad +\e^2\R^2 \sum_mB^{(2)2m}X^{2m}_{AB}+O(\e^3)\Bigg],
\end{align}
along with the exact results $g_{IUR}=-1$ and $g_{IRR}=g_{IRA}=0$. Although it is written in a slightly different form, this is the usual metric of a tidally perturbed black hole. It is characterized by (i) having only quadrupole perturbations, and (ii) those perturbations scaling as $\R^2$ for large $\R$.

To write this metric in terms of a pair of tidal fields $\tilde\etide_{ab}$ and $\tilde\btide_{ab}$, I follow Appendix~A of Ref.~\cite{Eric_Igor}; I also make use of notation and simple identities from the Sec.~3.3.7 of Ref.~\cite{Eric_review}. First, I define the quantities
\begin{align}
\tilde\etide^* &\equiv -3\sum_m A_5^{(2)2m}Y^{2m},\\
\tilde\btide^* &\equiv -3\sum_m B^{(2)2m}Y^{2m}.
\end{align}
Next, I define the derived quantities
\begin{align}\label{tide_A def}
\tilde\etide^*_A &\equiv \tfrac{1}{2}D_A\tilde\etide^*,\\
\tilde\btide^*_A &\equiv -\tfrac{1}{2}\epsilon_A{}^B D_B\tilde\btide^*,
\end{align}
and
\begin{align}\label{tide_AB def}
\tilde\etide^*_{AB} & \equiv \left(D_AD_B+3\Omega_{AB}\right)\tilde\etide^*,\\
\tilde\btide^*_{AB} &\equiv -\tfrac{1}{2}\left(\epsilon_A{}^C D_B+\epsilon_B{}^C D_A\right)D_C\tilde\btide^*.
\end{align}
Using the definitions of $\tilde\etide^*$, $\tilde\btide^*$, $Y^{\ell m}_A$, $Y^{\ell m}_{AB}$, $X^{\ell m}_A$, and $X^{\ell m}_{AB}$, we can express the derived quantities as
\begin{align}
\tilde\etide^*_A &\equiv -\tfrac{3}{2}\sum_m A_5^{(2)2m}Y^{2m}_A,\\
\tilde\btide^*_A &\equiv -\tfrac{3}{2}\sum_m B^{(2)2m}X^{2m}_A,
\end{align}
and
\begin{align}
\tilde\etide^*_{AB} &\equiv -3\sum_m A_5^{(2)2m}Y^{2m}_{AB},\\
\tilde\btide^*_{AB} &\equiv -3\sum_m B^{(2)2m}X^{2m}_{AB}.
\end{align}
In terms of these quantities, we can write the metric as
\begin{align}
g_{IUU} &= -f-\e^2\R^2f^2\tilde\etide^* +O(\e^3),\\
g_{IUA} &= R\left[\tfrac{2}{3}\e^2\R^2f\left(\tilde\etide^*+\tilde\btide^*\right)+O(\e^3)\right],\\
g_{IAB} &= R^2\bigg[\Omega_{AB}-\tfrac{1}{3}\e^2\R^2\left(1-\frac{2M^2}{\R^2}\right)\tilde\etide^*_{AB}\nonumber\\
&\quad -\tfrac{1}{3}\e^2\R^2 \tilde\btide^*_{AB}+O(\e^3)\bigg].
\end{align}
This is the usual form of the metric of a tidally perturbed black hole.

Now, because $\tilde\etide^*$ and $\tilde\btide^*_a$ are, respectively, even- and odd-parity quadrupole terms, they can be written in terms of an STF decomposition:
\begin{equation}
\tilde\etide^* = \tilde\etide_{ab}\Omega^{\langle ab\rangle},\quad \tilde\btide^*_a = \epsilon_{acd}\tilde\btide^d_b\Omega^{\langle bc\rangle}.\label{Estar}
\end{equation}
By applying the definitions \eqref{tide_A def} and \eqref{tide_AB def}, we find
\begin{equation}
\tilde\etide^*_A = \Omega^a_A\tilde\etide_{ab}\Omega^b,\quad \tilde\btide^*_A = \Omega^a_A\epsilon_{acd}\tilde\btide^d_b\Omega^{\langle bc\rangle},
\end{equation}
and
\begin{align}
\tilde\etide^*_{AB} &= 2\Omega^a_A\Omega^b_B\tilde\etide_{ab}+\Omega_{AB}\tilde\etide_{ab}\Omega^{\langle ab\rangle},\\
\tilde\btide^*_{AB} &= \Omega^a_A\Omega^b_B\epsilon_{acd}\tilde\btide^d_b\Omega^c+\Omega^a_A\Omega^b_B\epsilon_{bcd}\tilde\btide^d_a\Omega^c.\label{BstarAB}
\end{align}
When written in more explicit form, these expressions agree with those in Ref.~\cite{Eric_review}. In order to arrive at these results, one requires the identities
\begin{align}
\epsilon_{AB} &= \epsilon_{abc}\Omega^a_A\Omega^b_B\Omega^c,\\
\epsilon_A{}^B\Omega^b_B &= -\Omega^a_A\epsilon_{ac}{}^b\Omega^c,\\
D_AD_B\Omega^a &= -\Omega^a\Omega_{AB}.
\end{align}

Finally, I convert to Cartesian coordinates $(U,Y^a)$, adapting the identities of Sec.~3.3.7 of Ref.~\cite{Eric_review}. The result is
\begin{align}
g_{IUU} &= -f-\e^2 f^2\R^2\ein^*+O(\e^3),\\
g_{IUa} &= -\Omega_a+\tfrac{2}{3}\e^2\R^2 f(\ein^*_a+\bin^*_a)+O(\e^3),\\
g_{Iab} &= \delta_{ab}-\Omega_{ab}-\tfrac{1}{3}\e^2\R^2\left(1-\frac{2 M^2}{\R^2}\right)\ein^*_{ab}\nonumber\\
&\quad-\tfrac{1}{3}\e^2\R^2\bin^*_{ab}+O(\e^3).
\end{align}
where $\ein_a^*=\ein_A^*\Omega^A_a$, $\bin_a^*=\bin_A^*\Omega^A_a$, $\ein_{ab}^*=\ein_{AB}^*\Omega^A_a\Omega^B_b$, and $\bin_{ab}^*=\bin_{AB}^*\Omega^A_a\Omega^B_b$. When this is expanded in the buffer region, by rewriting it in terms of the unscaled radial function $R$ and then re-expanding for small $\e$, it becomes
\begin{align}
g_{IUU} & = -1+\e\frac{2M}{R}-R^2\etide^*+4\e MR\etide^*\nonumber\\
&\quad +\order{\e^2,\e R^2,R^3}, \\
g_{IUa} & = -N_a+\tfrac{2}{3}R^2(\etide_a^*+\btide_a^*)-\tfrac{4}{3}MR(\etide_a^*+\btide_a^*)\nonumber\\
&\quad+\order{\e^2,\e R^2,R^3}\\
g_{Iab} & = \delta_{ab}-N_{ab}-\tfrac{1}{3}R^2(\etide^*_{ab}+\btide^*_{ab})\nonumber\\
&\quad+\order{\e^2,\e R^2,R^3}.
\end{align}

In order to agree with the external Lorenz gauge, I switch to harmonic coordinates via the transformation $Y^a = X^{a'}+\e MN^{a'}$, where $N^{a'}=X^{a'}/R'$ (note that, in the buffer region at first order in $\e$, this transformation to harmonic coordinates cannot be distinguished from a transformation to isotropic coordinates). I then switch from retarded coordinates to Fermi-type coordinates $(T,X^a)$ via the transformation
\begin{align}
U & = T-R-2\e M(\ln R + \tfrac{1}{6}R^2\etide_{ij}N^{ij}) \nonumber\\
&\quad+\tfrac{1}{6}R^3\etide_{ij}N^{ij}, \\
X^{a'} & = X^a - \tfrac{1}{3}R^3R^a{}_{b0c}N^{bc}+\tfrac{1}{6}R^3\etide^a_bN^b,
\end{align}
where $N^a=X^a/R$. After performing these transformations and decomposing the result into irreducible STF pieces, we arrive at  
\begin{align}
g_{ITT} & = -1+\e\frac{2M}{R}+\tfrac{5}{3}\e MR\ein_{ij}\hat N^{ij}-R^2\ein_{ij}\hat N^{ij}\nonumber\\
&\quad+\order{\e^2,\e R^2, R^3}, \\
g_{ITa} & = 2\e MR\ein_{ai}N^i+\tfrac{2}{3}\e MR\epsilon_{aij}\bin^j_k\hat N^{ik}\nonumber\\
&\quad+\tfrac{2}{3}R^2\epsilon_{aik}\bin^k_j\hat N^{ij}+\order{\e^2,\e R^2, R^3}, \\
g_{Iab} & = \delta_{ab}\left(1+\e\frac{2M}{R}-\tfrac{5}{9}\e MR\ein_{ij}\hat N^{ij}-\tfrac{1}{9}R^2\ein_{ij}\hat N^{ij}\right)\nonumber\\
&\quad +\tfrac{64}{21}\e MR\ein_{i\langle a}\hat N_{b\rangle}{}^i-\tfrac{46}{45}\e MR\ein_{ab}-\tfrac{1}{9}R^2\ein_{ab}\nonumber\\
&\quad +\tfrac{2}{3}\e MR\ein_{ij}\hat N_{ab}{}^{ij}+\tfrac{2}{3}R^2\ein_{i\langle a}\hat N^i_{b\rangle}\nonumber\\
&\quad -\tfrac{4}{3}\e MR\epsilon_{jk(a}\bin_{b)}^kN^j+\order{\e^2,\e R^2, R^3}.
\end{align}

\end{document}